\documentclass[mreferee,extra]{gji}
\usepackage[utf8]{inputenc}
\usepackage{timet,color}
\usepackage{graphicx}

\usepackage{xr}
\usepackage[urlcolor=blue,citecolor=black,linkcolor=black]{hyperref}

\usepackage[dvipsnames,usenames]{xcolor}

\usepackage[normalem]{ulem}

\title[]
  {Crustal and upper mantle model of the Middle East based on full-waveform inversion}
\author[Örsvuran et al.]
  {Rıdvan Örsvuran$^1$, Ebru Bozda{\u g}$^{1,2}$, Daniel Peter$^3$, Andrea Chiang$^{4}$, Rengin Gök$^{4}$, \\
    \LARGE{\rm Yahya M. Tarabulsi$^{5}$, Ahmed Hosny$^5$, Khalid Yousef$^5$, Abdullah Mousa$^5$} \\
    $^1$ Department of Applied Mathematics and Statistics, Colorado School of Mines, Golden, CO 80401, USA \\
    $^2$ Department of Geophysics, Colorado School of Mines, Golden, CO 80401, USA \\
    $^3$ St. Gallen, Switzerland \\
    $^4$ Lawrence Livermore National Laboratory, Livermore, CA, USA \\
    $^5$ Saudi Geological Survey, Jeddah, Saudi Arabia
  }
\date{Date}
\pagerange{\pageref{firstpage}--\pageref{lastpage}}
\volume{xxx}
\pubyear{2026}

\begin{document}

\label{firstpage}

\maketitle

\begin{summary}

We present {\bf MEAD-M20} ({\bf M}iddle {\bf E}ast {\bf AD}joint-{\bf M}odel{\bf 20}), a new tomographic model of the Middle East and its surrounding regions, including Anatolia, Iran, and the Caucasus. The model is developed within a full-waveform inversion framework, based on 3D numerical wavefield simulations and the adjoint method, after 20 conjugate gradient iterations, utilizing an extensive dataset from permanent and temporary stations available from EarthScope and regional networks, such as Kandilli Observatory, the Mesopotamian Seismological Network of Iraq, the International Institute of Earthquake Engineering and Seismology of Iran and additional data made available from the Saudi Geological Survey.

Starting from the global full-waveform inversion model GLAD-M25 \citep{lei_global_2020} on a $60^\circ \times 60^\circ$ regional mesh, we invert 210 regional earthquakes recorded by 1,215 stations to obtain the P- and S-wave model with shear-wave transverse isotropy confined to the upper mantle. During the first 12 iterations, we use two period bands, combining multitaper traveltime measurements of $15$–$50$~s body waves and $50$–$100$~s body and surface waves on three components. We use a refined crustal mesh to better sample the crust after the 12th iteration and gradually decrease the minimum surface-wave period to 30~s.

MEAD-M20 provides a self-consistent P- and S-wave model ready for seismic wave simulations, which is essential for accurate earthquake location, source parameter estimation, and seismic hazard assessment in regions such as the Middle East, where complex geology and tectonics prevail. MEAD-M20 reveals several important geodynamical and tectonic features, including local mantle plumes beneath the Arabian Plate, Jordan, and the Levant, characterized by low-velocity anomalies and likely associated with volcanism in the Harrats, Jordan, and Karacadağ regions. In addition to the active subduction and rifting in the area, the model clearly identifies remnants of the Tethys Ocean beneath Eastern Anatolia, which become progressively shallower toward the Makran region in the south, consistent with the subduction history along the Bitlis–Zagros suture zone. We also observe lithospheric-scale low-velocity anomalies associated with the North and East Anatolian faults, extending to depths of approximately 200~km.

\end{summary}


\section{Introduction}

High-resolution tomographic images of the crust and mantle are essential for understanding the tectonics and geology of the region of interest and for connecting the mantle and lithospheric dynamics to surface processes. Moreover, they are crucial for monitoring seismicity \citep[e.g.,][]{akcelik_high_2003} and estimating ground motion to mitigate seismic hazard in earthquake-prone regions \citep[e.g.,][]{pasyanos_case_2011,bethoux_earthquake_2016}, as well as providing a more accurate determination of the passive and active source parameters \citep{waldhauser_high-resolution_2002,rawlinson_simultaneous_2006}.

The Middle East and its surrounding region encompass all major types of plate boundaries, including spreading ridges, subduction zones, and continental collisions that form high mountain ranges. Active fault systems in this area frequently generate catastrophic earthquakes, particularly in Anatolia and Iran \citep[e.g.,][]{khodaverdian_seismicity_2016,karabulut_long_2023}. Therefore, many studies address the importance of accurate earthquake locations \citep[e.g.,][]{nissen_new_2011,nissen_zagros_2014,guvercin_active_2022} and source parameters \citep[e.g.,][]{adams_source_2009,gok_moment_2016,karasozen_normal_2016,karasozen_seismotectonics_2019,chiang_seismic_2021,tan_homogeneous_2021}, and improved seismic hazard estimations \citep[e.g.,][]{giardini_seismic_2018,akkar_ground-motion_2018,kiuchi_groundmotion_2019,kalafat_seismicity_2021} for mitigating earthquake risk and advancing our understanding of regional tectonics where 3D wave-propagation effects must be accounted for to address the region’s complex geology and tectonic setting \citep[e.g.,][]{braunmiller_regional_2002}.

So far, various P- and S-wave models of the Middle East are constructed based on teleseismic and regional body waves \citep[e.g.,][]{benoit_upper_2003,al-damegh_crustal_2005,park_upper_2007,park_s_2008,biryol_segmented_2011,amini_tomographic_2012,tang_lithospheric_2016,lim_asthenospheric_2020}, regional surface waves and their combination with body waves \citep[e.g.,][]{hansen_seismic_2008,chang_joint_2010,chang_new_2012,tang_shear_2019,celli_african_2020,el-sharkawy_slab_2020} and ambient noise \citep[e.g.,][]{movaghari_3-d_2020, kaviani_crustal_2020} to constrain the crustal and mantle structure. Most recent tomographic models leverage 3D numerical simulations with the adjoint method \citep{tarantola_inversion_1984, tromp_seismic_2005, fichtner_adjoint_2006, virieux_overview_2009} within the full-waveform inversion (FWI) framework.
The main advantages of FWI come from 1) taking the full 3D complexity of wave propagation into account in seismic tomography, which is crucial to increase the resolution of tomographic models of regions with complex geology and tectonics, and 2) the ability to select any wiggles in three-component seismograms, resulting in combining body and surface waves naturally, which improves data coverage significantly taking 3D finite-frequency effects into account. There are successful applications of FWI in earthquake seismology from regional \citep[e.g.,][]{tape_adjoint_2009, fichtner_deep_2013} to continental \citep[e.g.,][]{fichtner_full_2010, ciardelli_adjoint_2022, van_herwaarden_full-waveform_2023, Ziyi-adjoint2024} and global scales \citep[e.g.,][]{bozdag_global_2016,cui_glad-m35_2024, thrastarson_reveal_2024}. Recently, \citet{rodgers_adjoint_2024} constructed an elastic FWI model of the region using publicly available seismic stations, where sparse data coverage, particularly in the Arabian Peninsula, is a potential factor affecting resolution. \citet{espindolacarmona_anelastic_2024} also demonstrate the construction of an anelastic FWI model for a smaller area in the Middle East, using seismic stations from the Saudi Geological Survey (SGS). \citet{bozdag_p-wave_2025} address data coverage in the region in a P-wave arrival-time tomography by assimilating the onsets of first-arriving P waveforms from the SGS in the Arabian Peninsula, Iran, and Iraq, combined with ISC-EHB \citep{engdahl_isc-ehb_2020} data from a broader region including Anatolia and the Caucasus.

In this study, our goal is to extend the work of \citet{bozdag_p-wave_2025} for a higher-resolution crustal and upper mantle model of the Middle East and the surrounding region to shed light onto key geological and tectonic questions as well as for accurate 3D wave simulations for source modeling and characterization, by combining the advantages of a larger data set, including waveforms from the SGS stations in the Arabian Peninsula in addition to those from the permanent and temporary networks of EarthScope, Kandilli Observatory, International Institute of Earthquake Engineering and Seismology of Iran (IIES) and Mesopotamian Seismological Network of Iraq (MSPN), with the 3D complexity of wave propagation in an FWI framework. In the following section, we describe the geological and tectonic setting of the region, highlighting the key questions. In Section~\ref{sec:data_and_method}, we briefly describe the data and method used in this study. We then present our results in Section~\ref{sec:results} and discuss them in comparison with other studies. We finally summarize our results in Section~\ref{sec:conclusion}.

\section{Geological and tectonic setting}

\begin{figure}
  \centering
  \includegraphics[width=0.8\textwidth]{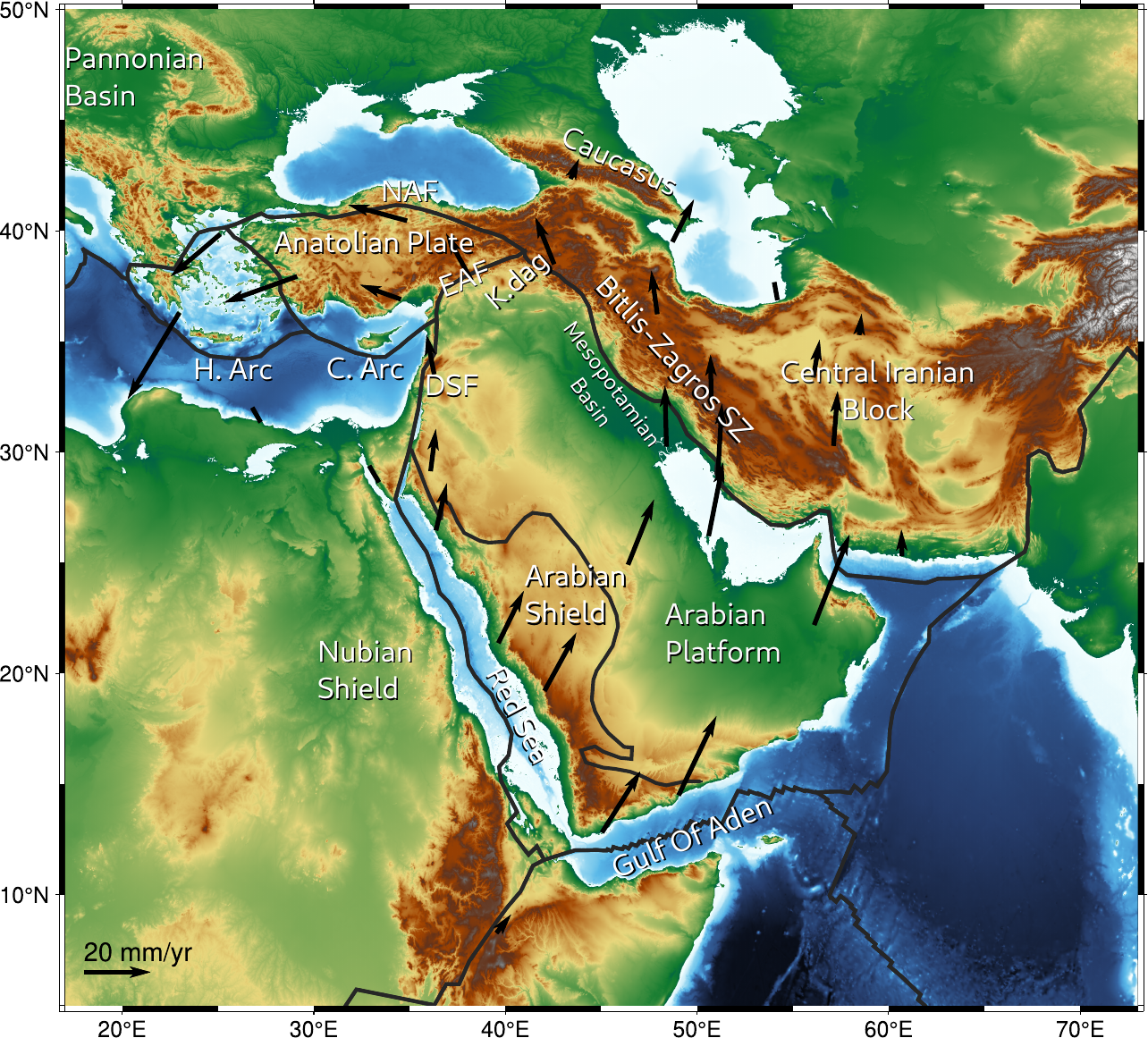}
  \caption{Topographic map of the study region with major tectonic units. NAF: North Anatolian Fault, EAF: East Anatolian Fault, DSF: Dead Sea Fault, H. Arc: Hellenic Arc, C. Arc: Cyprus Arc, K.dag: Karacadag. Black arrows denote GPS velocities relative to Eurasia \citep{reilinger_gps_2006}. Bottom left arrow shows the scale of a 20 mm/yr velocity. Black lines denote plate boundaries \citep{bird_updated_2003} and the boundary between the Arabian shield and the Arabian platform \citep{mokhtar_shear_1994}. The scale for the GPS vectors is shown in the bottom-left corner.}
  \label{fig:region_features}
\end{figure}

The Middle East is one of the most interesting regions located on the Arabian plate, which is bounded by spreading ridges along the Red Sea and the Gulf of Aden on the west, continental collision along the Bitlis-Zagros suture zone on the east and north, and a transform boundary marked by the Dead Sea fault on the northwest due to northward motions of the Arabian and African plates with different velocities. The major tectonic units of the study area are shown on a topographic map in Fig.~\ref{fig:region_features}, along with GPS velocities relative to the Eurasia fixed reference frame.

The driving forces behind the plate motions and volcanism in the Arabian Peninsula, Anatolia, and Africa are actively debated \citep[e.g.,][]{mcclusky_global_2000, wortel_subduction_2000, hubert-ferrari_morphology_2002, reilinger_gps_2006}. The major tectonic movements in the region are controlled by the subduction of the African plate along the Hellenic and Cyprus Arcs, spreading along the Red Sea and the Gulf of Aden, and collision along the Bitlis-Zagros suture zone. The spreading along the Red Sea and the Gulf of Aden, and subduction along the Makran, cause the Arabian Plate to move northward with a counterclockwise rotation relative to Eurasia \citep[e.g.,][]{reilinger_gps_2006, viltres_present-day_2022}. The motion of the Arabian plate has led to the formation of the Dead Sea and East Anatolian faults, and the collision along the Zagros suture zone, which has given rise to the Zagros mountains \citep{reilinger_gps_2006, pichon_miocene--present_2010, mouthereau_building_2012}.
This tectonic regime causes the westward escape of the Anatolian plate along the North and East Anatolian faults \citep[e.g.,][]{akbayram_evidence_2016}, which, together with the continental collision along Zagros \citep{pichon_miocene--present_2010} following the closure of the Neo-Tethys ocean \citep[e.g.,][]{agard_convergence_2005,hatzfeld_comparisons_2010}, create destructive earthquakes in the region \citep{reilinger_nubiaarabiaeurasia_2011}. Moreover, the subduction of the African plate along the Hellenic and Cyprus Arcs \citep{sengor_tethyan_1981,royden_slab_2011,menant_kinematic_2016} and the related slab roll-back drive the counterclockwise rotational motion of the Anatolian plate  \citep{royden_tectonic_1993,mcclusky_global_2000} and are responsible for the extensional regime in the Aegean region \citep{bozkurt_neotectonics_2001, menant_3d_2016}.

The topographic variations of the Arabian plate are suggested to be connected to the upper mantle structure, where the Afar plume and possible other regional plumes \citep{park_upper_2007,chang_mantle_2011,civiero_complex_2022} are key mantle features controlling the dynamics of the region
\citep{daradich_mantle_2003}.
The formation of the rifting processes and related tectonics, as well as the source of volcanism in the region, are still largely debated. The Red Sea is one of the slowest-spreading ridges, resulting in thick sediment layers that conceal potential surface faults and obscure the ridge features \citep{hansen_mantle_2012, augustin_13_2021, parisi_first_2024}, making it challenging to understand the complete rifting process. Current theories for the cause of rifting are generally associated with the rotational motion of the Arabian plate, potential changes in plate velocities and motions due to subduction along Zagros \citep[e,g.,][]{reilinger_nubiaarabiaeurasia_2011} or potentially due to mantle plumes, either the Afar plume \citep[e.g.,][]{montelli_global_2004}, or some other potential local plumes \citep[e..g,][]{chang_mantle_2011}.

Similarly, the source of active volcanism in the Middle East and Anatolia is still not well understood. The Harrats located in the Arabian Shield on the west is the largest lava field in the region with active volcanism \citep[][]{camp_madinah_1987} and potentially associated with seismic swarms \citep[e.g.,][]{pallister_broad_2010, mukhopadhyay_incipient_2013}. Its formation is proposed to be related to the rifting processes along the Red Sea \citep{rashed_new_2020,abdelfattah_key_2021}. Another theory for the source of volcanism in the Harrats is the channeled mantle flow originating from the Afar plume in the south \citep[e.g.,][]{camp_upwelling_1992, duncan_timing_2016} or the potential regional mantle upwellings in the Arabian plate and Jordan \citep{weinstein_role_2006, chang_mantle_2011}. Some studies also report channeled mantle
flow in the eastward direction via Aden-Sheba ridge system
\citep{leroy_recent_2010} and in the northeast direction
\citep{chang_mantle_2011} to also explain the volcanism in Jordan. The closure of the Neo-Tethys ocean along the Bitlis-Zagros suture zone and the subduction of the Mediterranean along the Hellenic and Cyprus arcs are likely the leading cause of volcanism in Iran and Anatolia \citep{pichon_miocene--present_2010,agard_zagros_2011}. However, there are also alternative theories, such as to explain the volcanism near Karacada\u g, if the Afar plume might be the driving force of volcanism all the way up to Eastern Anatolia \citep[e.g.,][]{keskin_geochronology_2012,faccenna_mantle_2013}. This theory is challenged by \citet{ekici_polybaric_2012,ekici_foreland_2014}. Another potential explanation is a local plume proposed in the Levant area close to Karacada\u g \citep[e.g.,][]{civiero_complex_2022}.

\section{Adjoint Tomography Workflow}\label{sec:data_and_method}

In this study, we closely follow the adjoint tomography workflow described in \citet{bozdag_global_2016} and \citet{lei_global_2020} while performing 20 conjugate gradient iterations. We use the 3D global seismic wave propagation solver \texttt{SPECFEM3D Globe} package \citep{komatitsch_spectral-element_2002-1, komatitsch_spectral-element_2002} for numerical simulations, and the associated open-source FWI tools for pre-and post-processing steps. In this section, we provide brief information on the data, the adaptation of the adjoint tomography workflow to our study region, the measurements, and the optimization, along with the rationale for the decisions made to ensure transparency and reproducibility of our results.

\subsection{Data}

\begin{figure}
  \centering
    \includegraphics[width=1.0\textwidth]{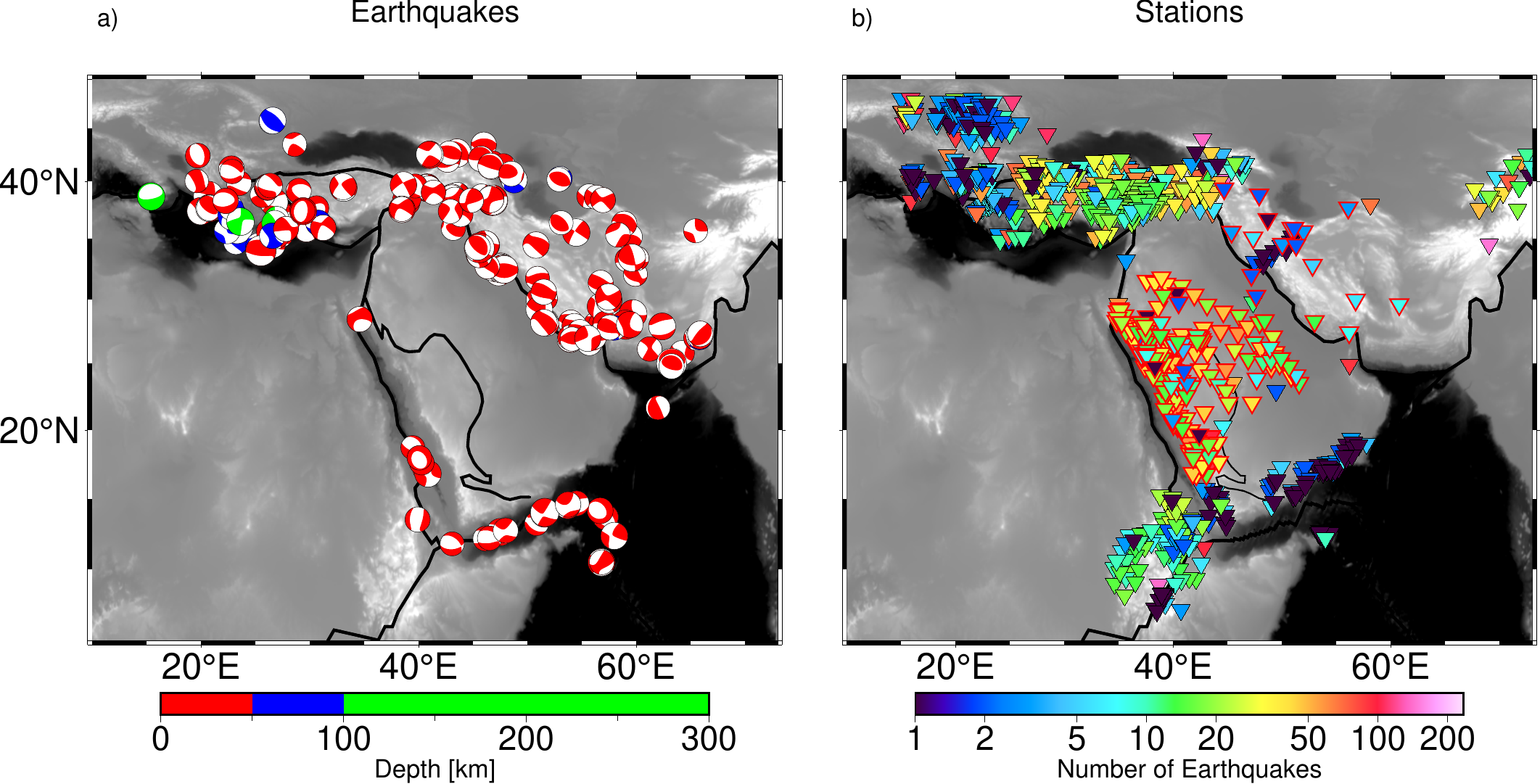}
    \caption{Distribution of a) 210 earthquakes selected from the GCMT catalog and b) permanent and temporary seismic stations used in this study. Publicly available and SGS stations are denoted by triangles with black and red frames, respectively.}
    \label{fig:source_and_stations}
\end{figure}

In this study, we gather broadband waveform data of 210 earthquakes (Fig.~\ref{fig:source_and_stations}a) from the \href{globalcmt.org}{GCMT} (Global Centroid Moment Tensor) catalog (see the Supplementary Material for the list of selected events). We select the events based on their geographical distribution and within the moment magnitude range of $M_w=$5.0-6.7, ensuring their half durations are smaller than the minimum period of our numerical simulations ($\sim 12$~s) to minimize finite-source effects. We select seismic stations from permanent and temporary broadband stations in our study region, considering their instrument responses (Fig.~\ref{fig:source_and_stations}b). We use data from 1215 stations in total. 956 of them are from 53 seismic networks whose data are publicly available from EarthScope and local networks such as Kandilli Observatory of T\"urkiye, providing good data coverage in Anatolia (see Appendix \ref{appendix:data} for references). In addition, to improve data coverage specifically in the Arabian Peninsula, we utilize waveform data recorded by 239 SGS stations, as well as 3 and 17 stations from the Mesopotamian Seismological Network of Iraq (MSPN), and the International Institute of Earthquake Engineering and Seismology of Iran (IIES), respectively.

\subsection{Starting model and mesh}
\label{startingmodel}

The starting model for our inversion is GLAD-M25 \citep{lei_global_2020}, which is a P- and S-wave model with transverse isotropy of shear waves confined to the upper mantle. GLAD-M25 is an FWI model constructed using 3D spectral-element simulations and the adjoint method. During the construction of GLAD-M25, the crust and mantle are inverted simultaneously, avoiding crustal corrections that are reported to potentially bias upper mantle structure and anisotropy \citep[e.g.,][]{bozdag_crustal_2008, ferreira_robustness_2010}. GLAD-M25 is sampled on a mesh of NEX=256 (NEX is the number of spectral elements at the surface on each side of each of six $90^{\circ}\times 90^{\circ}$ chunks that form the globe), where the size of each spectral element at the surface is $\sim 39$~km,
giving resolution in numerical simulations down to $\sim 17$~s. During the construction of GLAD-M25, frequency-dependent traveltime measurements of body waves down to $\sim 20$~s are combined with longer-period body waves and minor- and major-arc Rayleigh and Love-waves down to $\sim 40$~s on three components (vertical, radial, and transverse). We use a $60^{\circ}\times 60^{\circ}$~chunk of GLAD-M25, centered on the Middle East ($30.0^{\circ}$N $45.0^{\circ}$E) as the starting model for our continental-scale inversion, keeping the original resolution of the mesh (NEX=256), which leads to the surface spectral-element size of $\sim 26$~km. The size of the spectral elements is doubled first at Moho, then at the 660-km discontinuity. We retain the original parameterization of GLAD-M25 during our adjoint iterations over our study region (see Section~\ref{parameterization} for details).

GLAD-M25 uses the mesh originally designed for PREM \citep{dziewonski_preliminary_1981}, which honors first-order internal discontinuities \citep{komatitsch_spectral-element_2002-1,komatitsch_spectral-element_2002}. In the crustal part, the crustal thickness variations from Crust2.0 \citep{bassin_current_2000} are honored if the Moho depth is less than 15~km (predominantly the oceanic domain) and greater than 35~km (predominantly the continental domain) as described in \citet{tromp_near_2010}, where the oceanic and continental crusts are represented by one and two spectral elements in the vertical direction, respectively. The choice of such a mesh is primarily aimed at speeding up simulations while ensuring sufficient sampling in the thin oceanic crust for global simulations. We initiate our iterations with the same mesh. To obtain a better sampling of the continental crust, which dominates our study region, we increase the number of spectral elements in the vertical direction to three in the crust after 12 iterations (see Fig.~\ref{figS:mesh}).

\subsection{Parameterization}
\label{parameterization}

Following our starting model GLAD-M25, we use a transversely isotropic model parametrization for S waves, keeping the P-wave model isotropic, with 5 parameters: bulk sound speed ($c$), vertically ($\beta_v$) and horizontally ($\beta_h$) polarized shear wavespeeds, the dimensionless anisotropic parameter ($\eta$), and density ($\rho$). The gradient of the misfit function is given as

\begin{equation}
  \label{eq:parametrization}
  \delta \chi = \int_V K_c~\delta \ln{c} + K_{\beta_v}~\delta \ln{\beta_v} + K_{\beta_h}~\delta \ln{\beta_h} + K_{\eta}~\delta \ln{\eta}~dV~,
\end{equation}

\noindent where $K_c$, $K_{\beta_v}$, $K_{\beta_h}$ and $K_{\eta}$ are the Fr\'echet derivatives of the corresponding the model parameters. Note that, following the parameterizations of GLAD models \citep{bozdag_global_2016, lei_global_2020, cui_glad-m35_2024}, we prefer to scale the density from the isotropic shear wavespeed model ($\beta$) using the relation $\delta{\rm ln}\rho=0.33\delta{\rm ln}\beta$ \citep{montagner_petrological_1989} where $\beta=\sqrt{(2\beta_v^2+\beta_h^2)/3}$, to minimize the trade-off between wavespeeds and density throughout iterations.

\subsection{Numerical simulations}

We use the \texttt{SPECFEM3D Globe} software package \citep{komatitsch_spectral-element_2002-1} for forward and adjoint simulations to compute 3D synthetic seismograms and corresponding 3D data sensitivity kernels (also called Fr\' echet or adjoint kernels), respectively. As described in Section~\ref{startingmodel}, we perform continental-scale numerical simulations on a $60^{\circ} \times 60^{\circ}$ single chunk down to the core. During the numerical simulations, we account for topography, bathymetry, self-gravity (Cowling approximation), ellipticity, rotation, and 1D attenuation (QL6; \citet{durek_radial_1996}) to capture the 3D complexity of wave propagation. The minimum resolvable period for the NEX=256 mesh used in the study is approximately 12~s. We run our simulations on the Texas Advanced Computing Center (TACC)'s `Frontera' system, utilizing 256 CPU cores per event. We typically bundle simulations of 210 earthquakes in a single batch job to run forward and adjoint simulations for all events in our catalog, where forward and adjoint simulations take approximately 20~m and 45~m, respectively. We also include full attenuation in our adjoint simulations \citep{komatitsch_anelastic_2016}.

\subsection{Pre-processing stage}
\label{sec:measurements}

Following \citep{bozdag_global_2016}, our FWI workflow consists of three stages: numerical simulations, pre- and post-processing. The pre-processing stage involves data processing, measurements, and the computation of adjoint sources for adjoint simulations. Once we perform forward simulations, we first apply a set of data processing steps (i.e., tapering, removing the mean and trend, resampling, deconvolving the instrument response from the observed data, filtering, and windowing) to both the observed and corresponding synthetic data to ensure robust comparisons.

In this study, we make measurements on two-period bands: a shorter period band of 15-50~s for body-wave measurements and a longer period band of 50-100~s for both body- and surface-wave measurements. Although the minimum period of surface waves used in the construction of our starting model GLAD-M25 is 40~s, considering the sparse coverage in our study region, which degrades the resolution in the global inversion, we initiate the iterations starting with longer-period surface waves down to 50~s and gradually decrease the minimum period to 30~s as the iterations proceed to better deal with potential cycle skips in measurements. We use frequency-dependent (multitaper) cross-correlation traveltime misfit \citep{laske_constraints_1996}, normalized by the estimated standard deviations of the cross-correlations for each selected measurement window, which is a robust and well-demonstrated phase measurement in previous continental- and global-scale adjoint tomography studies \citep{zhu_seismic_2015,bozdag_global_2016, lei_global_2020}. Note that when the time windows are small, or the waveforms are not dispersive, the measurements become classical cross-correlation traveltimes \citep{luo_wave-equation_1991}. We repeat our measurements at every iteration by recomputing synthetics with the updated model. Following \citet{ruan_balancing_2019}, we balance our measurements with appropriate weightings:

\begin{equation}
  \label{eq:weighting}
  \phi =
  \sum_{s=1}^S\omega_s \sum_{c=1}^{C} \omega_c
  \sum_{r=1}^{R_{sc}} \omega_{scr} \sum_{w=1}^{N_{scr}} \omega_{scrw} \chi_{scrw}
\end{equation}

\noindent where $s$, $c$, $r$, and $w$ denote the source, category,
receiver, and measurement terms. Capital letters denote the total number of said parameters. $\omega$ values are the weighting terms and $\chi$ is the misfit value. We have six measurement categories ($c$) based
on two-period bands and three components (vertical, radial, and transverse), and the category weighting
$\omega_c$ is defined as the inverse of the number of
measurements in each category to balance body- and surface-wave measurements. $\omega_{s}$ and $\omega_{scr}$ address the uneven distribution of sources and receivers by accounting for their geographical distribution (i.e., regions with clusters of sources and receivers are down-weighted and vice versa). Note that the receiver weights $\omega_{scr}$ are calculated for each source and measurement category. Individual measurements at each station are weighted equally (i.e., $\omega_{scrw}$ is set to 1 for all measurements).

To maximize information from our data with the multitaper traveltime misfit function \citep{bozdag_misfit_2011}, we select measurement windows with the Python version of the automated window selection algorithm \texttt{FLEXWIN} \citep{maggi_automated_2009} on three components, which we also utilize to check data quality. Sample measurement windows selected from our three-component data at two period bands, for a sample ray path used in our inversions, are shown in Fig.~\ref{figS:meas_windows}. At the last iteration, we have a total of 87,399 measurements across six measurement categories (see Table~\ref{tab:n_meas}). As a final step in our pre-processing stage, we calculate multitaper traveltime adjoint sources at each iteration to initiate adjoint simulations for the gradient calculations \citep[e.g.,][]{tromp_seismic_2005, bozdag_misfit_2011}.

\begin{table}
\caption{Number of measurements in each measurement category by three components and two period bands throughout iterations. The shorter-period band (15-50~s) includes body waves only, whereas the longer-period band (50(30)-100~s) includes both body- and surface-wave measurements, with the minimum period gradually decreasing to 30~s at the 18th iteration. For the first two stages, the half-width of the horizontal and vertical smoothing is set to 80~km and 10~km, respectively. For the final stage, the horizontal half-width is reduced to 50 km while the vertical smoothing remains the same.}
\label{tab:n_meas}
\begin{tabular}{l|lll|l}
\textbf{M00-M12}   & \textbf{Vertical} & \textbf{Radial} & \textbf{Transverse} & \textbf{Total} \\ \hline
\textbf{T015-050s} & 16,658             &  9,946           & 13,120               & 39,724      \\
\textbf{T050-100s} & 11,117             &  1,856           &  4,521               & 17,494      \\ \hline
\textbf{Total}     & 27,775             & 11,802           & 17,641               & 57,218 \\

\textbf{M12-M18}   &  \\ \hline
\textbf{T015-050s} & 25,144             & 15,677           & 15,161               & 55,982      \\
\textbf{T035-100s} & 15,636             &  3,668           &  9,491               & 28,795      \\ \hline
\textbf{Total}     & 40,780             & 19,345           & 24,652               & 84,777 \\

\textbf{M18-M20}   &  \\ \hline
\textbf{T015-050s} & 25,240             & 15,709           & 15,167               & 56,116      \\
\textbf{T030-100s} & 15,733             & 4,368            & 11,182               & 31,283      \\ \hline
\textbf{Total}     & 40,973             & 20,077           & 26,349               & 87,399
\end{tabular}
\end{table}

\subsection{Post-processing stage}

This stage involves processing the computed gradient (e.g., smoothing and pre-conditioning), performing a line search, and updating the model at each iteration.

\subsubsection{Smoothing}
Similar to \citet{bozdag_global_2016}, we use a Gaussian
smoothing in horizontal and vertical directions, where we choose the
half-width, considering our target minimum wavelength and data coverage. We set the horizontal and vertical half-widths of the Gaussian operator to 80~km and 10~km, respectively, at the start of the iterations. After the 18th iteration, we decrease the horizontal half-width to 50~km while keeping the vertical half-width unchanged.

\subsubsection{Pre-conditioning}

We use the following pseudo-Hessian formulation from \citet{luo_seismic_2012} as a pre-conditioner to speed up convergence by further balancing our gradient as a result of uneven source and receiver distribution:

\begin{equation}
  \label{eq:preconditioner}
  P(\mathbf{x}) = \sum_s \int \partial_t^2 \mathbf{s}(\mathbf{x}, t) \cdot
                              \partial_t^2 \mathbf{s}^\dagger(\mathbf{x}, T-t) {\rm d}t~,
\end{equation}

\noindent where $\mathbf{s}$ and $\mathbf{s}^\dagger$ correspond to forward and adjoint displacements, respectively.

\subsubsection{Line search and optimization}

We use the conjugate-gradient optimization during our inversion, defined as

\begin{equation}
  \label{eq:update}
  \textbf{m}_{i+1} = \textbf{m}_{i} \exp{\left( \alpha \textbf{d}_i \right)} \, ,
\end{equation}

\noindent where $\mathbf{m}$ and $\mathbf{d_i}$ are the model parameters and the conjugate gradient direction, respectively.
The step length $\alpha$ is determined by line search, which is performed with various step lengths (at least 3 to capture the parabolic shape of the misfit function) to find an optimum value for the individual
misfit categories and the total misfit using a spatially
representative subset of our events (50 out of 210 events) to
decrease computational cost and speed up the process.

\noindent The gradient direction $\mathbf{d}$ is defined as

\begin{equation}
  \label{eq:direction}
  {\mathbf{d_i}} = - \mathbf{g}_i + \beta \mathbf{d_{i-1}} \, ,
\end{equation}

\noindent where $\mathbf{g}$ is the misfit gradient and $i$ denotes the iteration number. $\beta$ is given as \citep{fletcher_function_1964}

\begin{equation}
  \label{eq:conj_beta}
  \beta = \frac{\mathbf{g}_i^T \cdot \left( \mathbf{g}_i - \mathbf{g}_{i-1} \right)}
  {\mathbf{g}_{i-1}^T \cdot\mathbf{g}_{i-1}} \, .
\end{equation}


\section{Results}\label{sec:results}

\begin{figure}
    \centering
    \includegraphics[width=1.0\textwidth]{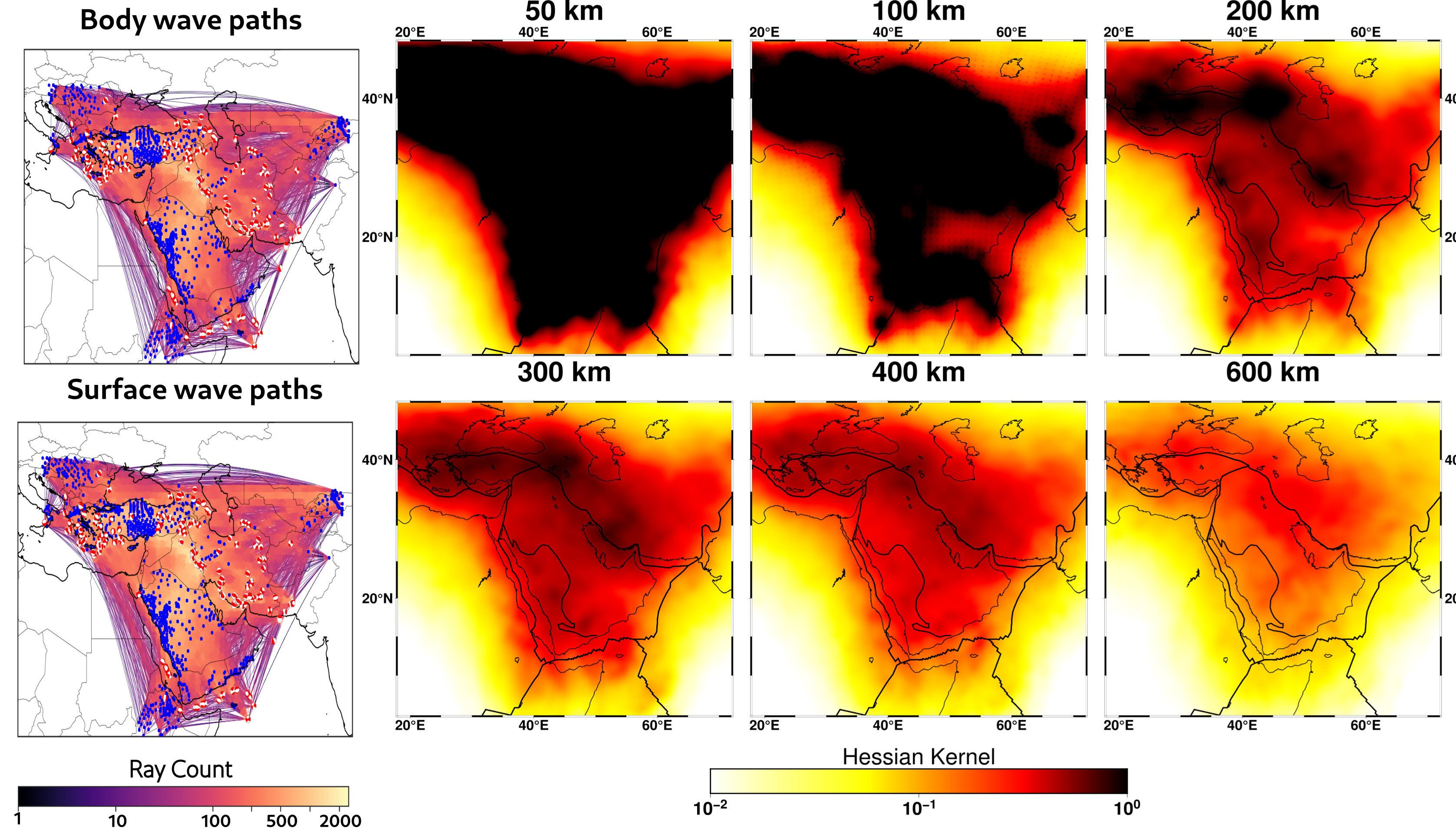}
    \caption{{\it Left:} Ray densities for body and surface waves, where at least one measurement is made for each source-receiver pair (blue dots: stations, red beachballs: earthquakes). {\it Right:} Normalized pseudo-Hessian kernels approximately showing data coverage, which decreases as a function of depth.}
    \label{fig:hess_kl}
\end{figure}

Starting from GLAD-M25, we initiate iterations with 57,218 body- and surface-wave measurements, which increase to 87,399 after 18 iterations due to gradual model improvement (see Table~\ref{tab:n_meas}). Fig.~\ref{fig:hess_kl} shows the body- and surface-wave ray paths, where at least one measurement window is selected, together with the pseudo-Hessian kernel as a function of depth computed by eq.~\ref{eq:preconditioner} by setting traveltime measurements to 1 at the 18th iteration, which mimics our approximate kernel coverage. The pseudo-Hessian kernel shows that we have the best source-receiver distribution in Anatolia. Data coverage in Iran is also reasonably good, largely due to the region's high seismic activity. The inclusion of the SGS stations improves sampling beneath the Arabian Peninsula, providing reasonable coverage but less than in the surrounding area. The coverage worsens to the south of the Gulf. The kernel coverage gradually decreases with depth, a characteristic of finite-frequency kernels, whose sensitivity is largest around the source and receivers and minimum in the middle. The coverage below $\sim 300$~km depth also decreases because sampling is primarily based on body waves. Note that we use the pseudo-Hessian kernels computed at each iteration as pre-conditioners, resulting in more balanced gradients between the crust and the upper mantle, as well as laterally.

\begin{figure}
    \centering
    \includegraphics[width=1.0\textwidth]{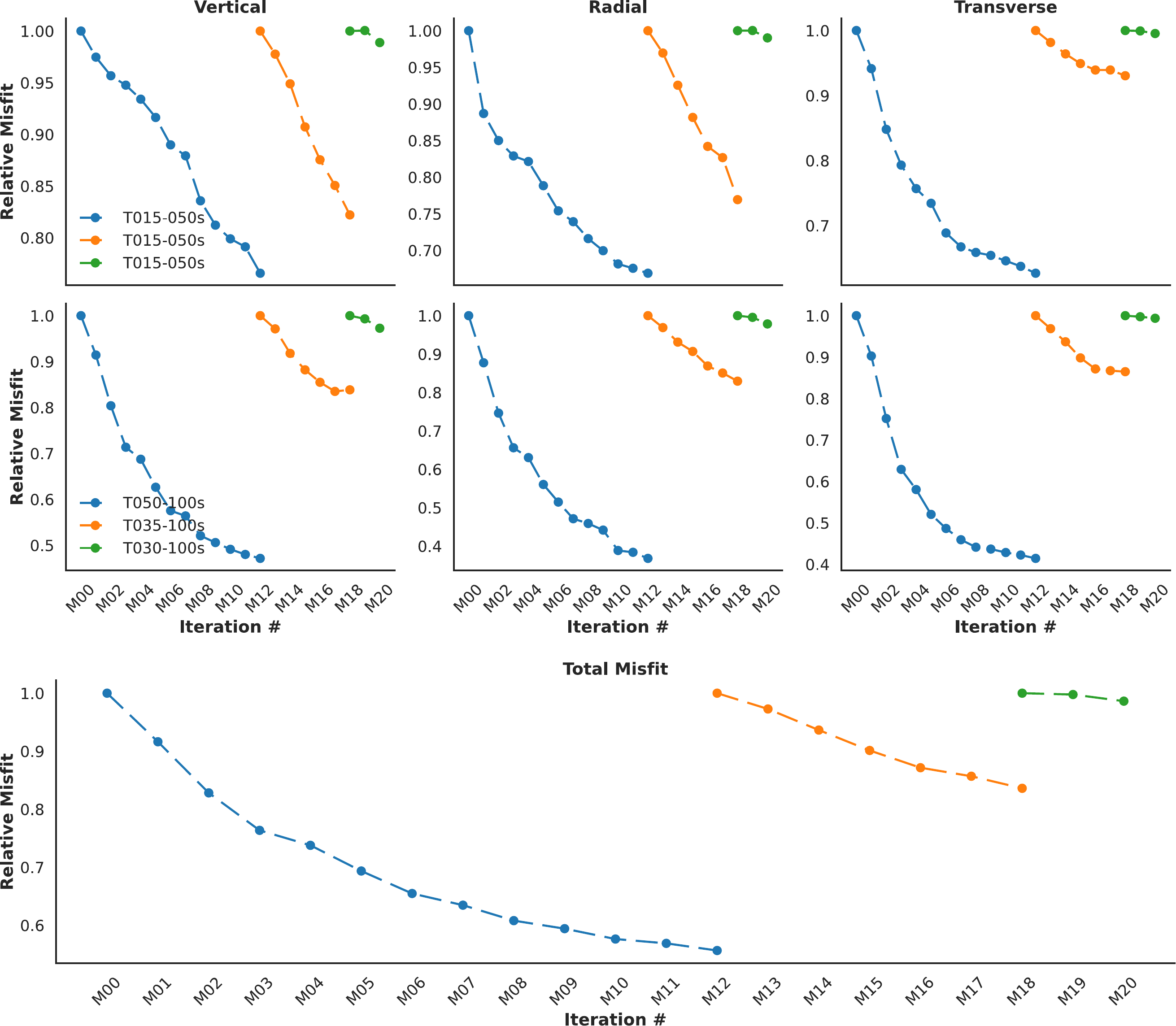}
    \caption{Misfit change across six measurement categories {\it (top)} and the total misfit reduction {\it (bottom)}. {\it Blue lines:} first 12 iterations with 15-50~s body waves and 50-100~s body and surface waves. {\it Orange lines:} next 6 iterations where the longer-period band is set to 35-100~s. {\it Green lines:} last 2 iterations when the longer-period band is set to 30-100~s. New measurement windows are included whenever the period band is changed. Each misfit value is normalized by the corresponding starting model for each period band, as denoted by blue, orange, and green colors.}
    \label{fig:misfit_change}
\end{figure}

\begin{figure}
    \centering
    \includegraphics[width=1.0\textwidth]{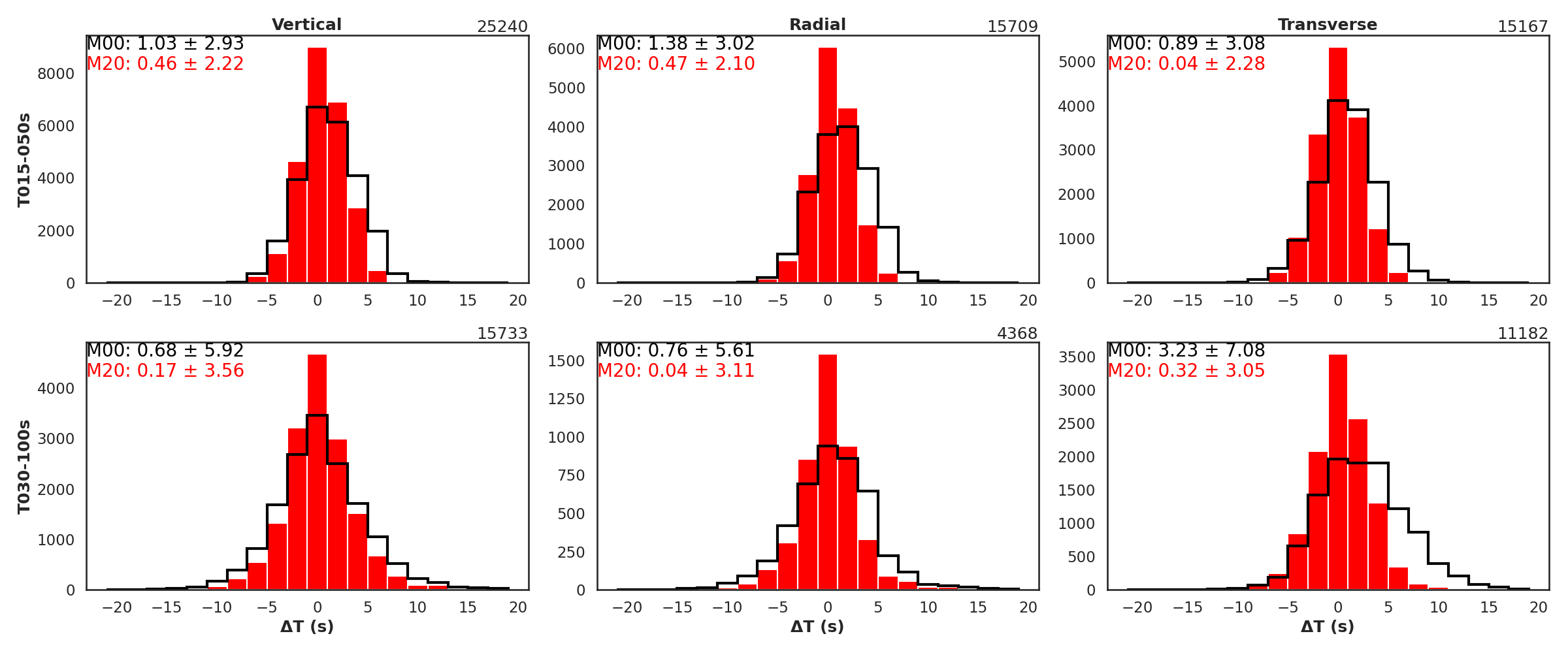}
    \caption{Cross-correlation traveltime histograms for six measurement categories for the initial model GLAD-M25 (M00, black) and the final model MEAD-M20 (M20, red). The mean and standard deviation of the histograms, along with the number of measurements, are displayed at the top left and top right of each plot, respectively.}
    \label{fig:dt_meas}
\end{figure}

\begin{figure}
    \centering
    \includegraphics[width=1.0\textwidth]{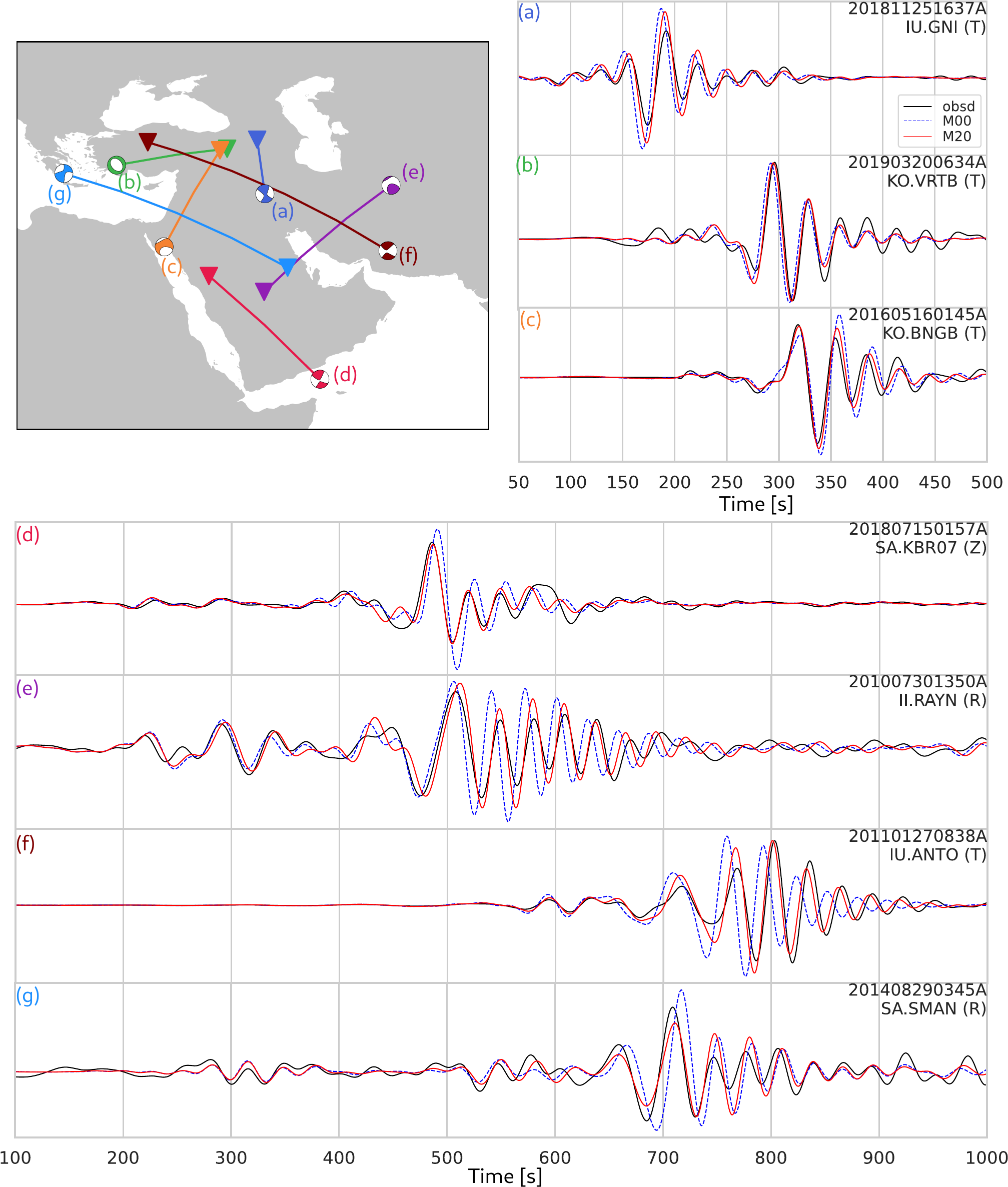}
    \caption{Sample selected waveforms (bandpassed between 30-100~s) of observed (black) and synthetics computed for the starting model GLAD-M25 (blue) and MEAD-M20 (red). The respective event IDs from the GCMT catalog and station name and components are given at the top
      right of the waveform plots. The ray paths for each event (beachballs) and station (triangles) are as shown on the map.}
    \label{fig:waveforms_selected}
\end{figure}

Fig.~\ref{fig:misfit_change} shows the total normalized misfit reduction with respect to the misfit of the starting model (GLAD-M25, M00) and misfits for 6 measurement categories (i.e., two period bands and three components) after 20 iterations. We prefer performing a line search at each iteration to ensure a reduction in misfit for each measurement category, thereby avoiding the dominance of a particular measurement group in the model updates. After 12 iterations, we observe a decrease of approximately $50\%$ in the misfit relative to the initial misfit. After the 12th iteration, the decrease in misfits gradually slows down. The misfit flattens after the 18th iteration, which is the reason for stopping iterations after 20 iterations with the current period range and inversion parameters. The improvement in waveforms at each measurement category can be seen in cross-correlation traveltime measurement histograms in Fig.~\ref{fig:dt_meas}. The traveltime histograms become noticeably narrower and centered around zero, particularly for the transverse component. Since we use traveltime measurements only in our iterations, we do not observe a drastic change in the amplitude histograms (see Fig.~\ref{figS:supp-dlna_meas}), which likely highlights the importance of addressing attenuation and source uncertainties as well as higher sensitivity of amplitudes to elastic heterogeneities \citep[e.g.,][]{laske_constraints_1996}. The improvement in waveforms is the most visible for surface waves across the Arabian Peninsula. In Fig.~\ref{fig:waveforms_selected}, we show waveforms from sample source-receiver pairs across our model, showing the improvement after 20 iterations (see Figs.~\ref{figS:supp-waveforms_body_waves}~-~\ref{figS:supp-waveforms-event3-all} for additional comparisons of body and surface waves).

\subsection{Resolution tests}

\subsubsection{Point-spread function tests}

\begin{figure}
    \centering
    \includegraphics[width=1.0\linewidth]{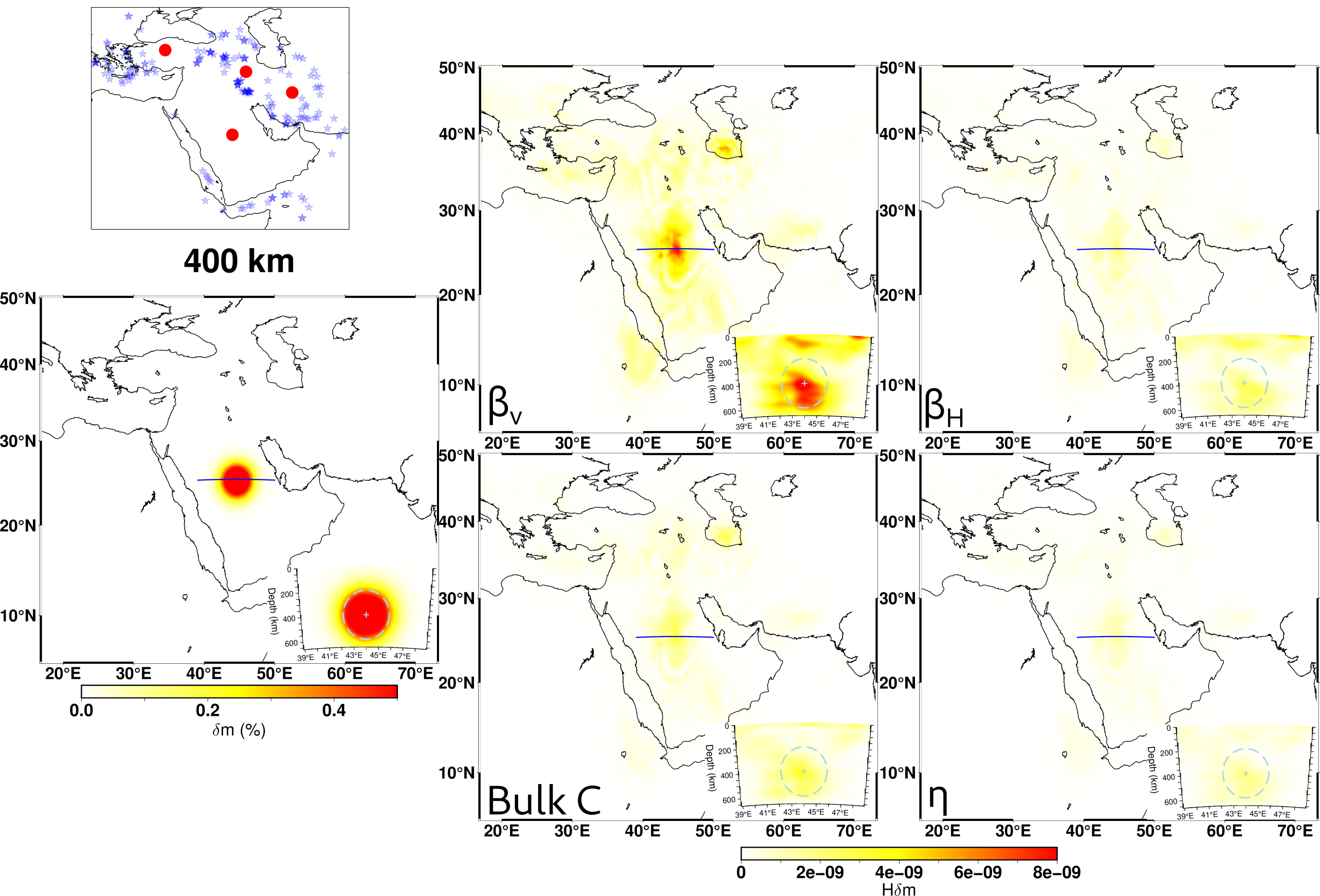}
    \caption{\label{fig:me_arabian_results} {\it Top left:} Locations of the PSF tests (red dots) conducted in this study. {\it Bottom left:} $\beta_v$ Gaussian perturbation at 400~km depth (half-width of 200~km) located on the Arabian Peninsula. {\it Right:} PSF test results showing the recovery of the Gaussian perturbation for $\beta_v$ and the trade-off with the $\beta_h$, $\textrm{bulk c}$, $\eta$ parameters. See the Supplementary Material for the PSF test results at all locations at 150~km, 400~km, and 600~km.}
    \label{fig:psf-arabia}
\end{figure}

To assess the resolution of our model and the trade-off between model parameters, we perform point-spread function (PSF) tests \citep{fichtner_resolution_2011} at four locations (see the top left map in Fig.~\ref{fig:psf-arabia} for PSF-test locations) at 150~km, 400~km, and 600~km depths with the dataset and measurements used during the construction of MEAD-M20. PSF tests are performed by calculating the difference between the perturbed and non-perturbed gradient using a finite-difference approximation to obtain the action of the Hessian at the location of the perturbation, given by

\begin{equation}
  \label{eq:psf}
  \mathbf{H}\cdot \delta \mathbf{m} \approx \textrm{g}(\mathbf{m} + \delta \mathbf{m}) -  \textrm{g}(\mathbf{m}) \, ,
\end{equation}

\noindent where $\mathbf{H}$ denotes the Hessian, $\delta \mathbf{m}$ is the perturbation in model parameters, and $\textrm{g}$ is the misfit gradient for both the unperturbed ($\textrm{g}(\mathbf{m})$) and perturbed
($\textrm{g}(\mathbf{m} + \delta \mathbf{m})$) models. In Fig.~\ref{fig:psf-arabia}, we show our PSF test results at 400~km in the Arabian platform close to the boundary with the Arabian shield (see also Table~\ref{tabS:supp-me_psf_loc}). We perturb the vertically-polarized shear-wave model ($\beta_v$) of MEAD-M20 with a Gaussian anomaly of 200~km half-width. The results show a decent recovery of the perturbation for $\beta_v$, with some smearing in the East-West direction, and a slight, though not significant, trade-off with other parameters, which provides confidence in the resolution of features and transverse isotropy in the region. We obtain better horizontal resolution in the PSF test at 150~km at the same location (see Fig.~\ref{figS:psf-arabia-150km}), where we observe some smearing to the crust. The PSF tests conducted at other locations in Anatolia, Zagros, and Iran at 150~km, 400~km, and 600~km are presented in the Supplementary Material, Figs.~\ref{figS:psf-iran-150km}-\ref{figS:psf-arabian-600km}, which give an idea of the expected resolution considering the data coverage in the region. Overall, we observe a similar trend at all locations, except in Zagros, where the results show a slightly more scattered pattern at 150~km. The PSF tests at 600~km show decent coverage down to the lower mantle. Still, resolution decreases as expected, since the coverage relies solely on body waves. At all locations where we conduct PSF tests, we do not observe significant trade-offs among seismic parameters.

\subsubsection{Tests with an independent data set}

\begin{figure}
  \centering
    \includegraphics[width=1.0\textwidth]{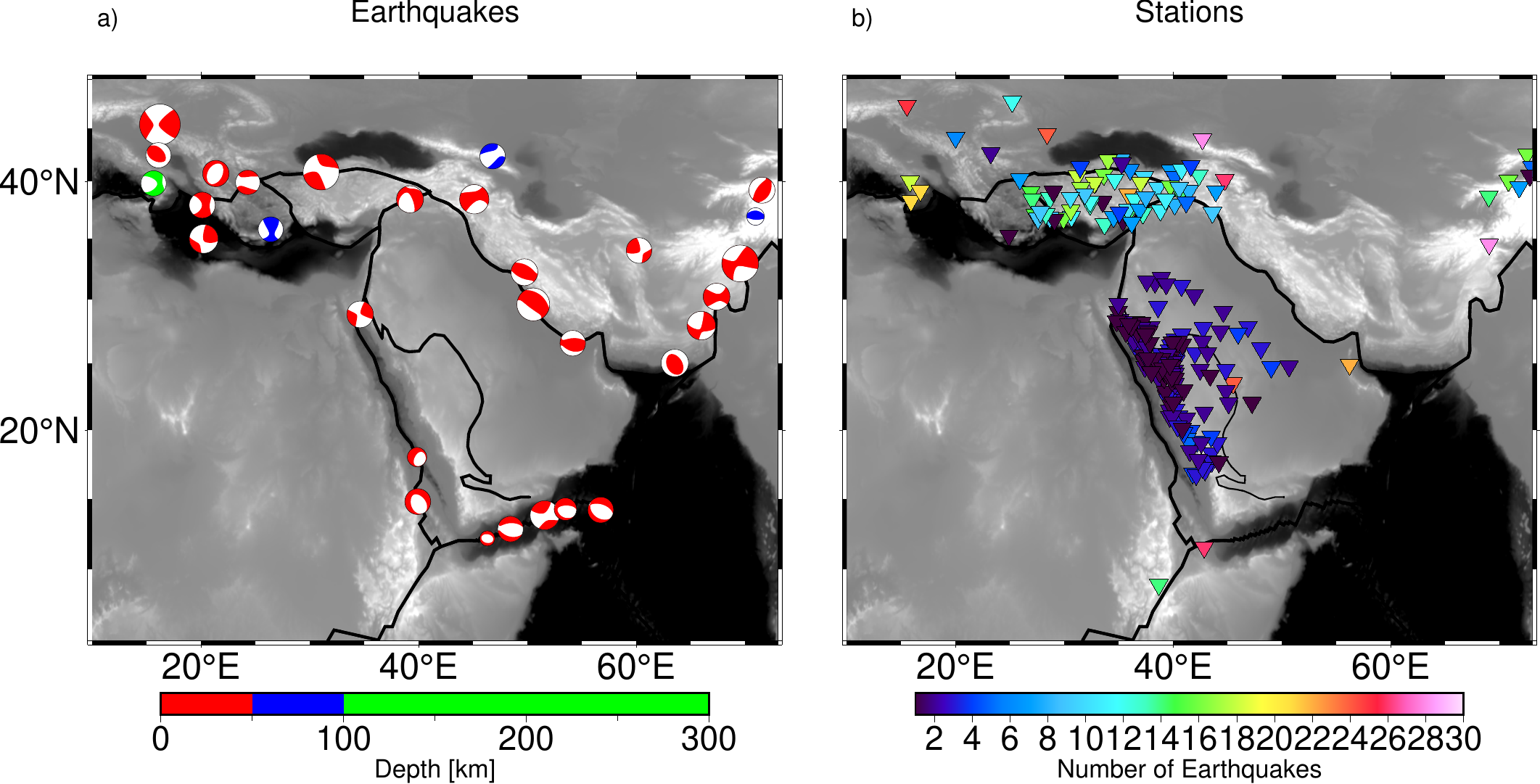}
    \caption{Distribution of 30 earthquakes {\it (a)} and stations {\it (b)} that are not used in the inversion to assess the improvement in travelltimes in Fig.~\ref{fig:benchmark_dt_hist}.}
    \label{fig:benchmark_source_and_stations}
\end{figure}

\begin{figure}
  \centering
    \includegraphics[width=1.0\textwidth]{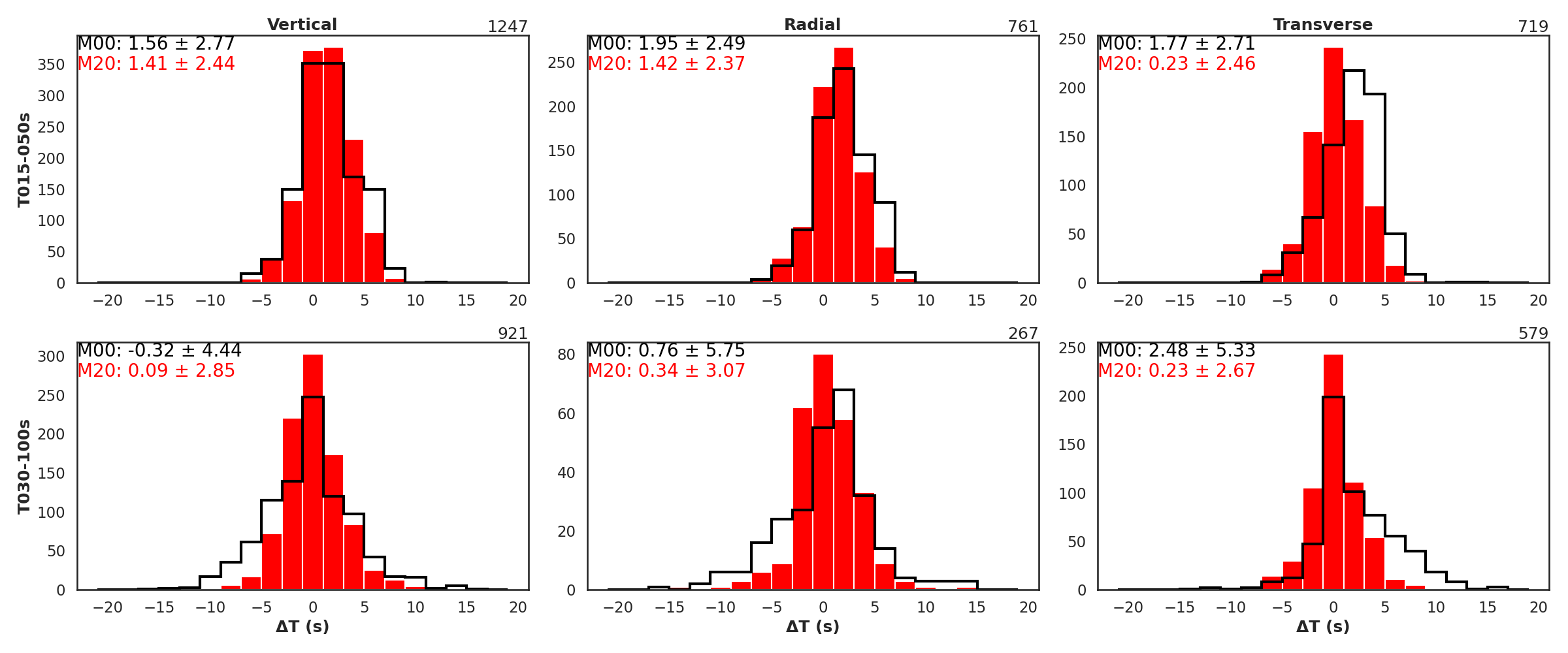}
    \caption{Cross-correlation traveltime histograms with the independent dataset (Fig.~\ref{fig:benchmark_source_and_stations}) for six measurement categories for the initial model GLAD-M25 (M00, black) and the final model MEAD-M20 (M20, red). The mean and standard deviation of the histograms, and the number of measurements, are displayed at the top left and top right of each plot, respectively.}
    \label{fig:benchmark_dt_hist}
\end{figure}

To further evaluate our model, we make cross-correlation traveltime measurements at our 6 measurement categories using waveforms from 30 independent earthquakes selected from the GCMT catalog that are not used in our inversion (Fig.~\ref{fig:benchmark_source_and_stations}). The traveltime histograms are promising, showing improvement for data from 30 independent earthquakes, particularly for surface waves. The Rayleigh and Love wave histograms of the initial model show negative and positive biases, respectively, highlighting the importance of radial anisotropy in the study area (Fig.~\ref{fig:benchmark_dt_hist}).

\subsection{Model MEAD-M20}
\begin{figure}
  \centering
  \includegraphics[width=1.0\textwidth]{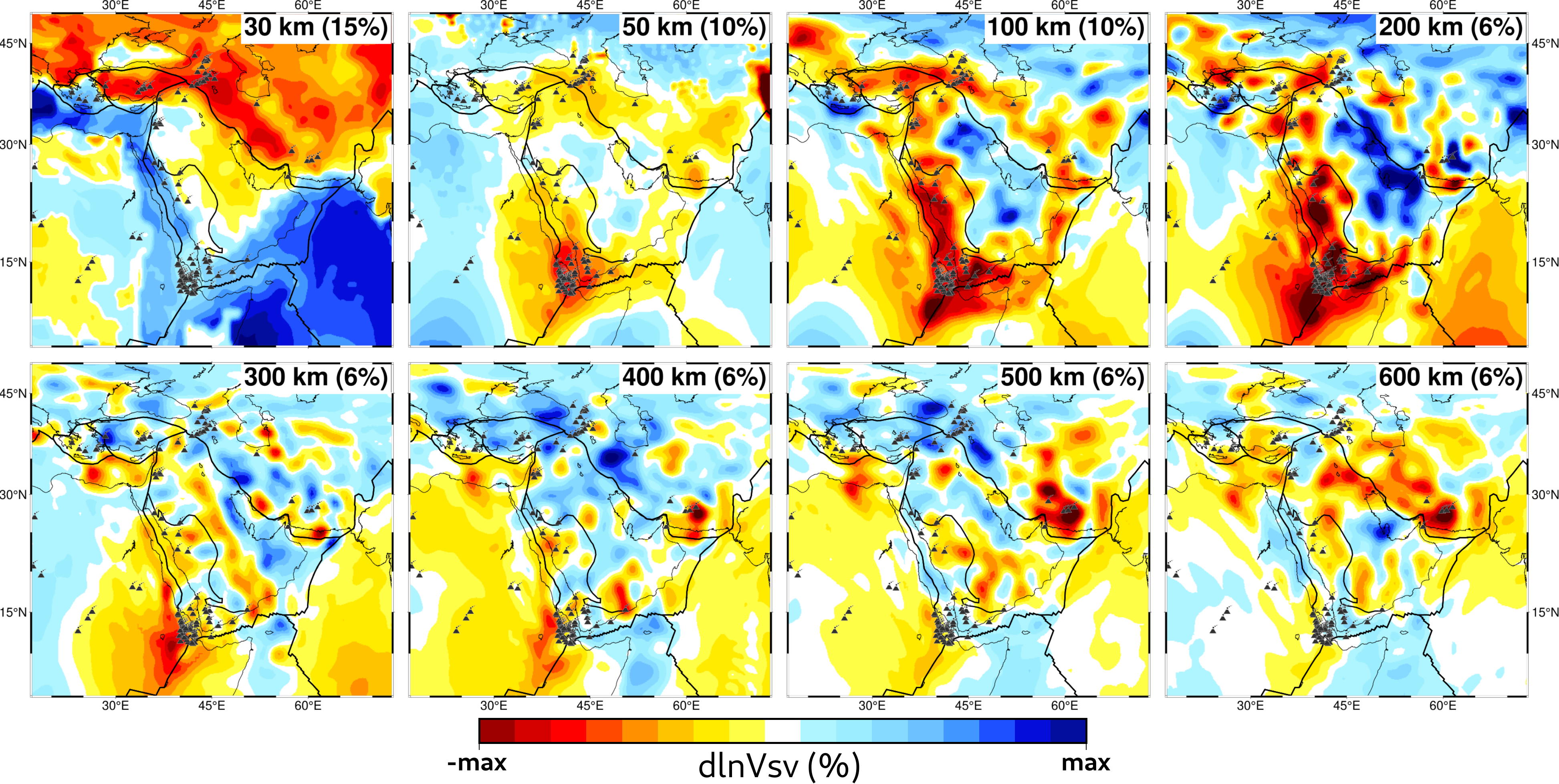}
  \caption{Vertically polarized shear-wavespeed ($V_{sv}$) perturbations with respect to the mean of MEAD-M20 from 30~km to 600~km depths. Maximum perturbations are denoted on the top right of each plot. }
  \label{fig:model_vsv}
\end{figure}

\begin{figure}
  \centering
  \includegraphics[width=1.0\textwidth]{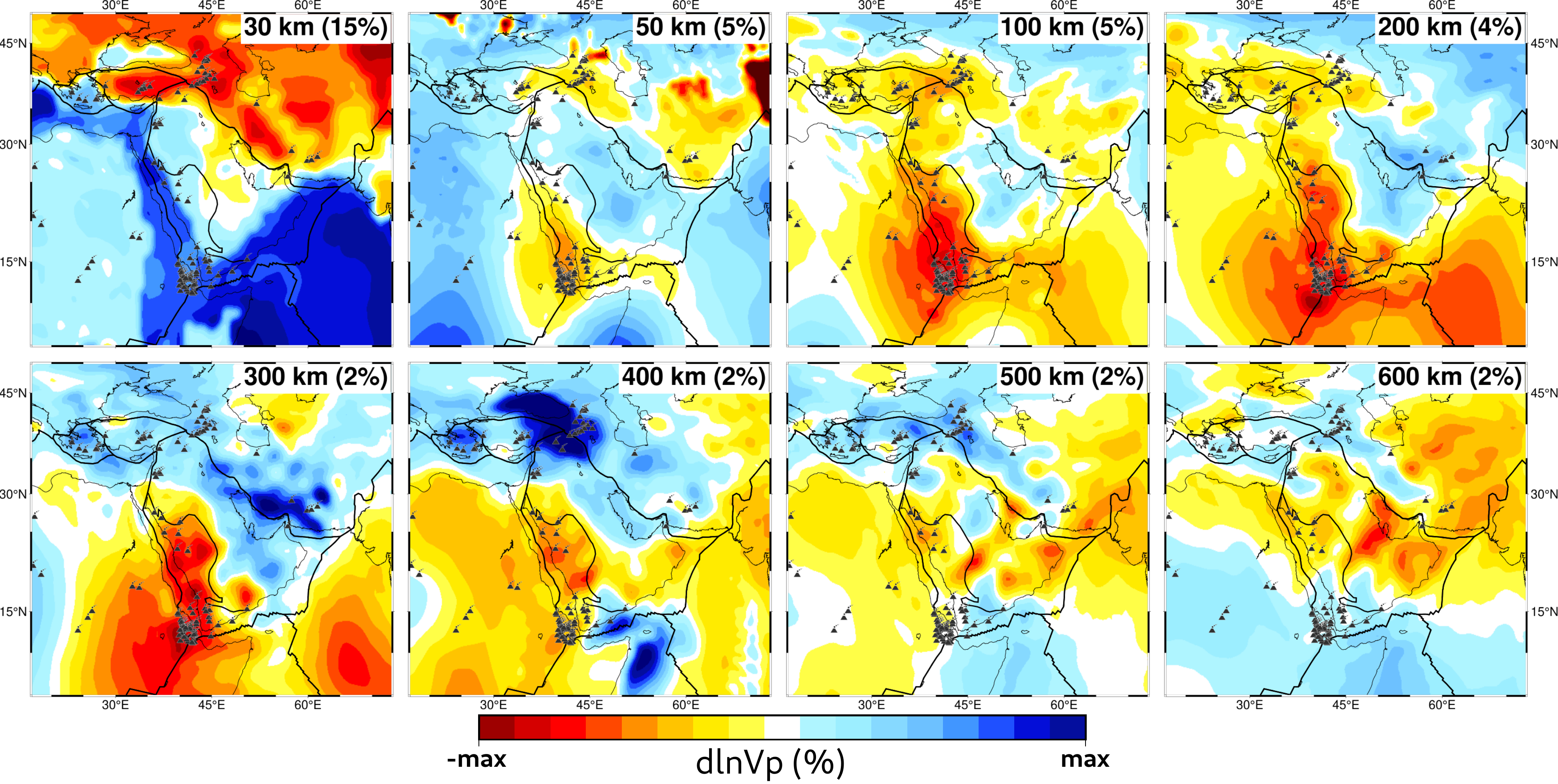}
  \caption{Same as Fig. \ref{fig:model_vsv} but for the isotropic compressional-wavespeed ($V_{p}$) model.}
  \label{fig:model_vp}
\end{figure}

Figs.~\ref{fig:model_vsv}~\&~\ref{fig:model_vp} show the depth sections of vertically-polarized shear (${\rm dln}V_{sv}$) and compressional (${\rm dln}V_{p}$) wavespeed perturbations, respectively, of our final model after 20 iterations in the lithosphere and upper mantle down to 600~km. As expected, the perturbations gradually decrease with depth, with up to 20\% and 8\% deviations from the mean shear wavespeeds at 30~km in the continental crust and 600~km, respectively. The perturbations are lower for P waves, which decrease from about 15\% in the crust to $\sim 2$\% in the transition zone. The boundary between the Arabian shield and the Arabian platform is clearly observed in the upper mantle within the 100-250~km depth range characterized by low and high perturbations, respectively, in both shear and compressional wavespeed models, which was also reported in previous studies \citep[e.g.,][]{hansen_imaging_2007,tang_shear_2019,celli_african_2020,lim_asthenospheric_2020,kaviani_crustal_2020,rodgers_adjoint_2024}. On the other hand, the shield and the platform are characterized by high and low wave speed perturbations in the crust, respectively, where the signal is reversed below 30 km. At 30~km, the Arabian platform is also sharply separated from the Iranian block along the Bitlis-Zagros Suture Zone (BZSZ) with significantly lower wavespeeds in the East and Anatolia, likely related to subduction-related volcanism \citep{omrani_arc-magmatism_2008}. The low-velocity belt along the BZSZ is shifted north in Anatolia and northeast in Iran, parallel to the suture zone at 200~km, where the slab remnants of the Tethys Ocean under Anatolia and Iran are visible as high-wavespeed perturbations which follow the same trend as the low-wavespeed perturbations and move away from the suture zone by depth. The Red Sea is characterized by low wavespeeds in the upper mantle down to 400~km, where perturbations are stronger in the south around Afar and strongest below 50~km down to 200~km. Lower perturbations in the Harrats, Eastern Anatolia, and Makran also characterize the volcanic regions. Shear and compressional models are consistent with each other. However, the P-wave resolution is less than that of S waves. Overall, there is a significant improvement in both shear and compressional models compared to the starting model, GLAD-M25, which shows the long-wavelength structure in the region. Due to insufficient data coverage in the global datasets, the change in GLAD-M25 relative to the starting model of GLAD models, S362ANI \citep{kustowski_anisotropic_2008}, is also minor in the Middle East (see Figs.~\ref{figS:supp-model_comp_vs_global}~\&~\ref{figS:supp-model_comp_vp_global}).

\begin{figure}
  \centering
  \includegraphics[width=1.0\textwidth]{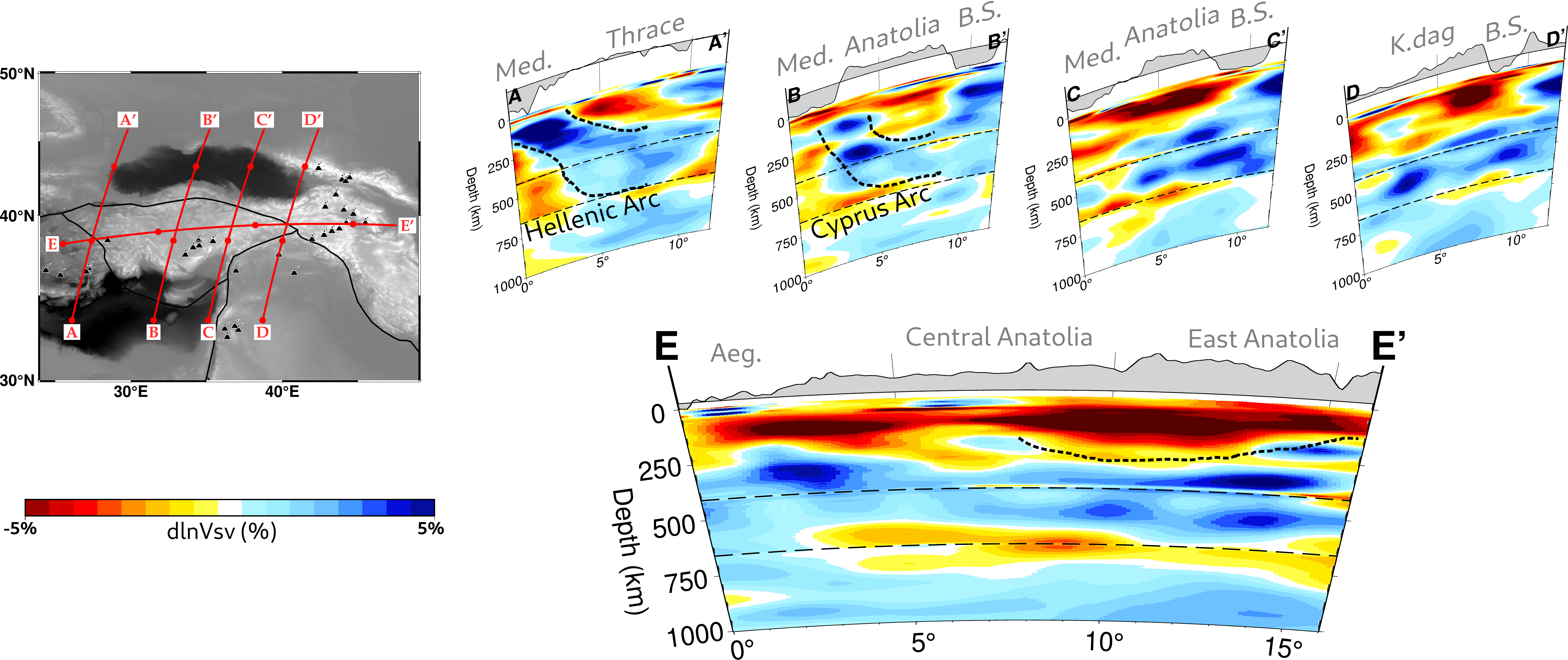}
  \caption{Vertical cross sections of the vertically polarized shear-wavespeed perturbations with respective to the mean of MEAD-M20 ($V_{sv}$) in Anatolia (Aeg.: Aegean Sea, Anatolia: Anatolia, Zagros Mountains, Med: Mediterranean Sea, B.S.: Black Sea, K.dag: Karacadag).}
  \label{fig:model_selected_vert_xsections}
\end{figure}

Subduction along the Hellenic and Cyprus arcs is better observed in the vertical sections shown in Fig.~\ref{fig:model_selected_vert_xsections}. The subducted slab along the Hellenic arc extends at least to a depth of 660~km (section AA'). The slab along the Cyprus arc seems to be steeper and flattens in the transition zone (section BB'). Indeed, towards Eastern Anatolia, high-wavespeed structures are clearly visible, lying over the 660-km discontinuity, which is likely associated with the closure of the Tethys Ocean (sections CC'-EE') \citep{biryol_segmented_2011}. Meanwhile, the lithosphere in Eastern Anatolia is characterized by low-wavespeed perturbations.

\begin{figure}
  \centering
  \includegraphics[width=1.0\textwidth]{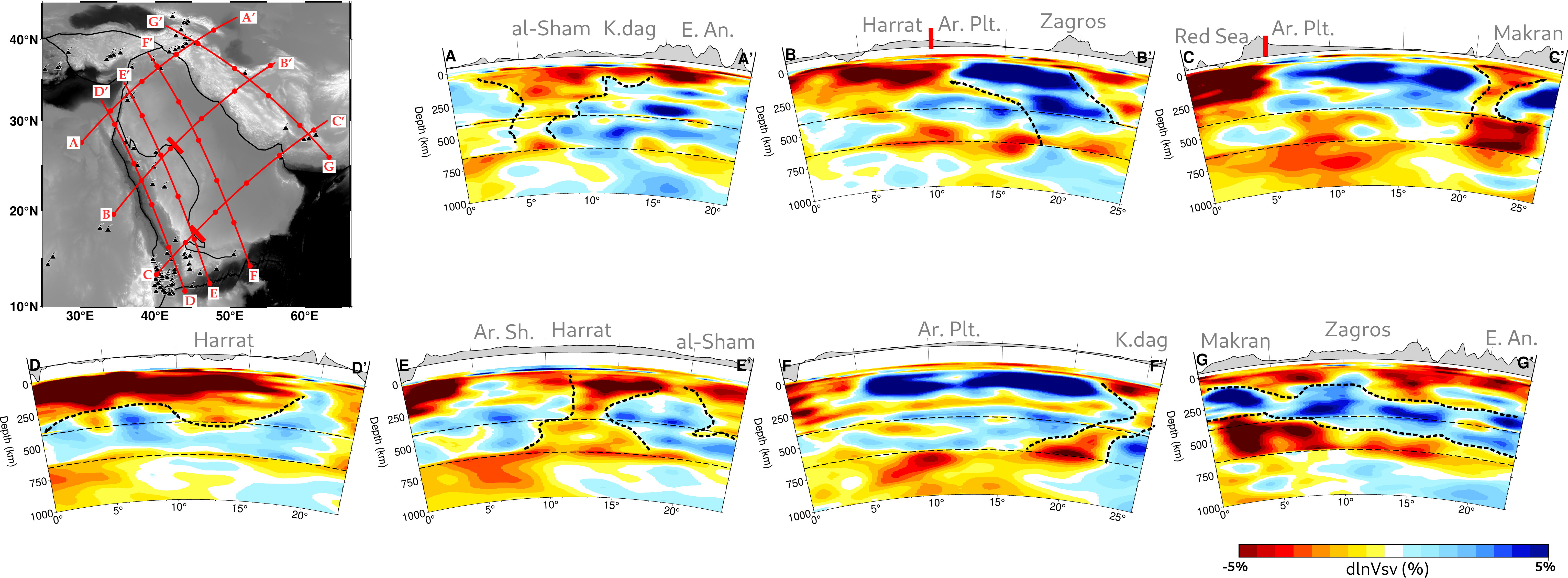}
  \caption{ Same as Fig.~\ref{fig:model_selected_vert_xsections} but for cross-sections across the Arabian Peninsula and Iran (Ar. Sh.: Arabian Shield, Ar. Plt.: Arabian Platform, Ea. An.: East Anatolia, Ir. Plt: Iranian Plateau, al-Sham: Harrat al-Sham region, K.dag: Karacadag). In BB' and CC' cross sections, the boundary between the Arabian shield and Arabian platform is indicated by thick red lines in the cross sections and on the map.}
  \label{fig:model_selected_vert_xsections2}
\end{figure}

Fig.~\ref{fig:model_selected_vert_xsections2} shows various vertical cross-sections across the Arabian Peninsula and Iran. Section AA' cuts through the volcanic region al-Sham in Jordan, showing low-wavespeed perturbations down to around 500~km. Closer to the volcanic mountain Karacada\u g in Eastern Anatolia, the lithosphere is also characterized by low wavespeeds. The southern sections from the Red Sea towards the Zagros suture zone (BB'-CC') clearly show the separation of the Arabian shield and the platform in the upper mantle. Section BB' also indicates the subduction along the Zagros. The eastern edge of section CC' shows the low-wavespeed perturbations in the volcanic Makran region in the east of the Gulf. In section DD', a channeled low-velocity zone is observed along the Red Sea. In the middle of section EE' a plume-like low-velocity feature down to 660-km depth is clearly observed. The end of the section shows strong low velocities in Jordan, associated with the potential plume reported in previous studies \citep{chang_mantle_2011}. Section FF' crosses the Arabian platform characterized by high upper-mantle wavespeeds from south to Karacada\u g, where a low-velocity zone is observed with deep roots down to 660~km. Section GG', parallel to the Zagros suture zone, nicely shows almost continuous high-wavespeed perturbations between 200-500~km, likely associated with the remnant of the subduction. The corresponding P-wavespeed sections are in general in agreement with those of S wavespeeds with relatively lower resolution (see Figs.~\ref{figS:supp-model_selected_vert_xsections_vp}~\&~\ref{figS:supp-model_selected_vert_xsections2_vp}).

\begin{figure}
  \centering
  \includegraphics[width=1.0\textwidth]{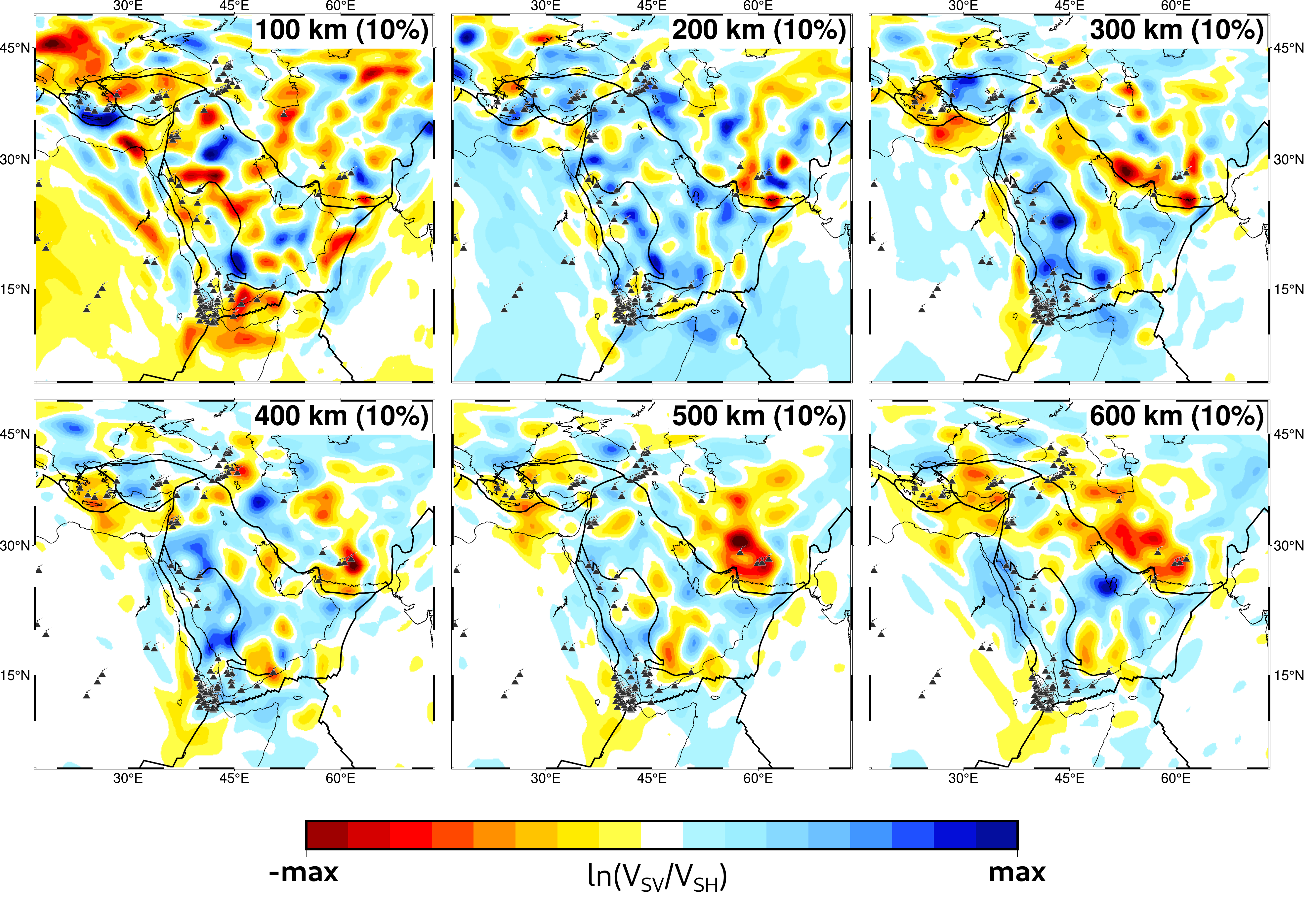}
  \caption{Same as Fig.~\ref{fig:model_vsv}, but for transverse isotropy ($\ln{\left( V_{sv}/V_{sh} \right)}$), with the mean transverse isotropy subtracted.}
  \label{fig:model_ti}
\end{figure}

Fig.~\ref{fig:model_ti} presents transverse isotropy as the logarithmic ratio of vertically-polarized shear waves ($V_{sv}$) to horizontally-polarized shear waves (${V_{sh}}$), with the mean transverse isotropy subtracted. A sign change in transverse isotropy is observed from 50~km to 200~km, where the Arabian plate is characterized by high $\rm{V_{sh}}$ at 50~km, consistent with plate motions, and by higher ${V_{sv}}$ at 200~km. We generally observe high ${V_{sv}}$ in the Red Sea, which likely indicates that the flow beneath the ridge or very slow spreading along the ridge may conceal the expected high $\rm{V_{sh}}$ at the spreading ridge. At 100~km, subducted slabs, such as the Hellenic Arc, typically exhibit high $\rm{V_{sh}}$. In contrast, volcanic regions, including Harrats, the Makran region, East Anatolia, and Jordan, show high $\rm{V_{sh}}$. High $\rm{V_{sh}}$ is also observed in the Pannonian basin, the Gulf of Aden, and to a lesser extent along the Red Sea. At the same time, the Makran region generally exhibits a high $\rm{V_{sh}}$ with a maximum around 500~km (see Figs.~\ref{figS:supp-model_selected_vert_xsections_ti}~\&~\ref{figS:supp-model_selected_vert_xsections2_ti} for transverse isotropy on vertical sections). ${V_p/V_s}$ ratios highlight the slab signatures with high $\rm{Vs}$ and volcanic regions with generally low $\rm{Vs}$, such as in Eastern Anatolia, Makran, etc., potentially suggesting partial melting (Figs.~\ref{figS:supp-model_selected_vert_xsections_vpvs}~\&~\ref{figS:supp-model_selected_vert_xsections2_vpvs}).

\subsection{Model comparisons}

\begin{figure}
  \centering
  \includegraphics[width=1.0\textwidth]{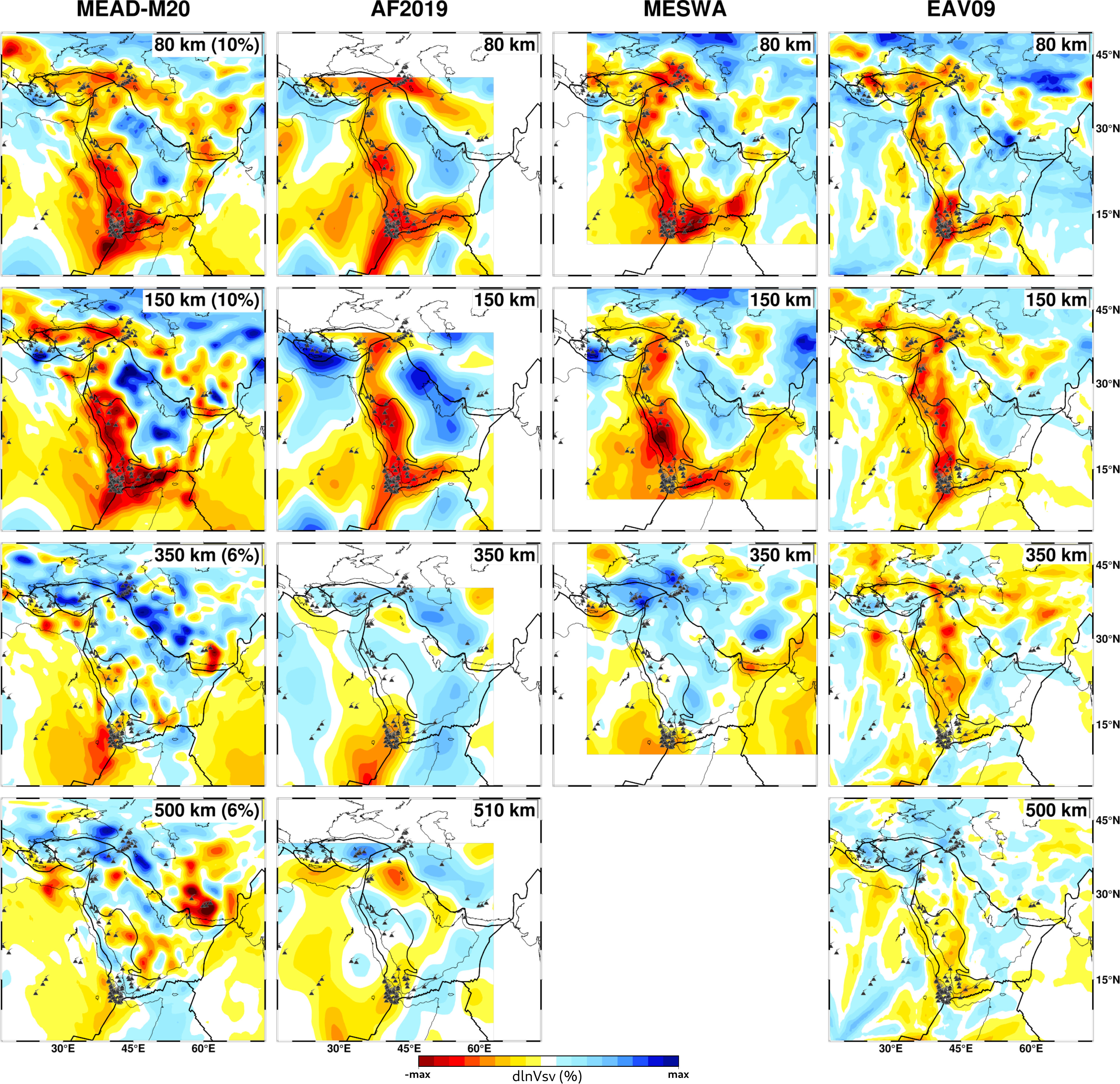}
  \caption{$V_{s}$ model comparisons of MEAD-M20, AF2019 \citep{celli_african_2020}, MESWA \citep{rodgers_adjoint_2024}, and EAV-09 \citep{chang_joint_2010}. Each model is plotted with respect to its own mean, and the maximum perturbations are denoted in the top-right corner of MEAD-M20 plots.}
  \label{fig:model_comp_vs}
\end{figure}

In Fig.~\ref{fig:model_comp_vs}, we compare our model MEAD-M20 to a collection of S-wavespeed models from other groups: AF2019 \citep{celli_african_2020} resulting in an automated multimode inversion 
and constructed based on surface and S waveforms; a full-waveform inversion model constructed based on adjoint iterations similar to the current study, MESWA \citep{rodgers_adjoint_2024}, whose starting model is the 3D transversely isotropic global mantle model SPiRaL \citep{simmons_spiral_2021} and assimilated 30-100~s waveforms at the final stage of iterations; an S-wave arrival-time model EAV09 \citep{chang_joint_2010}. The main long-wavelength structures are common in all models. At 150~km, MEAD-M20 shows larger perturbations that sharply mark the boundary between the Arabian platform and the shield, as observed in other models. At 350~km, perturbations decrease and we observe a significantly high-wavespeed structure along the Zagros suture zone, which is much sharper in MEAD-M20 than in other models. MEAD-M20, AF2019, and MESWA do not exhibit the north-south-trending low wavespeeds observed in EAV09. The perturbations remain large in the transition zone at 500~km in MEAD-M20, where we still observe high wavespeeds in Eastern Anatolia, the Caucasus, and along the Zagros, as well as significant low-velocity perturbations in Makran. MESWA's resolution is confined to the top 350-400~km only due to its longer-period data (30~s) dominated by surface waves during its construction.

\begin{figure}
  \centering
  \includegraphics[width=1.0\textwidth]{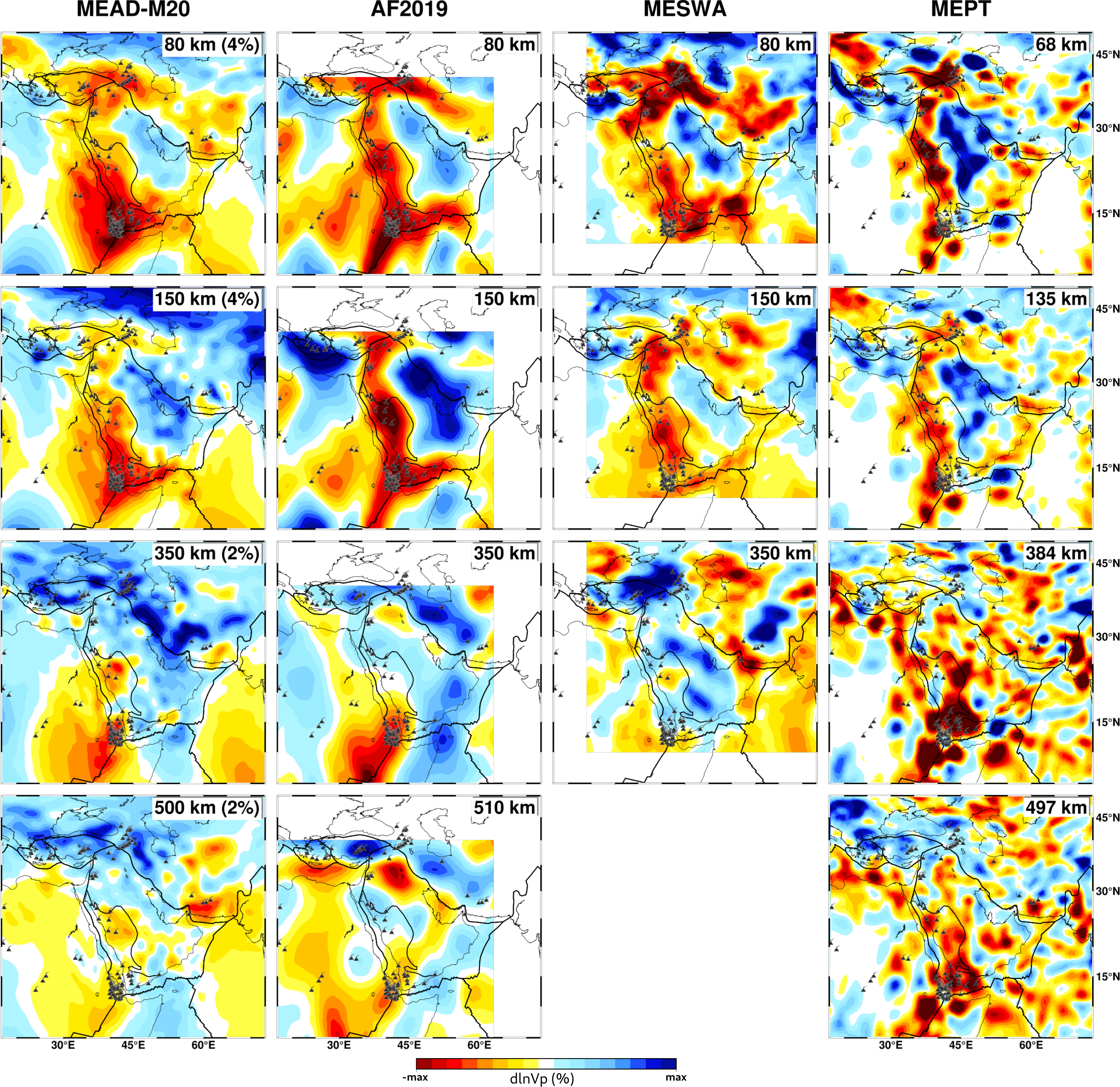}
  \caption{$V_{p}$ model comparisons of MEAD-M20, MESWA \citep{rodgers_adjoint_2024}, AF2019 \citep{celli_african_2020}, and MEPT \citep{bozdag_p-wave_2025}. Each model is plotted with respect to its own mean, and the maximum perturbations are denoted in the top-right corner of MEAD-M20 plots.}
  \label{fig:model_comp_vp}
\end{figure}

In Figure \ref{fig:model_comp_vp}, we compare the $V_p$ model of MEAD-M20 with those of AF2019 and MESWA, as well as a recent P-wave arrival-time tomography model, MEPT \citep{bozdag_p-wave_2025}. MEPT uses ISC data and first-arrival P-wave onset-time readings from the same waveforms recorded by the SGS stations in the Arabian Peninsula used in this study. Although there is agreement on the longer-wavelength structures, the perturbations from MEAD-M20 and AF2019 are lower than those from MESWA and MEPT, where MEPT shows the largest perturbations and smaller-scale features, likely related to the regularizations used in the model constructions. While higher-wavespeed perturbations of the Arabian platform are observed in MESWA at 50~km, they decrease significantly at 150~km, unlike other models. MEAD-M20 and AF2019 clearly show the remnant of subduction along the Zagros suture zone at 350~km, which is observed at deeper depths in MEPT but not observed in MESWA. Low-wavespeed perturbations along the Red Sea are common in all models at 150~km, with the largest perturbations centered around Afar in MEAD-M20. Significantly low-wavespeed perturbations extend down to 500~km in MEPT, a teleseismic P-wave arrival-time model, around the Afar plume, which is less visible in other models and might be associated with decreased data sensitivity with depth in continental-scale tomographic studies.

\begin{figure}
  \centering
  \includegraphics[width=0.85\textwidth]{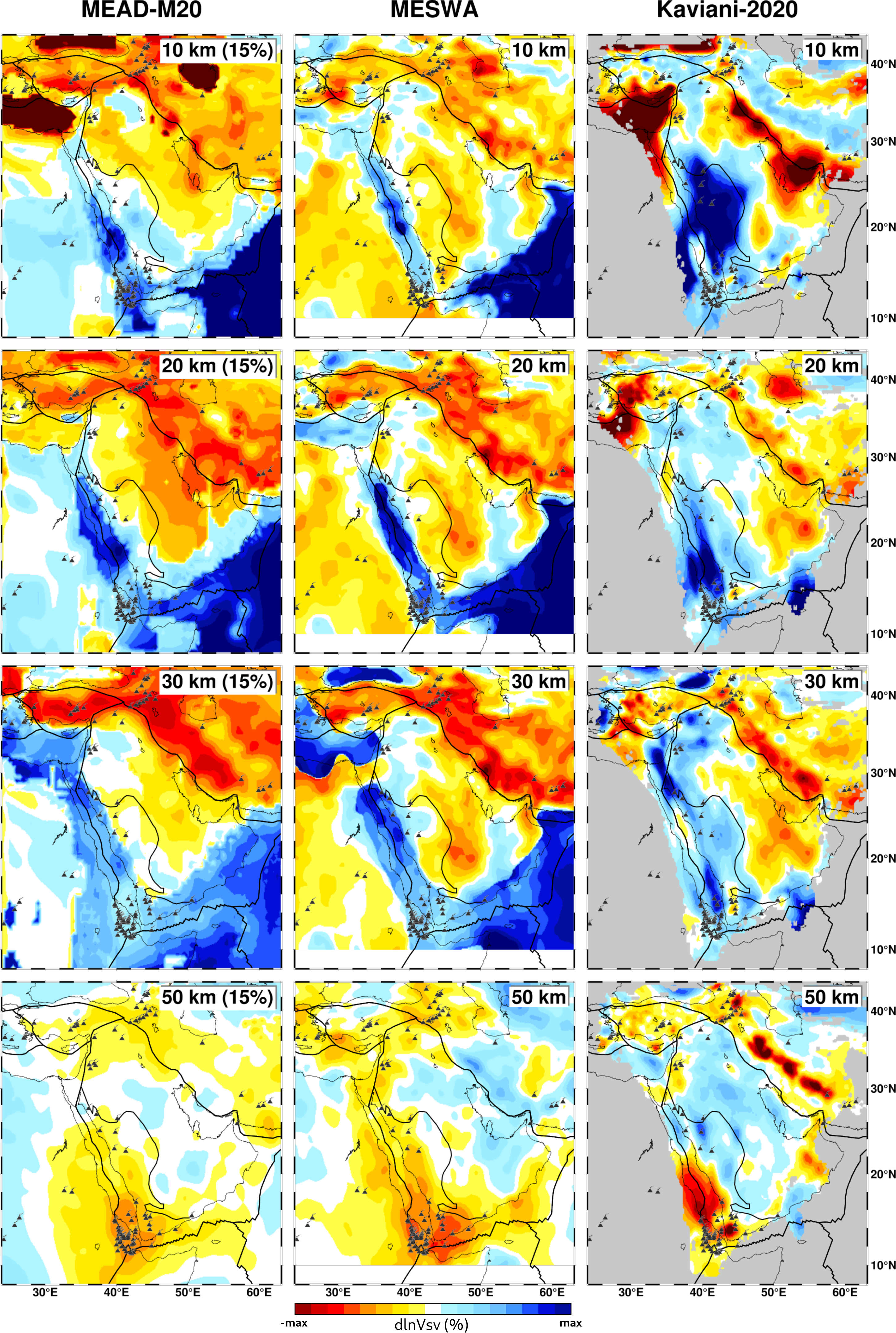}
  \caption{Crustal $V_{s}$ model comparisons of MEAD-M20, MESWA \citep{rodgers_adjoint_2024}, and KAVIANI-2020 \citep{kaviani_crustal_2020}. Each model is plotted with respect to its own mean, and the maximum perturbations are denoted in the top-right corner of MEAD-M20 plots.}
  \label{fig:model_comp_vs_crust_pert}
\end{figure}

In Fig.~\ref{fig:model_comp_vs_crust_pert}, we compare the crustal S-wave structure for MEAD-M20, MESWA, and the KAVIANI-2020 model by \citet{kaviani_crustal_2020}, which is based on an ambient-noise tomography study. Overall, the crustal parts of MEAD-M20 and MESWA are consistent with varying perturbations. The crustal parts of both models also show little difference relative to their starting models (see Fig. \ref{figS:supp-model_comp_vs_crust_abs}), likely due to the limited sensitivity of relatively long-period surface waves to shallow structures used during model constructions. KAVIANI-2020 differs most at 10~km with significantly higher wavespeeds in the Arabian shield.

\begin{figure}
  \centering
  \includegraphics[width=1.0\textwidth]{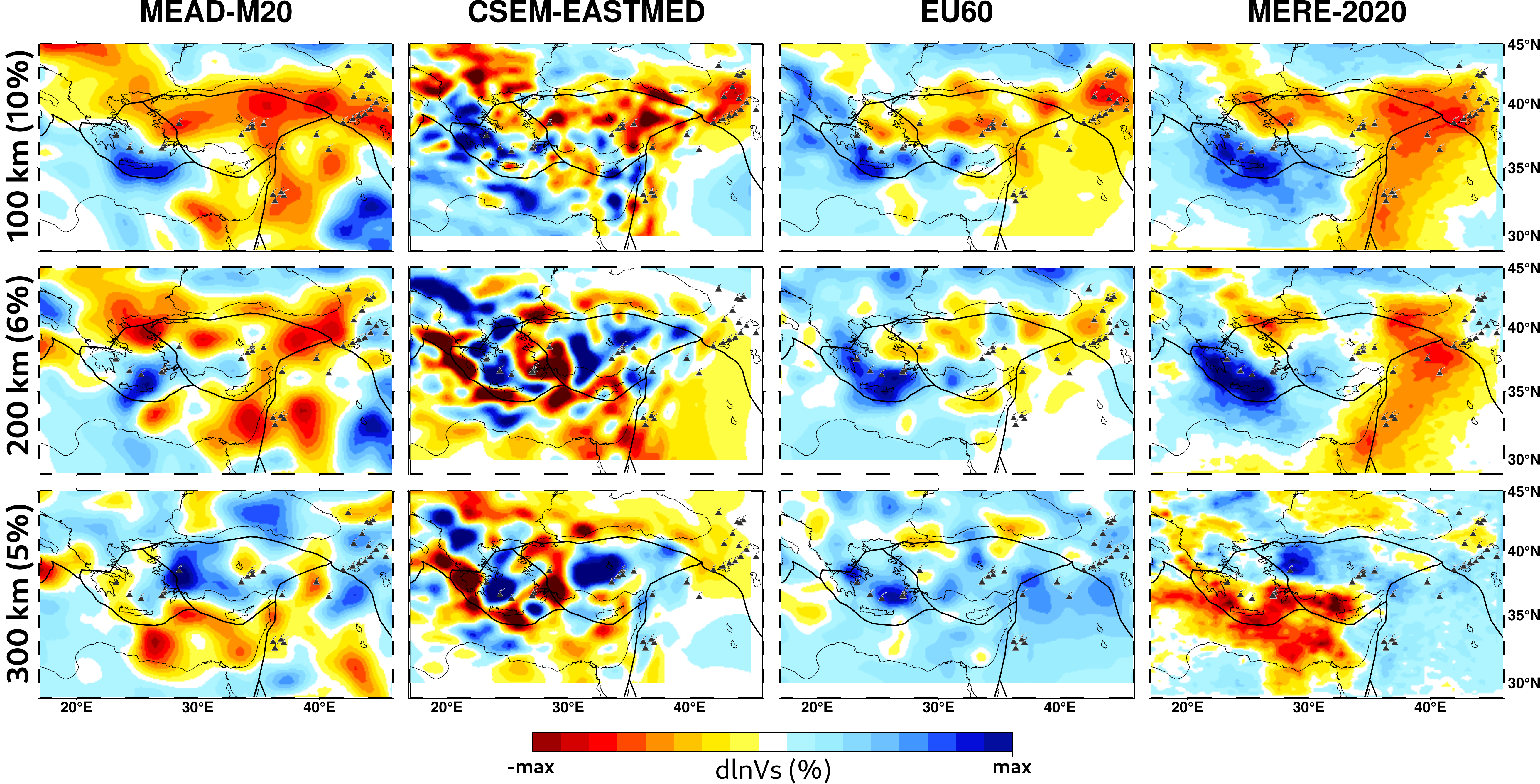}
  \caption{$V_{s}$ model comparisons in the Eastern Mediterranean and Anatolia between MEAD-M20, CSEM-EASTMED \citep{blom_seismic_2020}, EU60 \citep{zhu_seismic_2015} and MERE-2020 \citep{el-sharkawy_slab_2020}. Each model is plotted with respect to its own mean, and the maximum perturbations are denoted within parentheses on the left.}
  \label{fig:model_comp_vs_eastmed}
\end{figure}

In Fig.~\ref{fig:model_comp_vs_eastmed}, we focus on Anatolia and compare our model with other models inverted Anatolia as part of their larger-scale European or Eastern Mediterranean models: FWI model CSEM-EASTMED \citep{blom_seismic_2020} which uses the Collaborative Seismic Earth Model \citep{afanasiev_foundations_2016} as the starting model (constructed based on 28-150~s waveform data at the final stage); FWI model EU60 \citep{zhu_seismic_2015} which uses S362ANI \citep{kustowski_anisotropic_2008} and EPCrust \citep{molinari_epcrust_2011} as the starting model with waveform data down to 15~s for body waves and 25~s for surface waves at the final iteration; MERE-2020 \citep{el-sharkawy_slab_2020} which is constructed using Rayleigh-wave phase-velocity curves. The main features are consistent, such as low wave speeds in Anatolia at 100~km, Hellenic arc, etc., but with significant differences in perturbations, scale length of heterogeneities, and slab geometry. The North Anatolian fault's signature is visible down to 200~km in MEAD-M20.

\section{Discussions}\label{sec:discussions}

High-resolution tomographic models are crucial for ground motion simulations and seismic hazard assessments \citep{akcelik_high_2003,rodgers_broadband_2008,tape_adjoint_2009}, where self-consistent P- and S-wavespeed models, together with their attenuation models, are essential for accurate wave simulations. Further improvements in tomographic images require coupling the advances in 3D wave simulations together with data in the inverse problem \citep[e.g.,][]{bozdag_global_2016}. To this end, the main strengths of MEAD-M20 are 1) the additional waveform data from the Arabian Peninsula used during its construction, 2) incorporating the full complexity of wave propagation in an FWI framework, 3) jointly inverting for P and S wavespeeds, taking the transverse isotropy into account in the upper mantle, making it ready for wave simulations with its own numerical solver (\texttt{SPECFEM3D Globe}).

We observe the value of additional waveforms from the Arabian Peninsula in our resolution tests performed at four locations (i.e., Anatolia, Arabian Peninsula, Iran, and Zagros) with varying levels of data coverage, showing reasonably good resolution of heterogeneities (half-width of 100-200~km) without significant trade-off between the parameters.

The checkerboard tests of MEPT, the P-wave arrival tomography model by \citet{bozdag_p-wave_2025} using ISC data and the first-arrival P-wave onset-time readings from the same waveform data recorded in the Arabian Peninsula we use in this study, show improved resolution in the lithosphere beneath the Arabian Peninsula, underscoring the value of assimilation of the additional data from the SGS network. Note that in this study, we assimilate more information into the inversion, using traveltimes from different body-wave phases and surface waves. Resolution tests primarily provide information about the resolving power of data used in the inversion and are therefore closely linked to data coverage. Consequently, the PSF tests performed at four locations with varying levels of coverage provide an indication of the expected resolution when considered together with the corresponding kernel coverage. Another independent test of our model is performed by \citep{cardozo_gji_2025}, using MEAD-M20 to compute 3D source Green's functions and report promising results in the updated moment tensor solutions of earthquakes in Iran, specifically in constraining their double-couple components, highlighting the importance of self-consistent models ready for 3D wave simulations for regions with complex geology and tectonics like our study region, the Middle East.

MEAD-M20 shows all the known characteristics of the tectonic evolution of the region consistent with the previous models such as the subducted slabs along the Hellenic and Cyprus arcs, the slab detachment underneath upon the closure of the eastern part of the Tethys Ocean, slab remnants observed along Zagros suture zone, low wavespeeds along the Red Sea, from Afar all the way up to the Eastern Anatolia, as well as other volcanic regions in the Arabian plate, Anatolia and Iran, etc. (Figure \ref{fig:model_comp_vs}). MEAD-M20 shows sharper images of the P- and S-wavespeed structures in the Arabian Peninsula. The sharp separation between the Arabian platform and the shield down to at least 250~km depth is clearly observed (see Fig. \ref{fig:model_vsv} \& \ref{fig:model_vp}). This boundary has also been reported in previous tomographic studies of the region \citep{hansen_imaging_2007,tang_shear_2019,celli_african_2020,lim_asthenospheric_2020,kaviani_crustal_2020,rodgers_adjoint_2024}, albeit at lower resolution due to limited station coverage, or imaged only as part of a P- or S-wavespeed tomographic model. The recent P-wavespeed model, MEPT \citep{bozdag_p-wave_2025} reports this discontinuity down to $\sim 350$-km depth, deeper than other models. Our P-wave model is consistent with MEPT, whereas the sharp discontinuity is observed only down to about 250~km in the S-wavespeed model, which might indicate higher sensitivity of S waves to thermal variations \citep{koulakov_evidence_2016}. Similar observations on P- and S-wavespeed models are reported by \citep{chang_new_2012} that the difference in P- and S-wavespeed anomalies might be due to partial melt, water, or other compositional factors. Moreover, the previous crustal and mantle models of the region suggest polarity reversal in velocity perturbations in the Arabian Peninsula from the crust to the upper mantle \citep[e.g.,][]{al-damegh_crustal_2005,kaviani_crustal_2020,lim_asthenospheric_2020,rodgers_adjoint_2024,bozdag_p-wave_2025} where the Arabian shield and platform are characterized by high- and low-velocity perturbations in the crust which are reversed below 50~km depth. Note that we invert the crust and mantle simultaneously in our FWI framework, where the velocity polarity reversal from the crust to the mantle persists. The low-wavespeed perturbations in the crust of the Arabian platform are likely associated with sedimentary basins, such as the Mesopotamian Basin, whereas the high-wavespeed perturbations in the upper mantle indicate colder and denser material subducting along the Zagros suture zone. The low-wavespeed perturbations in the mantle beneath the Arabian shield are consistent with spreading along the Red Sea and volcanic regions at the Harrats. The results are also consistent with the suggested $Q$ (quality factor) model of the region by \citet{pasyanos_improved_2021}, in which the lithosphere and upper mantle beneath the shield and the platform are characterized by low and high $Q$, respectively.

The source of volcanism in the Harrats is a long-debated subject, whether it is related to spreading along the Red Sea and continental rifting at the Gulf of Aden, the northward flow of the Afar plume, or potential local plumes. Our model is consistent with previous models and shows the extension of a low-velocity channel along the Red Sea from Afar to the north (Fig. \ref{fig:model_selected_vert_xsections2} DD' cross-section) through Jordan all the way up to the Eastern Anatolia at 100-200 km (Fig. \ref{fig:model_selected_vert_xsections2} AA' cross-section). The stronger negative S-wavespeed perturbations around the Afar plume in both P- and S-wavespeed models may support the channeled flow from Afar \citep{ebinger_cenozoic_1998,park_upper_2007,chang_new_2012,lim_asthenospheric_2020}. While this observation is stronger for the P-wavespeed model, in the higher-resolution S-wavespeed model, the continuation of the channeled flow to the north becomes less pronounced, where localized low-velocity perturbations are highlighted, for instance, at the Harrats and Jordan, which seem to support the local plume theory to explain the volcanism \citep{chang_mantle_2011} (Fig. \ref{fig:model_selected_vert_xsections2} EE' cross-section).
Our model also depicts volcanism in the Makran region (Fig. \ref{fig:model_selected_vert_xsections2} CC' cross-section), denoted by low-velocity perturbations, which might link to the plume at the transition zone underneath the Arabian Plate, as also suggested by \citet{koulakov_evidence_2016}. We did not perform an additional PSF test in the Makran region. While the checkerboard tests performed in the same area for the MEPT model \citep{bozdag_p-wave_2025} show decent resolution, increasing confidence in the low-wavespeed perturbations, further analysis shall be performed to address the region's geodynamics in future studies. Another low-velocity region is present in our model down in the transition zone beneath Syria and Iraq (Fig. \ref{fig:model_selected_vert_xsections2} AA' cross-section), which is also observed by \citet{civiero_complex_2022} and defined as the Levant plume, which is likely related to the local volcanism. The low velocity perturbations characterizing Anatolia are also consistent with previous studies of the region, likely associated with the extensional regime on the west, subduction-related volcanism along the Hellenic arc (Fig. \ref{fig:model_selected_vert_xsections} AA' cross-section) and the Bitlis suture zone (Fig. \ref{fig:model_selected_vert_xsections} DD' \& EE' cross-section) \citep{pichon_miocene--present_2010,agard_zagros_2011}. The depletion of the mantle under the high plateau in Eastern Anatolia is confirmed by the lack of Lg waves \citep{gok_lithospheric_2007}. More strikingly, the low-velocity perturbations roughly follow the North and East Anatolian fault zones down to 200~km (Fig. \ref{fig:model_comp_vs_eastmed}), where the part along the North Anatolian fault was also reported by \citet{fichtner_deep_2013} but around 100~km, interpreted as the remnant of the closed Neo-Thetys ocean that enabled the development of the fault zone.

Another sharp observation in MEAD-M20 is the collision along the Zagros and related slab remnants, as well as volcanism. The slab remnant is observed progressively in the northwest direction by depth (Fig. \ref{fig:model_selected_vert_xsections2} GG'). Other studies also see the difference between the northwest and southeast directions. \citet{chang_joint_2010} argues for slab detachment in the north progressing southward due to the absence of dipping high-velocity anomalies. Our model indicates a steeply subducting slab in the northwest, which becomes shallower in the southeast (underthrusting). \citet{mouthereau_building_2012} offers a possible explanation by noting buoyant accretion, which could result in slower subduction in the northwest.

Overall, high $V_p/V_s$ ratios of MEAD-M20 well correlate with hotspot and volcanic regions at the Afar plume, the Harrats, the Red Sea spreading ridge, East Anatolia, etc., potentially highlighting partial melting (Fig. \ref{figS:supp-model_comp_vp_vs_ratio}).

Despite the decent correlation, it is important to note the difference in the resolution of the Vp and Vs models. Although we jointly invert Vp and Vs models, the resolution of P-wave models is lower than that of S waves. Moreover, there is no good agreement between different P-wave models, even at long wavelengths, unlike S-wave models, which have also been documented in a recent study by \citet{BozdagHouser2026}. The lower resolution and limited agreement of P-wave models are likely due to the following reasons: 1) lower P-wave amplitudes can make the selection of P waves challenging, 2) surface-wave sensitivities to P waves are drastically less than those to S waves, and 3) P wavelengths are longer than those of S wavelengths at the same frequency. To this end, the convergence of P-wave models is much slower, which might explain the significant differences between P-wave models, including the two FWI models, MEAD-M20 and MESWA, where the inclusion of shorter-period body waves (both P and S waves) in MEAD-M20 is crucial for the depth and P-wave resolutions. For a more robust comparison of $V_p/V_s$ ratios, a possible approach is to smooth the $V_s$ model to better correlate with the lower resolution of the Vp model, provided there is sufficient P-wave coverage. We observe that $V_p/V_s$ signals also persist with the smoothed Vs model. However, it is still challenging to make firm interpretations of the strength and size of regions with potential partial melting, etc., which is a good reminder to put more emphasis on P-wave selections in datasets and highlight them with appropriate weightings (e.g., \citet{cui_glad-m35_2024}). Another potential way to further improve P-wave resolution is to combine multiple datasets, such as P-wave arrival times with waveforms within the adjoint tomography framework. This idea was one of the main motivations for us to construct a P-wave model of the region \citep{bozdag_p-wave_2025} before our FWI inversions to demonstrate the combination of arrival times with waveforms. As mentioned in \citet{bozdag_p-wave_2025}, such a demonstration can also guide us in potentially combining arrival-time readings from floating acoustic robots, such as MERMAIDs \citep{sukhovich_seismic_2015,nolet_picking_2025}, with waveforms in a global adjoint tomography framework in future studies.

There are three FWI models of the Middle East: the elastic models MEAD-M20 and MESWA, and the anelastic FWI model for a smaller region focusing on the Arabian Peninsula by \citet{espindolacarmona_anelastic_2024}, which also involves source inversions before the structural inversion. The longer-wavelength structures from all three models are consistent in shear-wave models. In contrast, larger discrepancies are observed for smaller heterogeneities, particularly between P-wave models, as discussed above. \citet{espindolacarmona_anelastic_2024} uses a workflow similar to ours in this study, using the \texttt{SPECFEM3D Globe} package, except for introducing an anelastic parameterization in the inverse problem and performing L-BFGS optimization with body- and surface-wave measurements down to 25~s. It takes advantage of assimilating amplitude information in FWI, which is crucial not only to construct anelastic models but also to better resolve elastic heterogeneities \citep{laske_constraints_1996} as amplitudes are sensitive to the second derivative of phase speeds \citep{woodhouse_amplitude_1986}. Nevertheless, their model is relatively smooth, likely due to conservative model updates while demonstrating the sequential/simultaneous inversions of the elastic and anelastic parameters, a challenging task due to the potential trade-offs between parameters \citep[e.g.,][]{espindola-carmona_resolution_2024}. MESWA uses a different FWI workflow \citep{krischer_largescale_2015,thrastarson_lasif_2021} based on simulations with the numerical solver Salvus \citep{afanasiev_modular_2019-1}. Although the principles of the theory and wave simulations and the adjoint method are the same, the FWI workflows may show some notable differences in meshing the crust, optimization method (MESWA uses L-BFGS and MEAD-M20 conjugate gradient), data coverage (MEAD-M20 includes data from the SGS network in the Arabian Peninsula), measurements (MEAD-M20 includes 15-s body waves whereas MESWA is based on 30-100~s period band dominated by surface waves), etc. Also, MEAD-M20 includes a line search at each iteration to ensure reductions in misfit across all measurement categories. While L-BFGS provides more information by approximating the Hessian, which can also be useful for uncertainty estimations \citep[e.g.,][]{liu_pre-conditioned_2021}, within the number of iterations we perform, both methods likely show similar performance \citep{modrak_seismic_2016}. The inclusion of shorter-period body waves down to 15~s in MEAD-M20, given the dominant period of body waves, provides coverage below 300-400~km, which also improves the coverage in the upper mantle together with longer-period surface waves. Among all the differences in the FWI workflows, the main differences between MEAD-M20 and MESWA likely stem from the additional data from the Arabian Peninsula and the inclusion of shorter-period body waves in MEAD-M20.

Besides accounting for complex wave propagation in tomography, one should also address source and structural complexities in the inverse problem through appropriate parameterizations, as demonstrated by \citet{espindolacarmona_anelastic_2024} in the Arabian Peninsula. This study is the first step in our broader plan to improve the resolution of tomographic images and the source parameter estimates for our study area. As seen in the seismograms, there is still vast information to be explained in waveforms. Rather than continuing iterations by changing inversion parameters to fit all waveforms (i.e., decreasing smoothing, etc.), we release the first elastic model after 20 iterations when the misfit flattens to avoid overshooting the model and use it as a reference model to explore further the physics of the medium (i.e., general anisotropy, anelasticity) and source parameters while gradually increasing the resolution of the model of the study area. Specifically, in the next step, our goal is to address source parameters using MEAD-M20 before further addressing the structural complexities in the region, where sequentially updating source and structural parameters would be a good strategy for constructing high-resolution tomographic models. The residual traveltimes we observe in the amplitude histograms also support the importance of uncertainties in source and anelastic parameters, as well as the potential for general anisotropy in the region.


\section{Conclusions}\label{sec:conclusion}

We construct a new P- and S-wave model of the Middle East and its surrounding region, MEAD-M20, using a full-waveform inversion (FWI) framework with 20 conjugate gradient iterations. We assimilate an extensive dataset from 210 earthquakes recorded by 1,215 permanent and temporary global and local seismic networks, including waveforms from the Arabian Peninsula to enhance data coverage, performing the last three iterations with a total of 87,399 measurements of shorter-period body waves (15-50~s) and longer-period body and surface waves (30-100~s), leading to a self-consistent tomographic model of the region, ready for seismic wave simulations.

The moment tensor inversions in a subsequent study \citep{cardozo_gji_2025}, based on 3D Green's functions computed for MEAD-M20, highlight the role of 3D numerical simulations and higher-resolution models in regions like the Middle East, with complex geology and tectonics, in detecting and modeling seismic sources and mitigating seismic hazard. The enhanced resolution of the MEAD-M20 reveals distinct local mantle plumes beneath the Arabian Plate, Jordan, and the Levant, supporting a connection to volcanism in the Harrats, Jordan, and the Karacadağ region. Additionally, we identify sharper remnants of the Tethys Ocean in Eastern Anatolia, which become progressively shallower toward the Makran region, consistent with the subduction history along the Bitlis-Zagros suture zone. Moreover, low-velocity anomalies extending to $\sim 200$~km depth delineate the lithospheric expression of the North and East Anatolian faults.

Future work will focus on updating source parameters and refining the parameterization of the inverse problem to address general anisotropy and anelasticity, starting from MEAD-M20 and combining it with appropriate measurement techniques to achieve higher crustal resolution. Additionally, carefully selected and weighted P-wave datasets, or the integration of multiple data types such as arrival times with full waveforms, may further improve P-wave resolution.
\section*{Acknowledgments}
This is a pre-copyedited, author-produced PDF of an article accepted for publication in Geophysics Journal International following peer review.
This work was performed under the auspices of the U.S. Department of Energy by Lawrence Livermore National Laboratory under Contract DE-AC52-07NA27344. This is LLNL document LLNL-JRNL-2011368. The study was also supported by the National Science Foundation projects with grant numbers EAR-1945565 and OAC-2103621. We thank the Texas Advanced Computing Center (TACC) at the University of Texas at Austin for providing computational resources on `Frontera' system \citep{Frontera2020}. We also thank Cengiz Zabcı and Semih Ergintav for discussions on the geology of the region. We thank \href{https://ds.iris.edu/ds/nodes/dmc/data/}{Earthscope}, \href{https://www.koeri.boun.edu.tr/sismo/2/en/}{Kandilli Observatory and Earthquake Research Institute}, \href{https://sgs.gov.sa/en}{Saudi Geological Survey}, \href{http://irsc.ut.ac.ir/}{Iranian Seismological Survey} for hosting and providing seismic data. We thank the EarthScope Earth Model Collaboration project~\citep{dmc2011data} for hosting a wide range of Earth models which were used in this manuscript for model comparisons. Model cross sections and other maps are prepared using the Generic Mapping Tools \citep{wessel_generic_2019} which are available at \href{https://www.generic-mapping-tools.org/}{https://www.generic-mapping-tools.org/}. 3D global wave propagation solver \texttt{SPECFEM3D Globe} is available from \href{specfem.org}{specfem.org}.
\section*{DATA AVAILABILITY}
 Waveform data provided by Saudi Geological Survey is not open-access. Other waveform data is publicly available from the EarthScope Consortium, the Seismological Facility for Advancement of Geoscience (SAGE-IRIS), Kandilli Observatory and Earthquake Research Center, Mesopotamian Seismological Network of Iraq, and the International Institute of Earthquake Engineering and Seismology of Iran. The open-source spectral-element software package \texttt{SPECFEM3D Globe} used in this study is freely available via \href{specfem.org}{specfem.org}. Information about earthquakes and stations that are used in the study (\href{https://doi.org/10.5281/zenodo.19534092}{https://doi.org/10.5281/zenodo.19534092}) and processing software are available on Zenodo (\href{https://doi.org/10.5281/zenodo.19535099}{https://doi.org/10.5281/zenodo.19535099}). The model will be published on EarthScope Earth Model Collaboration project (IRIS-DMC).

\bibliography{mideast,seis_networks,manual}

@article{cardozo_gji_2025,
  title = {{{Regional Moment Tensor Estimation With 3D Velocity Models Application and Assessment to the 2017 Hojedk, Iran Sequence}}},
  author = {Félix Rodríguez-Cardozo and Jochen Braunmiller and Abdolreza Ghods and Lucas Sawade and Ridvan Örsvuran and Ebru Bozdag},
  year = {2026},
  journal = {Geophys. J. Int.},
  pages = {accepted}
}

@Misc{dmc2011data,
  title={Data Services Products: {EMC}, A repository of {E}arth models},
  author="{IRIS DMC}",
  year={2011},
  howpublished = {https://doi.org/10.17611/DP/EMC.1}}

@inproceedings{Frontera2020,
    author = {Stanzione, Dan and West, John and Evans, R. Todd and Minyard, Tommy and Ghattas, Omar and Panda, Dhabaleswar K.},
    title = {Frontera: The Evolution of Leadership Computing at the National Science Foundation},
    year = {2020},
    isbn = {9781450366892},
    publisher = {Association for Computing Machinery},
    address = {New York, NY, USA},
    url = {https://doi.org/10.1145/3311790.3396656},
    doi = {10.1145/3311790.3396656},
    booktitle = {Practice and Experience in Advanced Research Computing 2020: Catch the Wave},
    pages = {106–111},
    numpages = {6},
    location = {Portland, OR, USA},
    series = {PEARC '20}
}

@article{BozdagHouser2026,
title = {Survey of the Earth’s mantle with radially symmetric and tomographic P-wave models},
journal = {Physics of the Earth and Planetary Interiors},
volume = {374},
pages = {107526},
year = {2026},
issn = {0031-9201},
doi = {https://doi.org/10.1016/j.pepi.2026.107526},
url = {https://www.sciencedirect.com/science/article/pii/S0031920126000361},
author = {Ebru Bozdağ and Christine Houser}
}

@article{Ziyi-adjoint2024,
    author = {Xi, Ziyi and Chen, Min and Wei, Songqiao Shawn and Li, Jiaqi and Zhou, Tong and Wang, Baoshan and Kim, YoungHee},
    title = {EARA2024: a new radially anisotropic seismic velocity model for the crust and upper mantle beneath East Asia and Northwestern pacific subduction zones},
    journal = {Geophysical Journal International},
    volume = {239},
    number = {2},
    pages = {914-935},
    year = {2024},
    month = {11},
    doi = {10.1093/gji/ggae302}
}

@article{abdelfattah_key_2021,
  title = {The key role of conjugate fault system in importing earthquakes into the eastern flank of the {{Red Sea}}},
  author = {Abdelfattah, Ali K. and Jallouli, Chokri and Fnais, Mohamed and Qaysi, Saleh and Alzahrani, Hassan and Mogren, Saad},
  year = 2021,
  month = sep,
  journal = {Earth, Planets and Space},
  volume = {73},
  number = {1},
  pages = {178},
  issn = {1880-5981},
  doi = {10.1186/s40623-021-01513-1},
  urldate = {2025-08-26}
}

@article{adams_source_2009,
  title = {Source {{Parameters}} for {{Moderate Earthquakes}} in the {{Zagros Mountains}} with {{Implications}} for the {{Depth Extent}} of {{Seismicity}}},
  author = {Adams, Aubreya and Brazier, Richard and Nyblade, Andrew and Rodgers, Arthur and {Al-Amri}, Abdullah},
  year = 2009,
  month = jun,
  journal = {Bulletin of the Seismological Society of America},
  volume = {99},
  number = {3},
  pages = {2044--2049},
  issn = {0037-1106},
  doi = {10.1785/0120080314},
  urldate = {2026-06-19}
}

@article{afanasiev_foundations_2016,
  title = {Foundations for a multiscale collaborative {{Earth}} model},
  author = {Afanasiev, Michael and Peter, Daniel and Sager, Korbinian and Simut{\.e}, Saul{\.e} and Ermert, Laura and Krischer, Lion and Fichtner, Andreas},
  year = 2016,
  month = jan,
  journal = {Geophysical Journal International},
  volume = {204},
  number = {1},
  pages = {39--58},
  publisher = {Oxford Academic},
  issn = {0956-540X},
  doi = {10.1093/gji/ggv439},
  urldate = {2020-11-23},
  langid = {english}
}

@article{afanasiev_modular_2019-1,
  title = {Modular and flexible spectral-element waveform modelling in two and three dimensions},
  author = {Afanasiev, Michael and Boehm, Christian and {van~Driel}, Martin and Krischer, Lion and Rietmann, Max and May, Dave A and Knepley, Matthew G and Fichtner, Andreas},
  year = 2019,
  month = mar,
  journal = {Geophysical Journal International},
  volume = {216},
  number = {3},
  pages = {1675--1692},
  issn = {0956-540X},
  doi = {10.1093/gji/ggy469},
  urldate = {2026-04-19}
}

@article{agard_convergence_2005,
  title = {Convergence history across {{Zagros}} ({{Iran}}): constraints from collisional and earlier deformation},
  shorttitle = {Convergence history across {{Zagros}} ({{Iran}})},
  author = {Agard, P. and Omrani, J. and Jolivet, L. and Mouthereau, F.},
  year = 2005,
  month = jul,
  journal = {International Journal of Earth Sciences},
  volume = {94},
  number = {3},
  pages = {401--419},
  issn = {1437-3262},
  doi = {10.1007/s00531-005-0481-4},
  urldate = {2025-07-27},
  langid = {english}
}

@article{agard_zagros_2011,
  title = {Zagros orogeny: a subduction-dominated process},
  shorttitle = {Zagros orogeny},
  author = {Agard, P. and Omrani, J. and Jolivet, L. and Whitechurch, H. and Vrielynck, B. and Spakman, W. and Moni{\'e}, P. and Meyer, B. and Wortel, R.},
  year = 2011,
  month = nov,
  journal = {Geological Magazine},
  volume = {148},
  number = {5-6},
  pages = {692--725},
  issn = {1469-5081, 0016-7568},
  doi = {10.1017/S001675681100046X},
  urldate = {2025-07-29},
  langid = {english}
}

@article{akbayram_evidence_2016,
  title = {Evidence for a minimum 52 ~ \textpm{} ~ 1 ~ km of total offset along the northern branch of the {{North Anatolian Fault}} in northwest {{Turkey}}},
  author = {Akbayram, Kenan and Sorlien, Christopher C. and Okay, Aral I.},
  year = 2016,
  month = feb,
  journal = {Tectonophysics},
  volume = {668--669},
  pages = {35--41},
  issn = {0040-1951},
  doi = {10.1016/j.tecto.2015.11.026},
  urldate = {2024-07-03}
}

@inproceedings{akcelik_high_2003,
  title = {High resolution forward and inverse earthquake modeling on terascale computers},
  booktitle = {Proceedings of the 2003 {{ACM}}/{{IEEE}} conference on {{Supercomputing}}},
  author = {Akcelik, Vokan and Bielak, Jacobo and Biros, George and Epanomeritakis, Ioannis and Fernandez, Antonio and Ghattas, Omar and Kim, Eui Joong and Lopez, Julio and O'Hallaron, David and Tu, Tiankai},
  year = 2003,
  pages = {52},
  publisher = {ACM},
  doi = {10.1145/1048935.1050202}
}

@article{akkar_ground-motion_2018,
  title = {Ground-motion characterization for the probabilistic seismic hazard assessment in {{Turkey}}},
  author = {Akkar, Sinan and Kale, {\"O}zkan and Yakut, Ahmet and {\c C}eken, Ulubey},
  year = 2018,
  month = aug,
  journal = {Bulletin of Earthquake Engineering},
  volume = {16},
  number = {8},
  pages = {3439--3463},
  issn = {1573-1456},
  doi = {10.1007/s10518-017-0101-2},
  urldate = {2025-08-22},
  langid = {english}
}

@article{al-damegh_crustal_2005,
  title = {Crustal structure of the {{Arabian}} plate: new constraints from the analysis of teleseismic receiver functions},
  shorttitle = {Crustal structure of the {{Arabian}} plate},
  author = {{Al-Damegh}, Khaled and Sandvol, Eric and Barazangi, Muawia},
  year = 2005,
  month = mar,
  journal = {Earth and Planetary Science Letters},
  volume = {231},
  number = {3},
  pages = {177--196},
  issn = {0012-821X},
  doi = {10.1016/j.epsl.2004.12.020},
  urldate = {2024-10-05}
}

@article{amini_tomographic_2012,
  title = {Tomographic upper-mantle velocity structure beneath the {{Iranian Plateau}}},
  author = {Amini, Samar and Shomali, Z. Hossein and Koyi, Hemin and Roberts, Roland G.},
  year = 2012,
  month = jul,
  journal = {Tectonophysics},
  volume = {554--557},
  pages = {42--49},
  issn = {0040-1951},
  doi = {10.1016/j.tecto.2012.06.009},
  urldate = {2025-08-19}
}

@article{augustin_13_2021,
  title = {13 million years of seafloor spreading throughout the {{Red Sea Basin}}},
  author = {Augustin, Nico and {van der Zwan}, Froukje M. and Devey, Colin W. and Brandsd{\'o}ttir, Brynd{\'i}s},
  year = 2021,
  month = apr,
  journal = {Nature Communications},
  volume = {12},
  number = {1},
  pages = {2427},
  publisher = {Nature Publishing Group},
  issn = {2041-1723},
  doi = {10.1038/s41467-021-22586-2},
  urldate = {2026-06-19},
  copyright = {2021 The Author(s)},
  langid = {english}
}

@article{bassin_current_2000,
  title = {The current limits of resolution for surface wave tomography in {{North America}}},
  author = {Bassin, Chantal and Laske, Gabi and Masters, Guy},
  year = 2000,
  journal = {Eos Trans. AGU},
  volume = {81},
  pages = {F897},
  urldate = {2024-06-20}
}

@article{benoit_upper_2003,
  title = {Upper mantle {{P}} wave velocity structure and transition zone thickness beneath the {{Arabian Shield}}},
  author = {Benoit, Margaret H. and Nyblade, Andrew A. and VanDecar, John C. and Gurrola, Harold},
  year = 2003,
  journal = {Geophysical Research Letters},
  volume = {30},
  number = {10},
  issn = {1944-8007},
  doi = {10.1029/2002GL016436},
  urldate = {2026-04-13},
  copyright = {Copyright 2003 by the American Geophysical Union.},
  langid = {english}
}

@article{bethoux_earthquake_2016,
  title = {Earthquake relocation using a {{3D}} a-priori geological velocity model from the western {{Alps}} to {{Corsica}}: {{Implication}} for seismic hazard},
  shorttitle = {Earthquake relocation using a {{3D}} a-priori geological velocity model from the western {{Alps}} to {{Corsica}}},
  author = {B{\'e}thoux, Nicole and Theunissen, Thomas and Beslier, Marie-Odile and Font, Yvonne and Thouvenot, Fran{\c c}ois and Dessa, Jean-Xavier and Simon, Soazig and Courrioux, Gabriel and Guillen, Antonio},
  year = 2016,
  month = feb,
  journal = {Tectonophysics},
  volume = {670},
  pages = {82--100},
  issn = {0040-1951},
  doi = {10.1016/j.tecto.2015.12.016},
  urldate = {2025-08-19}
}

@article{bird_updated_2003,
  title = {An updated digital model of plate boundaries},
  author = {Bird, Peter},
  year = 2003,
  journal = {Geochemistry, Geophysics, Geosystems},
  volume = {4},
  number = {3},
  issn = {1525-2027},
  doi = {10.1029/2001GC000252},
  urldate = {2024-09-26},
  copyright = {Copyright 2003 by the American Geophysical Union.},
  langid = {english}
}

@article{biryol_segmented_2011,
  title = {Segmented {{African}} lithosphere beneath the {{Anatolian}} region inferred from teleseismic {{P-wave}} tomography},
  author = {Biryol, C. Berk and Beck, Susan L. and Zandt, George and {\"O}zacar, A. Arda},
  year = 2011,
  month = mar,
  journal = {Geophysical Journal International},
  volume = {184},
  number = {3},
  pages = {1037--1057},
  issn = {0956-540X},
  doi = {10.1111/j.1365-246X.2010.04910.x},
  urldate = {2023-08-09}
}

@article{blom_seismic_2020,
  title = {Seismic waveform tomography of the central and eastern {{Mediterranean}} upper mantle},
  author = {Blom, Nienke and Gokhberg, Alexey and Fichtner, Andreas},
  year = 2020,
  month = apr,
  journal = {Solid Earth},
  volume = {11},
  number = {2},
  pages = {669--690},
  publisher = {Copernicus GmbH},
  issn = {1869-9510},
  doi = {10.5194/se-11-669-2020},
  urldate = {2023-05-22},
  langid = {english}
}

@article{bozdag_crustal_2008,
  title = {On crustal corrections in surface wave tomography},
  author = {Bozda{\u g}, Ebru and Trampert, Jeannot},
  year = 2008,
  month = mar,
  journal = {Geophysical Journal International},
  volume = {172},
  number = {3},
  pages = {1066--1082},
  issn = {0956-540X},
  doi = {10.1111/j.1365-246X.2007.03690.x},
  urldate = {2018-04-24},
  langid = {english}
}

@article{bozdag_global_2016,
  title = {Global adjoint tomography: first-generation model},
  shorttitle = {Global adjoint tomography},
  author = {Bozda{\u g}, Ebru and Peter, Daniel and Lefebvre, Matthieu and Komatitsch, Dimitri and Tromp, Jeroen and Hill, Judith and Podhorszki, Norbert and Pugmire, David},
  year = 2016,
  month = dec,
  journal = {Geophysical Journal International},
  volume = {207},
  number = {3},
  pages = {1739--1766},
  issn = {0956-540X},
  doi = {10.1093/gji/ggw356},
  urldate = {2018-10-26},
  langid = {english}
}

@article{bozdag_misfit_2011,
  title = {Misfit functions for full waveform inversion based on instantaneous phase and envelope measurements},
  author = {Bozda{\u g}, Ebru and Trampert, Jeannot and Tromp, Jeroen},
  year = 2011,
  journal = {Geophysical Journal International},
  volume = {185},
  number = {2},
  pages = {845--870},
  doi = {10.1111/j.1365-246x.2011.04970.x},
  urldate = {2016-09-10}
}

@article{bozdag_p-wave_2025,
  title = {P-{{Wave Arrival-Time Tomography}} of the {{Middle East Using ISC-EHB}} and {{Waveform Data}}},
  author = {Bozda{\u g}, Ebru and Desilva, Susini and Nolet, Guust and Orsvuran, Ridvan and Gok, Rengin and Tarabulsi, Yahya M. and Hosny, Ahmed and Yousef, Khalid and Mousa, Abdullah},
  year = 2025,
  month = jun,
  journal = {Seismica},
  volume = {4},
  number = {1},
  issn = {2816-9387},
  doi = {10.26443/seismica.v4i1.1349},
  urldate = {2025-07-08},
  copyright = {Copyright (c) 2024 Ebru Bozda\u g, Susini Desilva, Guust Nolet, Ridvan Orsvuran, Rengin Gok, Yahya M. Tarabulsi, Ahmed Hosny, Khalid Yousef, Abdullah Mousa},
  langid = {english}
}

@article{bozkurt_neotectonics_2001,
  title = {Neotectonics of {{Turkey}} -- a synthesis},
  author = {Bozkurt, Erdin},
  year = 2001,
  month = jan,
  journal = {Geodinamica Acta},
  volume = {14},
  number = {1-3},
  pages = {3--30},
  publisher = {Taylor \& Francis},
  issn = {0985-3111},
  doi = {10.1080/09853111.2001.11432432},
  urldate = {2024-07-03}
}

@article{braunmiller_regional_2002,
  title = {Regional moment tensor determination in the {{European}}--{{Mediterranean}} area --- initial results},
  author = {Braunmiller, Jochen and Kradolfer, Urs and Baer, Manfred and Giardini, Domenico},
  year = 2002,
  month = oct,
  journal = {Tectonophysics},
  series = {Seismic {{Source Mechanism}} through {{Moment Tensors}}},
  volume = {356},
  number = {1},
  pages = {5--22},
  issn = {0040-1951},
  doi = {10.1016/S0040-1951(02)00374-8},
  urldate = {2026-06-19}
}

@article{camp_madinah_1987,
  title = {The {{Madinah}} eruption, {{Saudi Arabia}}: {{Magma}} mixing and simultaneous extrusion of three basaltic chemical types},
  shorttitle = {The {{Madinah}} eruption, {{Saudi Arabia}}},
  author = {Camp, Victor E. and Hooper, Peter R. and Roobol, M. John and White, D. L.},
  year = 1987,
  month = apr,
  journal = {Bulletin of Volcanology},
  volume = {49},
  number = {2},
  pages = {489--508},
  issn = {1432-0819},
  doi = {10.1007/BF01245475},
  urldate = {2026-06-19},
  langid = {english}
}

@article{camp_upwelling_1992,
  title = {Upwelling asthenosphere beneath western {{Arabia}} and its regional implications},
  author = {Camp, Victor E. and Roobol, M. John},
  year = 1992,
  journal = {Journal of Geophysical Research: Solid Earth},
  volume = {97},
  number = {B11},
  pages = {15255--15271},
  issn = {2156-2202},
  doi = {10.1029/92JB00943},
  urldate = {2025-08-22},
  copyright = {Copyright 1992 by the American Geophysical Union.},
  langid = {english}
}

@article{celli_african_2020,
  title = {African cratonic lithosphere carved by mantle plumes},
  author = {Celli, Nicolas Luca and Lebedev, Sergei and Schaeffer, Andrew J. and Gaina, Carmen},
  year = 2020,
  month = jan,
  journal = {Nature Communications},
  volume = {11},
  number = {1},
  pages = {92},
  publisher = {Nature Publishing Group},
  issn = {2041-1723},
  doi = {10.1038/s41467-019-13871-2},
  urldate = {2024-09-25},
  copyright = {2020 The Author(s)},
  langid = {english}
}

@article{chang_joint_2010,
  title = {Joint inversion for three-dimensional {{S}} velocity mantle structure along the {{Tethyan}} margin},
  author = {Chang, Sung-Joon and {van der Lee}, Suzan and Flanagan, Megan P. and Bedle, Heather and Marone, Federica and Matzel, Eric M. and Pasyanos, Michael E. and Rodgers, Arthur J. and Romanowicz, Barbara and Schmid, Christian},
  year = 2010,
  journal = {Journal of Geophysical Research: Solid Earth},
  volume = {115},
  number = {B8},
  issn = {2156-2202},
  doi = {10.1029/2009JB007204},
  urldate = {2023-05-18},
  langid = {english}
}

@article{chang_mantle_2011,
  title = {Mantle plumes and associated flow beneath {{Arabia}} and {{East Africa}}},
  author = {Chang, Sung-Joon and {Van der Lee}, Suzan},
  year = 2011,
  month = feb,
  journal = {Earth and Planetary Science Letters},
  volume = {302},
  number = {3},
  pages = {448--454},
  issn = {0012-821X},
  doi = {10.1016/j.epsl.2010.12.050},
  urldate = {2023-05-18},
  langid = {english}
}

@article{chang_new_2012,
  title = {A new {{{\emph{P}}}}-velocity model for the {{Tethyan}} margin from a scaled {{{\emph{S}}}}-velocity model and the inversion of {{{\emph{P}}}}- and {{{\emph{PKP}}}}-delay times},
  author = {Chang, Sung-Joon and {Van der Lee}, Suzan and Flanagan, Megan P.},
  year = 2012,
  month = nov,
  journal = {Physics of the Earth and Planetary Interiors},
  volume = {210--211},
  pages = {1--7},
  issn = {0031-9201},
  doi = {10.1016/j.pepi.2012.08.005},
  urldate = {2024-02-16}
}

@article{chiang_seismic_2021,
  title = {Seismic source characterization of the {{Arabian Peninsula}} and {{Zagros Mountains}} from regional moment tensor and coda envelopes},
  author = {Chiang, Andrea and G{\"o}k, Rengin and Tarabulsi, Yahya M. and {El-Hadidy}, Salah Y. and Raddadi, Wael W. and Mousa, Abdullah D.},
  year = 2021,
  month = jan,
  journal = {Arabian Journal of Geosciences},
  volume = {14},
  number = {1},
  pages = {9},
  issn = {1866-7538},
  doi = {10.1007/s12517-020-06266-x},
  urldate = {2026-06-19},
  langid = {english}
}

@article{ciardelli_adjoint_2022,
  title = {Adjoint {{Waveform Tomography}} of {{South America}}},
  author = {Ciardelli, Caio and Assump{\c c}{\~a}o, Marcelo and Bozda{\u g}, Ebru and {van der Lee}, Suzan},
  year = 2022,
  journal = {Journal of Geophysical Research: Solid Earth},
  volume = {127},
  number = {2},
  pages = {e2021JB022575},
  issn = {2169-9356},
  doi = {10.1029/2021JB022575},
  urldate = {2022-02-25},
  langid = {english}
}

@article{civiero_complex_2022,
  title = {A {{Complex Mantle Plume Head Below East Africa-Arabia Shaped}} by the {{Lithosphere-Asthenosphere Boundary Topography}}},
  author = {Civiero, Chiara and Lebedev, Sergei and Celli, Nicolas L.},
  year = 2022,
  journal = {Geochemistry, Geophysics, Geosystems},
  volume = {23},
  number = {11},
  pages = {e2022GC010610},
  issn = {1525-2027},
  doi = {10.1029/2022GC010610},
  urldate = {2025-08-18},
  copyright = {\copyright{} 2022 The Authors.},
  langid = {english}
}

@article{cui_glad-m35_2024,
  title = {{{GLAD-M35}}: {{A}} joint {{P}} and {{S}} global tomographic model with uncertainty quantification},
  shorttitle = {{{GLAD-M35}}},
  author = {Cui, Congyue and Lei, Wenjie and Liu, Qiancheng and Peter, Daniel and Bozda{\u g}, Ebru and Tromp, Jeroen and Hill, Judith and Podhorszki, Norbert and Pugmire, David},
  year = 2024,
  month = oct,
  journal = {Geophysical Journal International},
  volume = {239},
  number = {1},
  pages = {478--502},
  issn = {1365-246X},
  doi = {10.1093/gji/ggae270},
  urldate = {2024-08-18}
}

@article{daradich_mantle_2003,
  title = {Mantle flow, dynamic topography, and rift-flank uplift of {{Arabia}}},
  author = {Daradich, Amy and Mitrovica, Jerry X. and Pysklywec, Russell N. and Willett, Sean D. and Forte, Alessandro M.},
  year = 2003,
  month = oct,
  journal = {Geology},
  volume = {31},
  number = {10},
  pages = {901--904},
  issn = {0091-7613},
  doi = {10.1130/G19661.1},
  urldate = {2023-05-22}
}

@article{duncan_timing_2016,
  title = {Timing and composition of continental volcanism at {{Harrat Hutaymah}}, western {{Saudi Arabia}}},
  author = {Duncan, Robert A. and Kent, Adam J. R. and Thornber, Carl R. and Schlieder, Tyler D. and {Al-Amri}, Abdullah M.},
  year = 2016,
  month = mar,
  journal = {Journal of Volcanology and Geothermal Research},
  volume = {313},
  pages = {1--14},
  issn = {0377-0273},
  doi = {10.1016/j.jvolgeores.2016.01.010},
  urldate = {2026-06-19}
}

@article{durek_radial_1996,
  title = {A radial model of anelasticity consistent with long-period surface-wave attenuation},
  author = {Durek, Joseph J. and Ekstr{\"o}m, G{\"o}ran},
  year = 1996,
  month = feb,
  journal = {Bulletin of the Seismological Society of America},
  volume = {86},
  number = {1A},
  pages = {144--158},
  issn = {0037-1106},
  doi = {10.1785/bssa08601a0144},
  urldate = {2018-05-23},
  langid = {english}
}

@article{dziewonski_preliminary_1981,
  title = {Preliminary reference {{Earth}} model},
  author = {Dziewonski, Adam M. and Anderson, Don L.},
  year = 1981,
  month = jun,
  journal = {Physics of the Earth and Planetary Interiors},
  volume = {25},
  number = {4},
  pages = {297--356},
  issn = {0031-9201},
  doi = {10.1016/0031-9201(81)90046-7},
  urldate = {2018-07-06}
}

@article{ebinger_cenozoic_1998,
  title = {Cenozoic magmatism throughout east {{Africa}} resulting from impact of a single plume},
  author = {Ebinger, C. J. and Sleep, N. H.},
  year = 1998,
  month = oct,
  journal = {Nature},
  volume = {395},
  number = {6704},
  pages = {788--791},
  publisher = {Nature Publishing Group},
  issn = {1476-4687},
  doi = {10.1038/27417},
  urldate = {2023-05-25},
  copyright = {1998 Macmillan Magazines Ltd.},
  langid = {english}
}

@article{ekici_foreland_2014,
  title = {Foreland {{Magmatism}} during the {{Arabia}}--{{Eurasia Collision}}: {{Pliocene}}--{{Quaternary Activity}} of the {{Karacada\u g Volcanic Complex}}, {{SW Turkey}}},
  shorttitle = {Foreland {{Magmatism}} during the {{Arabia}}--{{Eurasia Collision}}},
  author = {Ekici, Taner and Macpherson, Colin G. and Otlu, Nazmi and Fontignie, Denis},
  year = 2014,
  month = sep,
  journal = {Journal of Petrology},
  volume = {55},
  number = {9},
  pages = {1753--1777},
  issn = {0022-3530},
  doi = {10.1093/petrology/egu040},
  urldate = {2024-10-31}
}

@article{ekici_polybaric_2012,
  title = {Polybaric melting of a single mantle source during the {{Neogene Siverek}} phase of the {{Karacada\u g Volcanic Complex}}, {{SE Turkey}}},
  author = {Ekici, Taner and Macpherson, Colin G. and Otlu, Nazmi},
  year = 2012,
  month = aug,
  journal = {Lithos},
  volume = {146--147},
  pages = {152--163},
  issn = {0024-4937},
  doi = {10.1016/j.lithos.2012.05.004},
  urldate = {2025-08-26}
}

@article{el-sharkawy_slab_2020,
  title = {The {{Slab Puzzle}} of the {{Alpine-Mediterranean Region}}: {{Insights From}} a {{New}}, {{High-Resolution}}, {{Shear Wave Velocity Model}} of the {{Upper Mantle}}},
  shorttitle = {The {{Slab Puzzle}} of the {{Alpine-Mediterranean Region}}},
  author = {{El-Sharkawy}, Amr and Meier, Thomas and Lebedev, Sergei and Behrmann, Jan H. and Hamada, Mona and Cristiano, Luigia and Weidle, Christian and K{\"o}hn, Daniel},
  year = 2020,
  journal = {Geochemistry, Geophysics, Geosystems},
  volume = {21},
  number = {8},
  pages = {e2020GC008993},
  issn = {1525-2027},
  doi = {10.1029/2020GC008993},
  urldate = {2023-08-09},
  copyright = {\copyright 2020. The Authors.},
  langid = {english}
}

@article{engdahl_isc-ehb_2020,
  title = {{{ISC-EHB}} 1964--2016, an {{Improved Data Set}} for {{Studies}} of {{Earth Structure}} and {{Global Seismicity}}},
  author = {Engdahl, E. R. and Di Giacomo, D. and Sakarya, B. and Gkarlaouni, C. G. and Harris, J. and Storchak, D. A.},
  year = 2020,
  month = jan,
  journal = {Earth and Space Science},
  volume = {7},
  number = {1},
  pages = {e2019EA000897},
  publisher = {John Wiley \& Sons, Ltd},
  issn = {2333-5084},
  doi = {10.1029/2019EA000897},
  urldate = {2024-06-17}
}

@article{espindola-carmona_resolution_2024,
  title = {Resolution and trade-offs in global anelastic full-waveform inversion},
  author = {{Espindola-Carmona}, Armando and {\"O}rsvuran, R{\i}dvan and Mai, P Martin and Bozda{\u g}, Ebru and Peter, Daniel B},
  year = 2024,
  month = feb,
  journal = {Geophysical Journal International},
  volume = {236},
  number = {2},
  pages = {952--966},
  issn = {0956-540X},
  doi = {10.1093/gji/ggad462},
  urldate = {2024-06-14}
}

@article{espindolacarmona_anelastic_2024,
  title = {Anelastic {{Tomography}} of the {{Arabian Plate}}},
  author = {{Espindola-Carmona}, Armando and Peter, Daniel B. and Parisi, Laura and Mai, P. Martin},
  year = 2024,
  month = may,
  journal = {Bulletin of the Seismological Society of America},
  volume = {114},
  number = {3},
  pages = {1347--1364},
  issn = {0037-1106},
  doi = {10.1785/0120230216},
  urldate = {2024-06-13}
}

@article{faccenna_mantle_2013,
  title = {Mantle convection in the {{Middle East}}: {{Reconciling Afar}} upwelling, {{Arabia}} indentation and {{Aegean}} trench rollback},
  shorttitle = {Mantle convection in the {{Middle East}}},
  author = {Faccenna, Claudio and Becker, Thorsten W. and Jolivet, Laurent and Keskin, Mehmet},
  year = 2013,
  month = aug,
  journal = {Earth and Planetary Science Letters},
  volume = {375},
  pages = {254--269},
  issn = {0012-821X},
  doi = {10.1016/j.epsl.2013.05.043},
  urldate = {2024-10-31}
}

@article{ferreira_robustness_2010,
  title = {On the robustness of global radially anisotropic surface wave tomography},
  author = {Ferreira, A. M. G. and Woodhouse, J. H. and Visser, K. and Trampert, J.},
  year = 2010,
  journal = {Journal of Geophysical Research: Solid Earth},
  volume = {115},
  number = {B4},
  issn = {2156-2202},
  doi = {10.1029/2009JB006716},
  urldate = {2024-06-17},
  copyright = {Copyright 2010 by the American Geophysical Union.},
  langid = {english}
}

@article{fichtner_adjoint_2006,
  title = {The adjoint method in seismology: {{I}}. {{Theory}}},
  shorttitle = {The adjoint method in seismology},
  author = {Fichtner, A. and Bunge, H. -P. and Igel, H.},
  year = 2006,
  month = aug,
  journal = {Physics of the Earth and Planetary Interiors},
  volume = {157},
  number = {1},
  pages = {86--104},
  issn = {0031-9201},
  doi = {10.1016/j.pepi.2006.03.016},
  urldate = {2019-09-06}
}

@article{fichtner_deep_2013,
  title = {The deep structure of the {{North Anatolian Fault Zone}}},
  author = {Fichtner, Andreas and Saygin, Erdinc and Taymaz, Tuncay and Cupillard, Paul and Capdeville, Yann and Trampert, Jeannot},
  year = 2013,
  month = jul,
  journal = {Earth and Planetary Science Letters},
  volume = {373},
  pages = {109--117},
  issn = {0012-821X},
  doi = {10.1016/j.epsl.2013.04.027},
  urldate = {2020-11-23},
  langid = {english}
}

@article{fichtner_full_2010,
  title = {Full waveform tomography for radially anisotropic structure: {{New}} insights into present and past states of the {{Australasian}} upper mantle},
  shorttitle = {Full waveform tomography for radially anisotropic structure},
  author = {Fichtner, Andreas and Kennett, Brian L. N. and Igel, Heiner and Bunge, Hans-Peter},
  year = 2010,
  month = feb,
  journal = {Earth and Planetary Science Letters},
  volume = {290},
  number = {3},
  pages = {270--280},
  issn = {0012-821X},
  doi = {10.1016/j.epsl.2009.12.003},
  urldate = {2026-06-19}
}

@article{fichtner_resolution_2011,
  title = {Resolution analysis in full waveform inversion},
  shorttitle = {Resolution analysis in full waveform inversion},
  author = {Fichtner, Andreas and Trampert, Jeannot},
  year = 2011,
  month = dec,
  journal = {Geophysical Journal International},
  volume = {187},
  number = {3},
  pages = {1604--1624},
  issn = {0956540X},
  doi = {10.1111/j.1365-246X.2011.05218.x},
  urldate = {2017-05-18},
  langid = {english}
}

@article{fletcher_function_1964,
  title = {Function minimization by conjugate gradients},
  author = {Fletcher, R. and Reeves, C. M.},
  year = 1964,
  month = jan,
  journal = {The Computer Journal},
  volume = {7},
  number = {2},
  pages = {149--154},
  issn = {0010-4620},
  doi = {10.1093/comjnl/7.2.149},
  urldate = {2017-07-05}
}

@article{giardini_seismic_2018,
  title = {Seismic hazard map of the {{Middle East}}},
  author = {Giardini, Domenico and Danciu, Laurentiu and Erdik, Mustafa and {\c S}e{\c s}etyan, Karin and Demircio{\u g}lu T{\"u}msa, Mine B. and Akkar, Sinan and G{\"u}len, Levent and Zare, Mehdi},
  year = 2018,
  month = aug,
  journal = {Bulletin of Earthquake Engineering},
  volume = {16},
  number = {8},
  pages = {3567--3570},
  issn = {1573-1456},
  doi = {10.1007/s10518-018-0347-3},
  urldate = {2025-08-22},
  langid = {english}
}

@article{gok_lithospheric_2007,
  title = {Lithospheric structure of the continent---continent collision zone: eastern {{Turkey}}},
  shorttitle = {Lithospheric structure of the continent---continent collision zone},
  author = {G{\"o}k, Rengin and Pasyanos, Michael E. and Zor, Ekrem},
  year = 2007,
  month = jun,
  journal = {Geophysical Journal International},
  volume = {169},
  number = {3},
  pages = {1079--1088},
  issn = {0956-540X},
  doi = {10.1111/j.1365-246X.2006.03288.x},
  urldate = {2023-05-22}
}

@article{gok_moment_2016,
  title = {Moment {{Magnitudes}} of {{Local}}/{{Regional Events}} from {{1D Coda Calibrations}} in the {{Broader Middle East Region}}},
  author = {G{\"o}k, Rengin and Kaviani, Ayoub and Matzel, Eric M. and Pasyanos, Michael E. and Mayeda, Kevin and Yetirmishli, Gurban and El-Hussain, Issa and Al-Amri, Abdullah and Al-Jeri, Farah and Godoladze, Tea and Kalafat, Dogan and Sandvol, Eric A. and Walter, William R.},
  year = 2016,
  month = sep,
  journal = {Bulletin of the Seismological Society of America},
  volume = {106},
  number = {5},
  pages = {1926--1938},
  issn = {0037-1106},
  doi = {10.1785/0120160045},
  urldate = {2025-08-22}
}

@article{guvercin_active_2022,
  title = {Active {{Seismotectonics}} of the {{East Anatolian Fault}}},
  author = {G{\"u}vercin, Sezim Ezgi and Karabulut, Hayrullah and Konca, A {\"O}zg{\"u}n and Do{\u g}an, U{\u g}ur and Ergintav, Semih},
  year = 2022,
  month = feb,
  journal = {Geophysical Journal International},
  pages = {ggac045},
  issn = {0956-540X},
  doi = {10.1093/gji/ggac045},
  urldate = {2022-02-09}
}

@article{hansen_imaging_2007,
  title = {Imaging ruptured lithosphere beneath the {{Red Sea}} and {{Arabian Peninsula}}},
  author = {Hansen, Samantha E. and Rodgers, Arthur J. and Schwartz, Susan Y. and {Al-Amri}, Abdullah M. S.},
  year = 2007,
  month = jul,
  journal = {Earth and Planetary Science Letters},
  volume = {259},
  number = {3},
  pages = {256--265},
  issn = {0012-821X},
  doi = {10.1016/j.epsl.2007.04.035},
  urldate = {2025-08-01}
}

@article{hansen_mantle_2012,
  title = {Mantle structure beneath {{Africa}} and {{Arabia}} from adaptively parameterized {{P-wave}} tomography: {{Implications}} for the origin of {{Cenozoic Afro-Arabian}} tectonism},
  shorttitle = {Mantle structure beneath {{Africa}} and {{Arabia}} from adaptively parameterized {{P-wave}} tomography},
  author = {Hansen, Samantha E. and Nyblade, Andrew A. and Benoit, Margaret H.},
  year = 2012,
  month = feb,
  journal = {Earth and Planetary Science Letters},
  volume = {319--320},
  pages = {23--34},
  issn = {0012-821X},
  doi = {10.1016/j.epsl.2011.12.023},
  urldate = {2023-05-26},
  langid = {english}
}

@article{hansen_seismic_2008,
  title = {Seismic velocity structure and depth-dependence of anisotropy in the {{Red Sea}} and {{Arabian}} shield from surface wave analysis},
  author = {Hansen, Samantha E. and Gaherty, James B. and Schwartz, Susan Y. and Rodgers, Arthur J. and {Al-Amri}, Abdullah M. S.},
  year = 2008,
  journal = {Journal of Geophysical Research: Solid Earth},
  volume = {113},
  number = {B10},
  issn = {2156-2202},
  doi = {10.1029/2007JB005335},
  urldate = {2025-08-01},
  copyright = {Copyright 2008 by the American Geophysical Union.},
  langid = {english}
}

@article{hatzfeld_comparisons_2010,
  title = {Comparisons of the kinematics and deep structures of the {{Zagros}} and {{Himalaya}} and of the {{Iranian}} and {{Tibetan}} plateaus and geodynamic implications},
  author = {Hatzfeld, Denis and Molnar, Peter},
  year = 2010,
  journal = {Reviews of Geophysics},
  volume = {48},
  number = {2},
  issn = {1944-9208},
  doi = {10.1029/2009RG000304},
  urldate = {2026-06-19},
  copyright = {Copyright 2010 by the American Geophysical Union.},
  langid = {english}
}

@article{hubert-ferrari_morphology_2002,
  title = {Morphology, displacement, and slip rates along the {{North Anatolian Fault}}, {{Turkey}}},
  author = {{Hubert-Ferrari}, Aur{\'e}lia and Armijo, Rolando and King, Geoffrey and Meyer, Bertrand and Barka, Aykut},
  year = 2002,
  journal = {Journal of Geophysical Research: Solid Earth},
  volume = {107},
  number = {B10},
  pages = {ETG 9-1-ETG 9-33},
  issn = {2156-2202},
  doi = {10.1029/2001JB000393},
  urldate = {2024-07-03},
  copyright = {Copyright 2002 by the American Geophysical Union.},
  langid = {english}
}

@article{kalafat_seismicity_2021,
  title = {Seismicity of {{Turkey}} and {{Real-Time Seismology Applications}} in {{Determining Earthquake Hazard}}},
  author = {Kalafat, Dogan and Z{\"u}lfikar, A. Can and Akcan, Seyhan Okuyan},
  year = 2021,
  month = dec,
  journal = {Academic Platform Journal of Natural Hazards and Disaster Management},
  volume = {2},
  number = {2},
  pages = {96--111},
  publisher = {Akademik Perspektif Derne\u gi},
  issn = {2717-8714},
  doi = {10.52114/apjhad.1039670},
  urldate = {2025-08-22},
  langid = {english}
}

@article{karabulut_long_2023,
  title = {Long silence on the {{East Anatolian Fault Zone}} ({{Southern Turkey}}) ends with devastating double earthquakes (6 {{February}} 2023) over a seismic gap: implications for the seismic potential in the {{Eastern Mediterranean}} region},
  shorttitle = {Long silence on the {{East Anatolian Fault Zone}} ({{Southern Turkey}}) ends with devastating double earthquakes (6 {{February}} 2023) over a seismic gap},
  author = {Karabulut, Hayrullah and G{\"u}vercin, Sezim Ezgi and Hollingsworth, James and Konca, Ali {\"O}zg{\"u}n},
  year = 2023,
  month = may,
  journal = {Journal of the Geological Society},
  volume = {180},
  number = {3},
  pages = {jgs2023-021},
  issn = {0016-7649},
  doi = {10.1144/jgs2023-021},
  urldate = {2025-08-22}
}

@article{karasozen_normal_2016,
  title = {Normal faulting in the {{Simav}} graben of western {{Turkey}} reassessed with calibrated earthquake relocations},
  author = {Karas{\"o}zen, Ezgi and Nissen, Edwin and Bergman, Eric A. and Johnson, Kendra L. and Walters, Richard J.},
  year = 2016,
  journal = {Journal of Geophysical Research: Solid Earth},
  volume = {121},
  number = {6},
  pages = {4553--4574},
  issn = {2169-9356},
  doi = {10.1002/2016JB012828},
  urldate = {2026-06-19},
  copyright = {\copyright 2016. American Geophysical Union. All Rights Reserved.},
  langid = {english}
}

@article{karasozen_seismotectonics_2019,
  title = {Seismotectonics of the {{Zagros}} ({{Iran}}) {{From Orogen-Wide}}, {{Calibrated Earthquake Relocations}}},
  author = {Karas{\"o}zen, Ezgi and Nissen, Edwin and Bergman, Eric A. and Ghods, Abdolreza},
  year = 2019,
  journal = {Journal of Geophysical Research: Solid Earth},
  volume = {124},
  number = {8},
  pages = {9109--9129},
  issn = {2169-9356},
  doi = {10.1029/2019JB017336},
  urldate = {2026-06-19},
  copyright = {\copyright 2019. American Geophysical Union. All Rights Reserved.},
  langid = {english}
}

@article{kaviani_crustal_2020,
  title = {Crustal and uppermost mantle shear wave velocity structure beneath the {{Middle East}} from surface wave tomography},
  author = {Kaviani, Ayoub and Paul, Anne and Moradi, Ali and Mai, Paul Martin and Pilia, Simone and Boschi, Lapo and R{\"u}mpker, Georg and Lu, Yang and Tang, Zheng and Sandvol, Eric},
  year = 2020,
  month = may,
  journal = {Geophysical Journal International},
  volume = {221},
  number = {2},
  pages = {1349--1365},
  issn = {0956-540X},
  doi = {10.1093/gji/ggaa075},
  urldate = {2023-05-18}
}

@article{keskin_geochronology_2012,
  title = {The geochronology and origin of mantle sources for late cenozoic intraplate volcanism in the frontal part of the {{Arabian}} plate in the {{Karacada\u g}} neovolcanic area of {{Turkey}}. {{Part}} 2. {{The}} results of geochemical and isotope ({{Sr-Nd-Pb}}) studies},
  author = {Keskin, M. and Chugaev, A. V. and Lebedev, V. A. and Sharkov, E. V. and Oyan, V. and Kavak, O.},
  year = 2012,
  month = nov,
  journal = {Journal of Volcanology and Seismology},
  volume = {6},
  number = {6},
  pages = {361--382},
  issn = {1819-7108},
  doi = {10.1134/S0742046312060048},
  urldate = {2024-10-31},
  langid = {english}
}

@article{khodaverdian_seismicity_2016,
  title = {Seismicity {{Parameters}} and {{Spatially Smoothed Seismicity Model}} for {{Iran}}},
  author = {Khodaverdian, A. and Zafarani, H. and Rahimian, M. and Dehnamaki, V.},
  year = 2016,
  month = may,
  journal = {Bulletin of the Seismological Society of America},
  volume = {106},
  number = {3},
  pages = {1133--1150},
  issn = {0037-1106},
  doi = {10.1785/0120150178},
  urldate = {2025-08-22}
}

@article{kiuchi_groundmotion_2019,
  title = {Ground-{{Motion Prediction Equations}} for {{Western Saudi Arabia}}},
  author = {Kiuchi, Ryota and Mooney, Walter D. and Zahran, Hani M.},
  year = 2019,
  month = oct,
  journal = {Bulletin of the Seismological Society of America},
  volume = {109},
  number = {6},
  pages = {2722--2737},
  issn = {0037-1106},
  doi = {10.1785/0120180302},
  urldate = {2026-06-19}
}

@article{komatitsch_anelastic_2016,
  title = {Anelastic sensitivity kernels with parsimonious storage for adjoint tomography and full waveform inversion},
  author = {Komatitsch, Dimitri and Xie, Zhinan and Bozda{\u g}, Ebru and {Sales de Andrade}, Elliott and Peter, Daniel and Liu, Qinya and Tromp, Jeroen},
  year = 2016,
  month = sep,
  journal = {Geophysical Journal International},
  volume = {206},
  number = {3},
  pages = {1467--1478},
  issn = {0956-540X},
  doi = {10.1093/gji/ggw224},
  urldate = {2018-05-21},
  langid = {english}
}

@article{komatitsch_spectral-element_2002,
  title = {Spectral-element simulations of global seismic wave propagation---{{II}}. {{Three-dimensional}} models, oceans, rotation and self-gravitation},
  author = {Komatitsch, Dimitri and Tromp, Jeroen},
  year = 2002,
  month = jul,
  journal = {Geophysical Journal International},
  volume = {150},
  number = {1},
  pages = {303--318},
  issn = {0956-540X},
  doi = {10.1046/j.1365-246X.2002.01716.x},
  urldate = {2018-05-21},
  langid = {english}
}

@article{komatitsch_spectral-element_2002-1,
  title = {Spectral-element simulations of global seismic wave propagation---{{I}}. {{Validation}}},
  author = {Komatitsch, Dimitri and Tromp, Jeroen},
  year = 2002,
  month = may,
  journal = {Geophysical Journal International},
  volume = {149},
  number = {2},
  pages = {390--412},
  issn = {0956-540X},
  doi = {10.1046/j.1365-246X.2002.01653.x},
  urldate = {2018-05-21},
  langid = {english}
}

@article{koulakov_evidence_2016,
  title = {Evidence for anomalous mantle upwelling beneath the {{Arabian Platform}} from travel time tomography inversion},
  author = {Koulakov, Ivan and Burov, Evgeniy and Cloetingh, Sierd and El Khrepy, Sami and {Al-Arifi}, Nassir and Bushenkova, Natalia},
  year = 2016,
  month = jan,
  journal = {Tectonophysics},
  volume = {667},
  pages = {176--188},
  issn = {0040-1951},
  doi = {10.1016/j.tecto.2015.11.022},
  urldate = {2024-11-10}
}

@article{krischer_largescale_2015,
  title = {Large-{{Scale Seismic Inversion Framework}}},
  author = {Krischer, Lion and Fichtner, Andreas and Zukauskaite, Saule and Igel, Heiner},
  year = 2015,
  month = jun,
  journal = {Seismological Research Letters},
  volume = {86},
  number = {4},
  pages = {1198--1207},
  issn = {0895-0695},
  doi = {10.1785/0220140248},
  urldate = {2026-04-19}
}

@article{kustowski_anisotropic_2008,
  title = {Anisotropic shear-wave velocity structure of the {{Earth}}'s mantle: {{A}} global model},
  shorttitle = {Anisotropic shear-wave velocity structure of the {{Earth}}'s mantle},
  author = {Kustowski, B. and Ekstr{\"o}m, G. and Dziewo{\'n}ski, A. M.},
  year = 2008,
  month = jun,
  journal = {Journal of Geophysical Research: Solid Earth},
  volume = {113},
  number = {B6},
  issn = {0148-0227},
  doi = {10.1029/2007JB005169},
  urldate = {2018-10-26}
}

@article{laske_constraints_1996,
  title = {Constraints on global phase velocity maps from long-period polarization data},
  author = {Laske, G. and Masters, G.},
  year = 1996,
  journal = {Journal of Geophysical Research: Solid Earth},
  volume = {101},
  number = {B7},
  pages = {16059--16075},
  issn = {2156-2202},
  doi = {10.1029/96JB00526},
  urldate = {2019-03-05},
  copyright = {Copyright 1996 by the American Geophysical Union.},
  langid = {english}
}

@article{lei_global_2020,
  title = {Global adjoint tomography---model {{GLAD-M25}}},
  author = {Lei, Wenjie and Ruan, Youyi and Bozda{\u g}, Ebru and Peter, Daniel and Lefebvre, Matthieu and Komatitsch, Dimitri and Tromp, Jeroen and Hill, Judith and Podhorszki, Norbert and Pugmire, David},
  year = 2020,
  month = oct,
  journal = {Geophysical Journal International},
  volume = {223},
  number = {1},
  pages = {1--21},
  issn = {0956-540X},
  doi = {10.1093/gji/ggaa253},
  urldate = {2021-01-25}
}

@article{leroy_recent_2010,
  title = {Recent off-axis volcanism in the eastern {{Gulf}} of {{Aden}}: {{Implications}} for plume--ridge interaction},
  shorttitle = {Recent off-axis volcanism in the eastern {{Gulf}} of {{Aden}}},
  author = {Leroy, Sylvie and {d'Acremont}, Elia and Tiberi, Christel and Basuyau, Cl{\'e}mence and Autin, Julia and Lucazeau, Francis and Sloan, Heather},
  year = 2010,
  month = apr,
  journal = {Earth and Planetary Science Letters},
  volume = {293},
  number = {1},
  pages = {140--153},
  issn = {0012-821X},
  doi = {10.1016/j.epsl.2010.02.036},
  urldate = {2023-05-22},
  langid = {english}
}

@article{lim_asthenospheric_2020,
  title = {Asthenospheric {{Flow}} of {{Plume Material Beneath Arabia Inferred From S Wave Traveltime Tomography}}},
  author = {Lim, Jung-A and Chang, Sung-Joon and Mai, P. Martin and Zahran, Hani},
  year = 2020,
  journal = {Journal of Geophysical Research: Solid Earth},
  volume = {125},
  number = {8},
  pages = {e2020JB019668},
  issn = {2169-9356},
  doi = {10.1029/2020JB019668},
  urldate = {2024-11-10},
  copyright = {\copyright 2020. The Authors.},
  langid = {english}
}

@article{liu_pre-conditioned_2021,
  title = {Pre-conditioned {{BFGS-based}} uncertainty quantification in elastic full-waveform inversion},
  author = {Liu, Qiancheng and Beller, Stephen and Lei, Wenjie and Peter, Daniel and Tromp, Jeroen},
  year = 2021,
  month = feb,
  journal = {Geophysical Journal International},
  volume = {228},
  number = {2},
  pages = {796--815},
  issn = {0956-540X},
  doi = {10.1093/gji/ggab375},
  urldate = {2021-11-02}
}

@phdthesis{luo_seismic_2012,
  title = {Seismic imaging and inversion based on spectral-element and adjoint methods},
  author = {Luo, Yang},
  year = 2012,
  urldate = {2018-03-01},
  langid = {english},
  school = {Princeton University}
}

@article{luo_wave-equation_1991,
  title = {Wave-equation traveltime inversion},
  author = {Luo, Yi and Schuster, Gerard T.},
  year = 1991,
  journal = {Geophysics},
  volume = {56},
  number = {5},
  pages = {645--653},
  doi = {10.1190/1.1889952},
  urldate = {2017-01-30}
}

@article{maggi_automated_2009,
  title = {An automated time-window selection algorithm for seismic tomography},
  author = {Maggi, Alessia and Tape, Carl and Chen, Min and Chao, Daniel and Tromp, Jeroen},
  year = 2009,
  journal = {Geophysical Journal International},
  volume = {178},
  number = {1},
  pages = {257--281},
  doi = {10.1111/j.1365-246x.2009.04099.x},
  urldate = {2016-09-10}
}

@article{mcclusky_global_2000,
  title = {Global {{Positioning System}} constraints on plate kinematics and dynamics in the eastern {{Mediterranean}} and {{Caucasus}}},
  author = {McClusky, S. and Balassanian, S. and Barka, A. and Demir, C. and Ergintav, S. and Georgiev, I. and Gurkan, O. and Hamburger, M. and Hurst, K. and Kahle, H. and Kastens, K. and Kekelidze, G. and King, R. and Kotzev, V. and Lenk, O. and Mahmoud, S. and Mishin, A. and Nadariya, M. and Ouzounis, A. and Paradissis, D. and Peter, Y. and Prilepin, M. and Reilinger, R. and Sanli, I. and Seeger, H. and Tealeb, A. and Toks{\"o}z, M. N. and Veis, G.},
  year = 2000,
  journal = {Journal of Geophysical Research: Solid Earth},
  volume = {105},
  number = {B3},
  pages = {5695--5719},
  issn = {2156-2202},
  doi = {10.1029/1999JB900351},
  urldate = {2024-07-03},
  copyright = {Copyright 2000 by the American Geophysical Union.},
  langid = {english}
}

@article{menant_3d_2016,
  title = {{{3D}} numerical modeling of mantle flow, crustal dynamics and magma genesis associated with slab roll-back and tearing: {{The}} eastern {{Mediterranean}} case},
  shorttitle = {{{3D}} numerical modeling of mantle flow, crustal dynamics and magma genesis associated with slab roll-back and tearing},
  author = {Menant, Armel and Sternai, Pietro and Jolivet, Laurent and {Guillou-Frottier}, Laurent and Gerya, Taras},
  year = 2016,
  month = may,
  journal = {Earth and Planetary Science Letters},
  volume = {442},
  pages = {93--107},
  issn = {0012-821X},
  doi = {10.1016/j.epsl.2016.03.002},
  urldate = {2024-07-03}
}

@article{menant_kinematic_2016,
  title = {Kinematic reconstructions and magmatic evolution illuminating crustal and mantle dynamics of the eastern {{Mediterranean}} region since the late {{Cretaceous}}},
  author = {Menant, Armel and Jolivet, Laurent and Vrielynck, Bruno},
  year = 2016,
  month = apr,
  journal = {Tectonophysics},
  volume = {675},
  pages = {103--140},
  issn = {0040-1951},
  doi = {10.1016/j.tecto.2016.03.007},
  urldate = {2023-05-22},
  langid = {english}
}

@article{modrak_seismic_2016,
  title = {Seismic waveform inversion best practices: regional, global and exploration test cases},
  shorttitle = {Seismic waveform inversion best practices},
  author = {Modrak, Ryan and Tromp, Jeroen},
  year = 2016,
  month = sep,
  journal = {Geophysical Journal International},
  volume = {206},
  number = {3},
  pages = {1864--1889},
  issn = {0956-540X},
  doi = {10.1093/gji/ggw202},
  urldate = {2017-06-27}
}

@article{mokhtar_shear_1994,
  title = {Shear wave velocity structures of the {{Arabian Peninsula}}},
  author = {Mokhtar, Talal A. and {Al-Saeed}, Mohammed M.},
  year = 1994,
  month = feb,
  journal = {Tectonophysics},
  volume = {230},
  number = {1},
  pages = {105--125},
  issn = {0040-1951},
  doi = {10.1016/0040-1951(94)90149-X},
  urldate = {2024-09-25}
}

@article{molinari_epcrust_2011,
  title = {{{EPcrust}}: a reference crustal model for the {{European Plate}}},
  shorttitle = {{{EPcrust}}},
  author = {Molinari, Irene and Morelli, Andrea},
  year = 2011,
  month = apr,
  journal = {Geophysical Journal International},
  volume = {185},
  number = {1},
  pages = {352--364},
  issn = {0956-540X},
  doi = {10.1111/j.1365-246X.2011.04940.x},
  urldate = {2025-07-01}
}

@article{montagner_petrological_1989,
  title = {Petrological constraints on seismic anisotropy},
  author = {Montagner, Jean-Paul and Anderson, Don L.},
  year = 1989,
  month = apr,
  journal = {Physics of the Earth and Planetary Interiors},
  volume = {54},
  number = {1},
  pages = {82--105},
  issn = {0031-9201},
  doi = {10.1016/0031-9201(89)90189-1},
  urldate = {2020-12-02},
  langid = {english}
}

@article{montelli_global_2004,
  title = {Global {{P}} and {{PP}} traveltime tomography: rays versus waves},
  shorttitle = {Global {{P}} and {{PP}} traveltime tomography},
  author = {Montelli, R. and Nolet, G. and Masters, G. and Dahlen, F. A. and Hung, S.-H.},
  year = 2004,
  month = aug,
  journal = {Geophysical Journal International},
  volume = {158},
  number = {2},
  pages = {637--654},
  issn = {0956-540X},
  doi = {10.1111/j.1365-246X.2004.02346.x},
  urldate = {2018-10-31},
  langid = {english}
}

@article{mouthereau_building_2012,
  title = {Building the {{Zagros}} collisional orogen: {{Timing}}, strain distribution and the dynamics of {{Arabia}}/{{Eurasia}} plate convergence},
  shorttitle = {Building the {{Zagros}} collisional orogen},
  author = {Mouthereau, F. and Lacombe, O. and Verg{\'e}s, J.},
  year = 2012,
  month = apr,
  journal = {Tectonophysics},
  volume = {532--535},
  pages = {27--60},
  issn = {0040-1951},
  doi = {10.1016/j.tecto.2012.01.022},
  urldate = {2024-07-03}
}

@article{movaghari_3-d_2020,
  title = {3-{{D}} crustal structure of the {{Iran}} plateau using phase velocity ambient noise tomography},
  author = {Movaghari, R and Doloei, G Javan},
  year = 2020,
  month = mar,
  journal = {Geophysical Journal International},
  volume = {220},
  number = {3},
  pages = {1555--1568},
  issn = {0956-540X},
  doi = {10.1093/gji/ggz537},
  urldate = {2023-05-26}
}

@article{mukhopadhyay_incipient_2013,
  title = {Incipient status of dyke intrusion in top crust -- evidences from the {{Al-Ays}} 2009 earthquake swarm, {{Harrat Lunayyir}}, {{SW Saudi Arabia}}},
  author = {Mukhopadhyay, Basab and Mogren, Saad and Mukhopadhyay, Manoj and Dasgupta, Sujit},
  year = 2013,
  month = mar,
  journal = {Geomatics, Natural Hazards and Risk},
  volume = {4},
  number = {1},
  pages = {30--48},
  publisher = {Taylor \& Francis},
  issn = {1947-5705},
  doi = {10.1080/19475705.2012.663794},
  urldate = {2026-06-19}
}

@article{nissen_new_2011,
  title = {New views on earthquake faulting in the {{Zagros}} fold-and-thrust belt of {{Iran}}},
  author = {Nissen, Edwin and Tatar, Mohammad and Jackson, James A. and Allen, Mark B.},
  year = 2011,
  month = sep,
  journal = {Geophysical Journal International},
  volume = {186},
  number = {3},
  pages = {928--944},
  issn = {0956-540X},
  doi = {10.1111/j.1365-246X.2011.05119.x},
  urldate = {2026-06-19}
}

@article{nissen_zagros_2014,
  title = {Zagros ``phantom earthquakes'' reassessed---{{The}} interplay of seismicity and deep salt flow in the {{Simply Folded Belt}}?},
  author = {Nissen, Edwin and Jackson, James and Jahani, Salman and Tatar, Mohammad},
  year = 2014,
  journal = {Journal of Geophysical Research: Solid Earth},
  volume = {119},
  number = {4},
  pages = {3561--3583},
  issn = {2169-9356},
  doi = {10.1002/2013JB010796},
  urldate = {2026-06-19},
  copyright = {\copyright 2014. American Geophysical Union. All Rights Reserved.},
  langid = {english}
}

@article{nolet_picking_2025,
  title = {Picking first arrivals in hydroacoustic seismograms from {{MERMAID}} floats},
  author = {Nolet, Guust and Hoang, Nguyen Ba and Bonnieux, S{\'e}bastien and Kondo, Yuko and Kong, Fanchao and Obayashi, Masayuki and Pipatprathanporn, Sirawich and Sigloch, Karin and Simon, Joel D. and Simons, Frederik J. and Sogioka, Hiroko and Yoshimitsu, Junko and Zhang, Qinling},
  year = 2025,
  month = apr,
  journal = {Seismica},
  volume = {4},
  number = {1},
  issn = {2816-9387},
  doi = {10.26443/seismica.v4i1.1505},
  urldate = {2025-05-21},
  copyright = {Copyright (c) 2024 Guust Nolet, Nguyen Ba Hoang, S\'ebastien Bonnieux, Yuko Kondo, Fanchao Kong, Masayuki Obayashi, Sirawich Pipatprathanporn, Karin Sigloch, Joel D. Simon, Frederik J. Simons, Hiroko Sogioka, Junko Yoshimitsu, Qinling Zhang},
  langid = {english}
}

@article{omrani_arc-magmatism_2008,
  title = {Arc-magmatism and subduction history beneath the {{Zagros Mountains}}, {{Iran}}: {{A}} new report of adakites and geodynamic consequences},
  shorttitle = {Arc-magmatism and subduction history beneath the {{Zagros Mountains}}, {{Iran}}},
  author = {Omrani, Jafar and Agard, Philippe and Whitechurch, Hubert and Benoit, Mathieu and Prouteau, Ga{\"e}lle and Jolivet, Laurent},
  year = 2008,
  month = dec,
  journal = {Lithos},
  volume = {106},
  number = {3},
  pages = {380--398},
  issn = {0024-4937},
  doi = {10.1016/j.lithos.2008.09.008},
  urldate = {2025-08-22}
}

@article{pallister_broad_2010,
  title = {Broad accommodation of rift-related extension recorded by dyke intrusion in {{Saudi Arabia}}},
  author = {Pallister, John S. and McCausland, Wendy A. and J{\'o}nsson, Sigurj{\'o}n and Lu, Zhong and Zahran, Hani M. and Hadidy, Salah El and Aburukbah, Abdallah and Stewart, Ian C. F. and Lundgren, Paul R. and White, Randal A. and Moufti, Mohammed R. H.},
  year = 2010,
  month = oct,
  journal = {Nature Geoscience},
  volume = {3},
  number = {10},
  pages = {705--712},
  publisher = {Nature Publishing Group},
  issn = {1752-0908},
  doi = {10.1038/ngeo966},
  urldate = {2026-06-19},
  copyright = {2010 Springer Nature Limited},
  langid = {english}
}

@article{parisi_first_2024,
  title = {The {{First Network}} of {{Ocean Bottom Seismometers}} in the {{Red Sea}} to {{Investigate}} the {{Zabargad Fracture Zone}}},
  author = {Parisi, Laura and Augustin, Nico and Trippanera, Daniele and Kirk, Henning and Dannowski, Anke and Matrau, R{\'e}mi and Fittipaldi, Margherita and Nobile, Adriano and Zielke, Olaf and Cano, Eduardo Valero and Hoogewerf, Guus and Aspiotis, Theodoros and {Manzo-Vega}, Sofia and Carmona, Armando Espindola and Barreto, Alejandra and Juchem, Marlin and Suhendi, Cahli and {Schmidt-Aursch}, Mechita and Mai, P. Martin and J{\'o}nsson, Sigurj{\'o}n},
  year = 2024,
  month = apr,
  journal = {Seismica},
  volume = {3},
  number = {1},
  issn = {2816-9387},
  doi = {10.26443/seismica.v3i1.729},
  urldate = {2026-06-19},
  copyright = {Copyright (c) 2024 Laura Parisi, Nico Augustin, Daniele Trippanera, Henning  Kirk, Anke Dannowski, R\'emi Matrau, Margherita Fittipaldi, Adriano  Nobile, Olaf Zielke, Eduardo Valero Cano, Guus Hoogewerf, Theodoros  Aspiotis, Sofia Manzo-Vega, Armando Espindola Carmona, Alejandra Barreto, Marlin  Juchem, Cahli Suhendi, Mechita Schmidt-Aursch, P. Martin Mai, Sigurj\'on J\'onsson},
  langid = {english}
}

@article{park_s_2008,
  title = {S wave velocity structure of the {{Arabian Shield}} upper mantle from {{Rayleigh}} wave tomography},
  author = {Park, Yongcheol and Nyblade, Andrew A. and Rodgers, Arthur J. and {Al-Amri}, Abdullah},
  year = 2008,
  journal = {Geochemistry, Geophysics, Geosystems},
  volume = {9},
  number = {7},
  issn = {1525-2027},
  doi = {10.1029/2007GC001895},
  urldate = {2026-04-13},
  copyright = {Copyright 2008 by the American Geophysical Union.},
  langid = {english}
}

@article{park_upper_2007,
  title = {Upper mantle structure beneath the {{Arabian Peninsula}} and northern {{Red Sea}} from teleseismic body wave tomography: {{Implications}} for the origin of {{Cenozoic}} uplift and volcanism in the {{Arabian Shield}}},
  shorttitle = {Upper mantle structure beneath the {{Arabian Peninsula}} and northern {{Red Sea}} from teleseismic body wave tomography},
  author = {Park, Yongcheol and Nyblade, Andrew A. and Rodgers, Arthur J. and {Al-Amri}, Abdullah},
  year = 2007,
  journal = {Geochemistry, Geophysics, Geosystems},
  volume = {8},
  number = {6},
  issn = {1525-2027},
  doi = {10.1029/2006GC001566},
  urldate = {2025-08-18},
  copyright = {Copyright 2007 by the American Geophysical Union.},
  langid = {english}
}

@article{pasyanos_case_2011,
  title = {A {{Case}} for the {{Use}} of {{3D Attenuation Models}} in {{Ground-Motion}} and {{Seismic-Hazard Assessment}}},
  author = {Pasyanos, Michael E.},
  year = 2011,
  month = aug,
  journal = {Bulletin of the Seismological Society of America},
  volume = {101},
  number = {4},
  pages = {1965--1970},
  issn = {0037-1106},
  doi = {10.1785/0120110004},
  urldate = {2025-08-19}
}

@article{pasyanos_improved_2021,
  title = {Improved lithospheric attenuation structure of the {{Arabian Peninsula}} through the use of national network data},
  author = {Pasyanos, Michael E. and Tarabulsi, Yahya M. and {Al-Hadidy}, Salah Y. and Raddadi, Wael W. and Mousa, Abdullah D. and {El-Hussain}, Issa and {Al-Jeri}, Farah and {Al-Shukri}, Haydar and G{\"o}k, Rengin},
  year = 2021,
  month = may,
  journal = {Arabian Journal of Geosciences},
  volume = {14},
  number = {10},
  pages = {914},
  issn = {1866-7538},
  doi = {10.1007/s12517-021-07294-x},
  urldate = {2023-05-18},
  langid = {english}
}

@article{pichon_miocene--present_2010,
  title = {The {{Miocene-to-Present Kinematic Evolution}} of the {{Eastern Mediterranean}} and {{Middle East}} and {{Its Implications}} for {{Dynamics}}},
  author = {Pichon, Xavier Le and Kreemer, Corn{\'e}},
  year = 2010,
  month = may,
  journal = {Annual Review of Earth and Planetary Sciences},
  volume = {38},
  number = {Volume 38, 2010},
  pages = {323--351},
  publisher = {Annual Reviews},
  issn = {0084-6597, 1545-4495},
  doi = {10.1146/annurev-earth-040809-152419},
  urldate = {2024-07-03},
  langid = {english}
}

@article{rashed_new_2020,
  title = {A new perspective of the 2009 {{Al-Ays}} earthquake episode at western {{Arabia}}},
  author = {Rashed, Mohamed and Zahran, Hani and Atef, Ali and Harbi, Hussein and {Al-Dhahry}, Maher and {Al-Hady}, Sherif},
  year = 2020,
  month = feb,
  journal = {Journal of Asian Earth Sciences},
  volume = {188},
  pages = {104101},
  issn = {1367-9120},
  doi = {10.1016/j.jseaes.2019.104101},
  urldate = {2025-08-26}
}

@article{rawlinson_simultaneous_2006,
  title = {Simultaneous inversion of active and passive source datasets for 3-{{D}} seismic structure with application to {{Tasmania}}},
  author = {Rawlinson, N. and Urvoy, M.},
  year = 2006,
  journal = {Geophysical Research Letters},
  volume = {33},
  number = {24},
  issn = {1944-8007},
  doi = {10.1029/2006GL028105},
  urldate = {2025-08-22},
  copyright = {Copyright 2006 by the American Geophysical Union.},
  langid = {english}
}

@article{reilinger_gps_2006,
  title = {{{GPS}} constraints on continental deformation in the {{Africa-Arabia-Eurasia}} continental collision zone and implications for the dynamics of plate interactions},
  author = {Reilinger, Robert and McClusky, Simon and Vernant, Philippe and Lawrence, Shawn and Ergintav, Semih and Cakmak, Rahsan and Ozener, Haluk and Kadirov, Fakhraddin and Guliev, Ibrahim and Stepanyan, Ruben and Nadariya, Merab and Hahubia, Galaktion and Mahmoud, Salah and Sakr, K. and ArRajehi, Abdullah and Paradissis, Demitris and {Al-Aydrus}, A. and Prilepin, Mikhail and Guseva, Tamara and Evren, Emre and Dmitrotsa, Andriy and Filikov, S. V. and Gomez, Francisco and {Al-Ghazzi}, Riad and Karam, Gebran},
  year = 2006,
  journal = {Journal of Geophysical Research: Solid Earth},
  volume = {111},
  number = {B5},
  issn = {2156-2202},
  doi = {10.1029/2005JB004051},
  urldate = {2024-07-03},
  copyright = {Copyright 2006 by the American Geophysical Union.},
  langid = {english}
}

@article{reilinger_nubiaarabiaeurasia_2011,
  title = {Nubia--{{Arabia}}--{{Eurasia}} plate motions and the dynamics of {{Mediterranean}} and {{Middle East}} tectonics},
  author = {Reilinger, Robert and McClusky, Simon},
  year = 2011,
  month = sep,
  journal = {Geophysical Journal International},
  volume = {186},
  number = {3},
  pages = {971--979},
  issn = {0956-540X},
  doi = {10.1111/j.1365-246X.2011.05133.x},
  urldate = {2024-07-03}
}

@article{rodgers_adjoint_2024,
  title = {Adjoint {{Waveform Tomography}} for {{Crustal}} and {{Upper Mantle Structure}} of the {{Middle East}} and {{Southwest Asia}} for {{Improved Waveform Simulations Using Openly Available Broadband Data}}},
  author = {Rodgers, Arthur J. and Krischer, Lion and Afanasiev, Michael and Boehm, Christian and Doody, Claire and Simmons, Nathan},
  year = 2024,
  month = feb,
  journal = {Bulletin of the Seismological Society of America},
  volume = {114},
  number = {3},
  pages = {1365--1391},
  issn = {0037-1106},
  doi = {10.1785/0120230248},
  urldate = {2024-03-06}
}

@article{rodgers_broadband_2008,
  title = {Broadband {{Waveform Modeling}} of {{Moderate Earthquakes}} in the {{San Francisco Bay Area}} and {{Preliminary Assessment}} of the {{USGS3D Seismic Velocity Model}}},
  author = {Rodgers, Arthur and Petersson, N. Anders and Nilsson, Stefan and Sj{\"o}green, Bj{\"o}rn and McCandless, Kathleen},
  year = 2008,
  month = apr,
  journal = {Bulletin of the Seismological Society of America},
  volume = {98},
  number = {2},
  pages = {969--988},
  issn = {0037-1106},
  doi = {10.1785/0120060407},
  urldate = {2025-08-18}
}

@article{royden_slab_2011,
  title = {Slab segmentation and late {{Cenozoic}} disruption of the {{Hellenic}} arc},
  author = {Royden, Leigh H. and Papanikolaou, Dimitrios J.},
  year = 2011,
  journal = {Geochemistry, Geophysics, Geosystems},
  volume = {12},
  number = {3},
  issn = {1525-2027},
  doi = {10.1029/2010GC003280},
  urldate = {2024-07-03},
  copyright = {Copyright 2011 by the American Geophysical Union.},
  langid = {english}
}

@article{royden_tectonic_1993,
  title = {The tectonic expression slab pull at continental convergent boundaries},
  author = {Royden, Leigh H.},
  year = 1993,
  journal = {Tectonics},
  volume = {12},
  number = {2},
  pages = {303--325},
  issn = {1944-9194},
  doi = {10.1029/92TC02248},
  urldate = {2025-08-22},
  copyright = {Copyright 1993 by the American Geophysical Union.},
  langid = {english}
}

@article{ruan_balancing_2019,
  ids = {ruan_balancing_nodate,ruan_balancing_nodate-1},
  title = {Balancing unevenly distributed data in seismic tomography: a global adjoint tomography example},
  shorttitle = {Balancing unevenly distributed data in seismic tomography},
  author = {Ruan, Youyi and Lei, Wenjie and Modrak, Ryan and {\"O}rsvuran, R{\i}dvan and Bozda{\u g}, Ebru and Tromp, Jeroen},
  year = 2019,
  month = nov,
  journal = {Geophysical Journal International},
  volume = {219},
  number = {2},
  pages = {1225--1236},
  publisher = {Oxford Academic},
  issn = {0956-540X},
  doi = {10.1093/gji/ggz356},
  urldate = {2020-08-29},
  langid = {english}
}

@article{sengor_tethyan_1981,
  title = {Tethyan evolution of {{Turkey}}: {{A}} plate tectonic approach},
  shorttitle = {Tethyan evolution of {{Turkey}}},
  author = {{\c S}eng{\"o}r, A. M. Cel{\^a}l and Yilmaz, Y{\"u}cel},
  year = 1981,
  month = jun,
  journal = {Tectonophysics},
  volume = {75},
  number = {3},
  pages = {181--241},
  issn = {0040-1951},
  doi = {10.1016/0040-1951(81)90275-4},
  urldate = {2023-05-22},
  langid = {english}
}

@article{simmons_spiral_2021,
  title = {{{SPiRaL}}: a multiresolution global tomography model of seismic wave speeds and radial anisotropy variations in the crust and mantle},
  shorttitle = {{{SPiRaL}}},
  author = {Simmons, N A and Myers, S C and Morency, C and Chiang, A and Knapp, D R},
  year = 2021,
  month = nov,
  journal = {Geophysical Journal International},
  volume = {227},
  number = {2},
  pages = {1366--1391},
  issn = {0956-540X},
  doi = {10.1093/gji/ggab277},
  urldate = {2023-07-31}
}

@article{sukhovich_seismic_2015,
  title = {Seismic monitoring in the oceans by autonomous floats},
  author = {Sukhovich, Alexey and Bonnieux, S{\'e}bastien and Hello, Yann and Irisson, Jean-Olivier and Simons, Frederik J. and Nolet, Guust},
  year = 2015,
  month = aug,
  journal = {Nature Communications},
  volume = {6},
  number = {1},
  pages = {8027},
  publisher = {Nature Publishing Group},
  issn = {2041-1723},
  doi = {10.1038/ncomms9027},
  urldate = {2025-08-23},
  copyright = {2015 The Author(s)},
  langid = {english}
}

@article{tan_homogeneous_2021,
  title = {A homogeneous earthquake catalogue for {{Turkey}}},
  author = {Tan, Onur},
  year = 2021,
  month = jul,
  journal = {Natural Hazards and Earth System Sciences},
  volume = {21},
  number = {7},
  pages = {2059--2073},
  publisher = {Copernicus GmbH},
  issn = {1561-8633},
  doi = {10.5194/nhess-21-2059-2021},
  urldate = {2025-08-22},
  langid = {english}
}

@article{tang_lithospheric_2016,
  title = {The lithospheric shear-wave velocity structure of {{Saudi Arabia}}: {{Young}} volcanism in an old shield},
  shorttitle = {The lithospheric shear-wave velocity structure of {{Saudi Arabia}}},
  author = {Tang, Zheng and Juli{\`a}, Jordi and Zahran, Hani and Mai, P. Martin},
  year = 2016,
  month = jun,
  journal = {Tectonophysics},
  volume = {680},
  pages = {8--27},
  issn = {0040-1951},
  doi = {10.1016/j.tecto.2016.05.004},
  urldate = {2024-09-25}
}

@article{tang_shear_2019,
  title = {Shear {{Velocity Structure Beneath Saudi Arabia From}} the {{Joint Inversion}} of {{P}} and {{S Wave Receiver Functions}}, and {{Rayleigh Wave Group Velocity Dispersion Data}}},
  author = {Tang, Zheng and Mai, P. Martin and Juli{\`a}, Jordi and Zahran, Hani},
  year = 2019,
  journal = {Journal of Geophysical Research: Solid Earth},
  volume = {124},
  number = {5},
  pages = {4767--4787},
  issn = {2169-9356},
  doi = {10.1029/2018JB017131},
  urldate = {2024-09-25},
  copyright = {\copyright 2019. The Authors.},
  langid = {english}
}

@article{tape_adjoint_2009,
  title = {Adjoint tomography of the southern {{California}} crust},
  author = {Tape, Carl and Liu, Qinya and Maggi, Alessia and Tromp, Jeroen},
  year = 2009,
  journal = {Science},
  volume = {325},
  number = {5943},
  pages = {988--992},
  doi = {10.1126/science.1175298},
  urldate = {2017-04-10}
}

@article{tarantola_inversion_1984,
  title = {Inversion of seismic reflection data in the acoustic approximation},
  author = {Tarantola, Albert},
  year = 1984,
  journal = {Geophysics},
  volume = {49},
  number = {8},
  pages = {1259--1266},
  doi = {10.1190/1.1441754},
  urldate = {2017-01-15}
}

@article{thrastarson_lasif_2021,
  title = {{{LASIF}}: {{LArge-scale}} seismic inversion framework, an updated version},
  shorttitle = {{{LASIF}}},
  author = {Thrastarson, S{\"o}lvi and {van Herwaarden}, Dirk-Philip and Krischer, Lion and Fichtner, Andreas},
  year = 2021,
  journal = {EarthArXiv},
  publisher = {California Digital Library (CDL)},
  doi = {10.31223/x5nc84},
  urldate = {2026-04-19}
}

@article{thrastarson_reveal_2024,
  title = {{{REVEAL}}: {{A Global Full}}-{{Waveform Inversion Model}}},
  shorttitle = {{{REVEAL}}},
  author = {Thrastarson, Solvi and {van Herwaarden}, Dirk-Philip and Noe, Sebastian and Josef Schiller, Carl and Fichtner, Andreas},
  year = 2024,
  month = apr,
  journal = {Bulletin of the Seismological Society of America},
  issn = {0037-1106},
  doi = {10.1785/0120230273},
  urldate = {2024-04-17}
}

@article{tromp_near_2010,
  ids = {tromp_near_2010-1},
  title = {Near real-time simulations of global {{CMT}} earthquakes},
  author = {Tromp, Jeroen and Komatitsch, Dimitri and Hj{\"o}rleifsd{\'o}ttir, Vala and Liu, Qinya and Zhu, Hejun and Peter, Daniel and Bozdag, Ebru and McRitchie, Dennis and Friberg, Paul and Trabant, Chad and Hutko, Alex},
  year = 2010,
  month = oct,
  journal = {Geophysical Journal International},
  volume = {183},
  number = {1},
  pages = {381--389},
  publisher = {Oxford Academic},
  issn = {0956-540X},
  doi = {10.1111/j.1365-246X.2010.04734.x},
  urldate = {2020-12-01},
  langid = {english}
}

@article{tromp_seismic_2005,
  title = {Seismic tomography, adjoint methods, time reversal and banana-doughnut kernels: {{Seismic}} tomography, adjoint methods, time reversal and banana-doughnut kernels},
  shorttitle = {Seismic tomography, adjoint methods, time reversal and banana-doughnut kernels},
  author = {Tromp, Jeroen and Tape, Carl and Liu, Qinya},
  year = 2005,
  month = dec,
  journal = {Geophysical Journal International},
  volume = {160},
  number = {1},
  pages = {195--216},
  issn = {0956540X, 1365246X},
  doi = {10.1111/j.1365-246X.2004.02453.x},
  urldate = {2016-09-10},
  langid = {english}
}

@article{van_herwaarden_full-waveform_2023,
  title = {Full-{{Waveform Tomography}} of the {{African Plate Using Dynamic Mini-Batches}}},
  author = {{van Herwaarden}, Dirk-Philip and Thrastarson, Solvi and Hapla, Vaclav and Afanasiev, Michael and Trampert, Jeannot and Fichtner, Andreas},
  year = 2023,
  journal = {Journal of Geophysical Research: Solid Earth},
  volume = {128},
  number = {6},
  pages = {e2022JB026023},
  issn = {2169-9356},
  doi = {10.1029/2022JB026023},
  urldate = {2026-01-30},
  copyright = {\copyright{} 2023. The Authors.},
  langid = {english}
}

@article{viltres_present-day_2022,
  title = {Present-{{Day Motion}} of the {{Arabian Plate}}},
  author = {Viltres, Renier and J{\'o}nsson, Sigurj{\'o}n and Alothman, Abdulaziz O. and Liu, Shaozhuo and Leroy, Sylvie and Masson, Fr{\'e}d{\'e}ric and Doubre, C{\'e}cile and Reilinger, Robert},
  year = 2022,
  journal = {Tectonics},
  volume = {41},
  number = {3},
  pages = {e2021TC007013},
  issn = {1944-9194},
  doi = {10.1029/2021TC007013},
  urldate = {2026-06-19},
  copyright = {\copyright{} 2022. American Geophysical Union. All Rights Reserved.},
  langid = {english}
}

@article{virieux_overview_2009,
  title = {An overview of full-waveform inversion in exploration geophysics},
  author = {Virieux, J. and Operto, S.},
  year = 2009,
  month = nov,
  journal = {Geophysics},
  volume = {74},
  number = {6},
  pages = {WCC1-WCC26},
  issn = {0016-8033, 1942-2156},
  doi = {10.1190/1.3238367},
  urldate = {2017-05-17},
  langid = {english}
}

@article{waldhauser_high-resolution_2002,
  title = {High-resolution teleseismic tomography of upper-mantle structure using an a priori three-dimensional crustal model},
  author = {Waldhauser, Felix and Lippitsch, Regina and Kissling, Edi and Ansorge, J{\"o}rg},
  year = 2002,
  month = aug,
  journal = {Geophysical Journal International},
  volume = {150},
  number = {2},
  pages = {403--414},
  issn = {0956-540X},
  doi = {10.1046/j.1365-246X.2002.01690.x},
  urldate = {2025-08-22}
}

@article{weinstein_role_2006,
  title = {The {{Role}} of {{Lithospheric Mantle Heterogeneity}} in the {{Generation}} of {{Plio-Pleistocene Alkali Basaltic Suites}} from {{NW Harrat Ash Shaam}} ({{Israel}})},
  author = {Weinstein, Y. and Navon, O. and Altherr, R. and Stein, M.},
  year = 2006,
  month = may,
  journal = {Journal of Petrology},
  volume = {47},
  number = {5},
  pages = {1017--1050},
  issn = {0022-3530},
  doi = {10.1093/petrology/egl003},
  urldate = {2026-06-19}
}

@article{wessel_generic_2019,
  title = {The {{Generic Mapping Tools Version}} 6},
  author = {Wessel, P. and Luis, J. F. and Uieda, L. and Scharroo, R. and Wobbe, F. and Smith, W. H. F. and Tian, D.},
  year = 2019,
  month = nov,
  journal = {Geochemistry, Geophysics, Geosystems},
  volume = {20},
  number = {11},
  pages = {5556--5564},
  publisher = {John Wiley \& Sons, Ltd},
  issn = {1525-2027},
  doi = {10.1029/2019GC008515},
  urldate = {2025-09-05},
  langid = {english}
}

@article{woodhouse_amplitude_1986,
  title = {Amplitude, phase and path anomalies of mantle waves},
  author = {Woodhouse, J. H. and Wong, Y. K.},
  year = 1986,
  month = dec,
  journal = {Geophysical Journal International},
  volume = {87},
  number = {3},
  pages = {753--773},
  issn = {0956-540X},
  doi = {10.1111/j.1365-246X.1986.tb01970.x},
  urldate = {2019-06-26},
  langid = {english}
}

@article{wortel_subduction_2000,
  title = {Subduction and {{Slab Detachment}} in the {{Mediterranean-Carpathian Region}}},
  author = {Wortel, M. J. R. and Spakman, W.},
  year = 2000,
  month = dec,
  journal = {Science},
  volume = {290},
  number = {5498},
  pages = {1910--1917},
  publisher = {American Association for the Advancement of Science},
  doi = {10.1126/science.290.5498.1910},
  urldate = {2023-05-23}
}

@article{zhu_seismic_2015,
  title = {Seismic structure of the {{European}} upper mantle based on adjoint tomography},
  author = {Zhu, Hejun and Bozda{\u g}, Ebru and Tromp, Jeroen},
  year = 2015,
  month = apr,
  journal = {Geophysical Journal International},
  volume = {201},
  number = {1},
  pages = {18--52},
  issn = {0956-540X},
  doi = {10.1093/gji/ggu492},
  urldate = {2019-03-26},
  langid = {english}
}

@misc{seis_network_2H,
  doi = {10.7914/SN/2H_2009},
  url = {https://www.fdsn.org/networks/detail/2H_2009/},
  author = {{Derek Keir} and {James O.S. Hammond}},
  title = {AFAR0911},
  publisher = {International Federation of Digital Seismograph Networks},
  year = {2009}
}

@misc{seis_network_4H,
  doi = {10.7914/SN/4H_2011},
  url = {https://www.fdsn.org/networks/detail/4H_2011/},
  author = {{James Hammond} and {Berhe Goitom} and {John Michael Kendall} and {Ghebrebrhan Ogubazghi}},
  title = {Nabro Urgency Array},
  publisher = {International Federation of Digital Seismograph Networks},
  year = {2011}
}

@misc{seis_network_5H,
  doi = {10.7914/SN/5H_2011},
  url = {https://www.fdsn.org/networks/detail/5H_2011/},
  author = {{James Hammond} and {Berhe Goitom} and {John Michael Kendall} and {Ghebrebrhan Ogubazghi}},
  title = {Eritrea Seismic Project},
  publisher = {International Federation of Digital Seismograph Networks},
  year = {2011}
}

@misc{seis_network_6G,
  doi = {10.7914/SN/6G_2013},
  url = {https://www.fdsn.org/networks/detail/6G_2013/},
  author = {{Prof. Semih Ergintav} and {Mustafa Comoglu} and {Nurcan Meral Özel}},
  title = {MARsite Project},
  publisher = {International Federation of Digital Seismograph Networks},
  year = {2013}
}

@misc{seis_network_AC,
  doi = {10.7914/SN/AC},
  url = {https://www.fdsn.org/networks/detail/AC/},
  author = {{Institute of GeoSciences, Polytechnic University of Tirana}},
  title = {Albanian Seismological Network},
  publisher = {International Federation of Digital Seismograph Networks},
  year = {2002}
}

@misc{seis_network_AF,
  doi = {10.7914/SN/AF},
  url = {https://www.fdsn.org/networks/detail/AF/},
  author = {{Penn State University}},
  title = {AfricaArray},
  publisher = {International Federation of Digital Seismograph Networks},
  year = {2004}
}

@misc{seis_network_CB,
  doi = {10.7914/SN/CB},
  url = {https://www.fdsn.org/networks/detail/CB/},
  author = {{Institute of Geophysics China Earthquake Administration (IGPCEA)}},
  title = {China National Seismic Network, Data Management Centre of China National Seismic Network at Institute of Geophysics, CEA},
  publisher = {International Federation of Digital Seismograph Networks},
  year = {2000}
}

@misc{seis_network_CR,
  doi = {10.7914/SN/CR},
  url = {https://www.fdsn.org/networks/detail/CR/},
  author = {{University of Zagreb}},
  title = {Croatian Seismograph Network},
  publisher = {International Federation of Digital Seismograph Networks},
  year = {2001}
}

@misc{seis_network_G,
  doi = {10.18715/GEOSCOPE.G},
  url = {http://geoscope.ipgp.fr/networks/detail/G/},
  author = {{Institut de physique du globe de Paris (IPGP)} and {École et Observatoire des Sciences de la Terre de Strasbourg (EOST)}},
  language = {en},
  title = {GEOSCOPE, French Global Network of broad band seismic stations},
  publisher = {Institut de physique du globe de Paris (IPGP), Université de Paris},
  year = {1982},
  copyright = {Creative Commons Attribution 4.0 International}
}

@misc{seis_network_GE,
  doi = {10.14470/TR560404},
  url = {http://geofon.gfz-potsdam.de/doi/network/GE},
  author = {{GEOFON Data Centre}},
  language = {en},
  title = {GEOFON Seismic Network},
  publisher = {Deutsches GeoForschungsZentrum GFZ},
  year = {1993}
}

@misc{seis_network_GO,
  doi = {10.7914/1YX9-8844},
  url = {https://www.fdsn.org/networks/detail/GO/},
  author = {{Ilia State University- Seismic Monitoring Centre of Georgia}},
  title = {National Seismic Network of Georgia},
  publisher = {International Federation of Digital Seismograph Networks},
  year = {1988}
}

@misc{seis_network_HL,
  doi = {10.7914/SN/HL},
  url = {https://www.fdsn.org/networks/detail/HL/},
  author = {{National Observatory of Athens, Institute of Geodynamics, Athens}},
  title = {National Observatory of Athens Seismic Network},
  publisher = {International Federation of Digital Seismograph Networks},
  year = {1975}
}

@misc{seis_network_HT,
  doi = {10.7914/SN/HT},
  url = {https://www.fdsn.org/networks/detail/HT/},
  author = {{Aristotle University of Thessaloniki}},
  title = {Aristotle University of Thessaloniki Seismological Network},
  publisher = {International Federation of Digital Seismograph Networks},
  year = {1981}
}

@misc{seis_network_II,
  doi = {10.7914/SN/II},
  url = {https://www.fdsn.org/networks/detail/II/},
  author = {{Scripps Institution of Oceanography}},
  title = {Global Seismograph Network - IRIS/IDA},
  publisher = {International Federation of Digital Seismograph Networks},
  year = {1986}
}

@misc{seis_network_IJ,
  doi = {10.7914/SN/IJ},
  url = {https://www.fdsn.org/networks/detail/IJ/},
  author = {{Kandilli Observatory And Earthquake Research Institute, Boğaziçi University}},
  title = {Istanbul University Seismic Network},
  publisher = {International Federation of Digital Seismograph Networks},
  year = {2013}
}

@misc{seis_network_IU,
  doi = {10.7914/SN/IU},
  url = {https://www.fdsn.org/networks/detail/IU/},
  author = {{Albuquerque Seismological Laboratory/USGS}},
  title = {Global Seismograph Network (GSN - IRIS/USGS)},
  publisher = {International Federation of Digital Seismograph Networks},
  year = {2014}
}

@misc{seis_network_KN,
  doi = {10.7914/SN/KN},
  url = {https://www.fdsn.org/networks/detail/KN/},
  author = {{Kyrgyz Institute of Seismology, IVTAN/KIS } and {University of California, San Diego}},
  title = {Kyrgyz Seismic Telemetry Network},
  publisher = {International Federation of Digital Seismograph Networks},
  year = {1991}
}

@misc{seis_network_KO,
  doi = {10.7914/SN/KO},
  url = {https://www.fdsn.org/networks/detail/KO/},
  author = {{Kandilli Observatory And Earthquake Research Institute, Boğaziçi University}},
  title = {Kandilli Observatory And Earthquake Research Institute (KOERI)},
  publisher = {International Federation of Digital Seismograph Networks},
  year = {1971}
}

@misc{seis_network_KR,
  doi = {10.7914/SN/KR},
  url = {https://www.fdsn.org/networks/detail/KR/},
  author = {{Kyrgyz Institute of Seismology, KIS }},
  title = {Kyrgyz Digital Network},
  publisher = {International Federation of Digital Seismograph Networks},
  year = {2007}
}

@misc{seis_network_KZ,
  doi = {10.7914/SN/KZ},
  url = {https://www.fdsn.org/networks/detail/KZ/},
  author = {{KNDC/Institute of Geophysical Research (Kazakhstan)}},
  title = {Kazakhstan Network},
  publisher = {International Federation of Digital Seismograph Networks},
  year = {1994}
}

@misc{seis_network_MN,
  doi = {10.13127/SD/FBBBTDTD6Q},
  url = {https://eida.ingv.it/en/network/MN},
  author = {{MedNet Project Partner Institutions}},
  title = {Mediterranean Very Broadband Seismographic Network (MedNet)},
  publisher = {Istituto Nazionale di Geofisica e Vulcanologia (INGV)},
  year = {1990},
  copyright = {Open Access}
}

@misc{seis_network_MP,
  doi = {10.7914/SN/MP},
  url = {https://www.fdsn.org/networks/detail/MP/},
  author = {{Seismological Laboratory of University of Basrah}},
  title = {Iraqi Seismic Observatory},
  publisher = {International Federation of Digital Seismograph Networks},
  year = {2014}
}
\bibliographystyle{gji}

\label{lastpage}

\clearpage
\appendix
\section{Data}
\label{appendix:data}

We have used 956 stations from 53 seismic networks whose data are
publicly available from EarthScope and local networks such as Kandilli
Observatory of T\"urkiye \citep{seis_network_2H, seis_network_4H,
  seis_network_5H, seis_network_6G, seis_network_AC, seis_network_AF,
  seis_network_CB, seis_network_CR, seis_network_G, seis_network_GE,
  seis_network_GO, seis_network_HL, seis_network_HT, seis_network_II,
  seis_network_IJ, seis_network_IU, seis_network_KN, seis_network_KO,
  seis_network_KR, seis_network_KZ, seis_network_MN, seis_network_MP,
  seis_network_OE, seis_network_ZH, seis_network_RO, seis_network_SL,
  seis_network_TJ, seis_network_TU, seis_network_XA, seis_network_XH,
  seis_network_XM, seis_network_XP, seis_network_XS, seis_network_XW,
  seis_network_XZ, seis_network_Y6, seis_network_YB, seis_network_YH,
  seis_network_YL, seis_network_YY, seis_network_ZE,
  seis_network_ZK}. You can find the full list of events and stations
on Zenodo
(\href{https://doi.org/10.5281/zenodo.19534092}{https://doi.org/10.5281/zenodo.19534092}).

\end{document}